%% file: qc.tex
\input epsf\input psfig\magnification=1000
\def\magnification=1#100{} 
\input harvmac \hsize=6true in\vsize=8.5true in \parskip=4pt plus 2pt minus 1pt
\input symbols \loadBigten{1200} \def\normalpoint{\Bigpoint \normalbaselines}
\normalpoint\nopagenumbers \def\Ninepoint{\ninepoint \baselineskip=11pt}
\def\Tenpoint{\tenpoint \baselineskip=12pt} \def\captionpoint{\Tenpoint}
\hoffset=3truemm\voffset=15truemm
\pageno=0\leftline{ADP-95-11/M28, gr-qc/0101003}\bigskip
\noindent in B. Robson, N. Visvanathan and W.S. Woolcock (eds.),
{\it ``Cosmology: The Physics of the Universe''}, Proceedings of the 8th
Physics Summer School, Australian National University, Canberra, Australia,
16 January -- 3 February, 1995, (World Scientific, Singapore, 1996),
pp.\ 473-531.\vfil
\noindent {\it Note added}: These lecture notes, as published above, date
from 1995. For more recent work on the issue of the prediction of the period
of inflation and the debate about boundary condition proposals (\S5.1), see:\br
N. Kontoleon and D.L. Wiltshire, \PR{D59} (1999) 063513 [= gr-qc/9807075].\br
D.L. Wiltshire, \GRG{32} (2000) 515 [= gr-qc/9905090].\br
M.J.W. Hall, K. Kumar, and M. Reginatto, \JP{A36} (2003) 9779
[= hep-th/0307259].
\eject \footline={\hss\tenrm\folio\hss}
\def\refskip{\leftskip=7truemm \parindent=20pt} \def\footpoint{\Ninepoint}
\def\pr{\par\vskip-2pt\noindent} \def\bomb{\omit&\hbox to5mm{\dotfill}\ }

\font\tenes=eusm10 \TpT \font\sevenes=eusm7 \TpT \font\fivees=eusm5 \TpT
\newfam\esfam \textfont\esfam=\tenes \scriptfont\esfam=\sevenes
\scriptscriptfont\esfam=\fivees \mathchardef\Cop"0943 \mathchardef\Rop"0952
\mathchardef\Vv"0B56 \mathchardef\pH"0B1E \mathchardef\PI"0B19
\mathchardef\DD"0A44 \mathchardef\Pt"0B40 \mathchardef\a"0B61
\mathchardef\scri"0C49 \mathchardef\JJ"0C4A \mathchardef\Zop"095A
\mathchardef\euA"0C41 \mathchardef\euB"0C42 \def\Rn{\mathchar"0B52_n}
\mathchardef\HYP"0B06 \mathchardef\pS"0B20 \mathchardef\xx"0B78 \def\Jn{\JJ_n}
\mathchardef\SS"0B53 \def\Sn{\SS_n} \def\One{\,^*\mathchar"0B31}
\mathchardef\cc"0B63\mathchardef\ii"0B69\def\cn{\cc_n}\def\PSB{\overline{\PS}}
\mathchardef\uu"0B55 \mathchardef\ff"0B66 \mathchardef\hh"0B48
\def\psuf{\ph\Ns{suff}}
\def\PP{{\cal P}} \def\Ta{a}\def\Tp{\ph} \def\Iz{{\rm I}\Z0}\def\Jz{{\rm J}\Z0}
\def\V{{\cal V}} \def\GG{{\cal G}} \def\HH{{\cal H}} \def\NN{{\cal N}}
\def\UU{{\cal U}}\def\qA{q\W A}\def\dqA{\dot\qA}\def\qB{q\W B}\def\dqB{\dot\qB}
\def\gABC{\mathchar"0B00\,\UD{\raise2pt\hbox{$\scrscr A$}}{\lower1pt\hbox{$
\scrscr BC$}}} \def\dOM{\left(\dd\th^2+\sin^2\th\dd\varphi^2\right)}
\def\hS#1{\hs\vphantom{#1}} 
\def\PIPH{\PI_{\lower.5pt\hbox{$\scriptscriptstyle\PH$}}} \def\PIz{\PI^0}
\def\ric#1{\,\!^#1\!R} \def\KA{\ka^2} \def\hz{\hat0} \def\hi{\hat\imath}
\def\qC{q\W C} \def\CC{{\cal C}} \def\An{{\cal A}_n} \def\PSn{\PS_n}

\lref\HalA{J. J. Halliwell, in {\it Quantum Cosmology and Baby Universes}, eds.
S. Coleman, J. B. Hartle T. Piran and S. Weinberg, (World Scientific,
Singapore, 1991), p.\ 159.}
\lref\PagA{D. N. Page, in {\it Proceedings of the Banff Summer Institute on
Gravitation, August 1990}, eds. R. B. Mann and P. S. Wesson, (World Scientific,
Singapore, 1991).}
\lref\vaNi{P. van Nieuwenhuizen, Phys.\ Rep.\ {\bf 68} (1981) 189.}
\lref\West{P. West, {\it Introduction to Supersymmetry and Supergravity},
(World Scientific, Singapore, 1986).}
\lref\SaSa{A. Salam and E. Sezgin (eds.), {\it Supergravities in Diverse
Dimensions}, (World Scientific, Singapore, 1989).}
\lref\Espo{G. Esposito, {\it Quantum Gravity, Quantum Cosmology and Lorentzian
Geometries}, (Springer {\it Lecture Notes in Physics} {\bf m12}, Berlin,
1992).}
\lref\GoSa{M. H. Goroff and A. Sagnotti, \PL{160B} (1985) 81; \NP{B266} (1986)
709.}
\lref\tHVe{G. 't Hooft and M. Veltman, Ann.\ Inst.\ H.\ Poincar\'e {\bf20}
(1974) 69.}
\lref\DeNi{S. Deser and P. van Nieuwenhuizen, \PR{D10} (1974) 411.}
\lref\DeNT{S. Deser, P. van Nieuwenhuizen and H. S. Tsao, \PR{D10} (1974) 3337.}
\lref\DeKS{S. Deser, J. H. Kay and K. S. Stelle, \PRL{38} (1977) 527.}
\lref\HoSt{P. S. Howe and K. S. Stelle, \IJMP{A4} (1989) 1871.}
\lref\GSW{M. B. Green, J. H. Schwarz and E. Witten, {\it Superstring theory},
(Cambridge University Press, 1987).}
\lref\GrPe{D. J. Gross and V. Periwal, \PRL{60} (1988) 2105.}
\lref\Da{S. Davis, Preprint DAMTP-R/94/27-rev (1994), hep-th/9503231.}
\lref\AshA{A. Ashtekar, {\it Lectures on Non-perturbative Canonical Gravity},
(World Scientific, Singapore, 1991).}
\lref\Rov{C. Rovelli, \CQG{8} (1991) 1613.}
\lref\Sen{A. Sen, \JMP{22} (1981) 1781.}
\lref\AshB{A. Ashtekar, \PRL{57} (1986) 2244; \PR{D36} (1987) 1587.}
\lref\RoSm{C. Rovelli and L. Smolin, \PRL{61} (1989) 1155; \NP{B331} (1990)
80.}
\lref\Blen{M. P. Blencowe, \NP{B341} (1990) 213.}
\lref\Kod{H. Kodama, \PTP{80} (1988) 1024; \PR{D42} (1990) 2548.}
\lref\SoCh{C. Soo and L. N. Chang, \IJMP{D3} (1994) 529.}
\lref\SmSo{L. Smolin and C. Soo, \NP{B449} (1995) 289.}
\lref\LouB{J. Louko, \PR{D51} (1995) 586.}
\lref\Re{T. Regge, \NC{19} (1961) 558.}
\lref\WiTu{R. M. Williams and P. A. Tuckey, \CQG{9} (1992) 1409.}
\lref\BLMS{L. Bombelli, J. Lee, D. Meyer and R. D. Sorkin, \PRL{59} (1987)
521.}
\lref\Ish{C. J. Isham, \CQG{6} (1989) 1509.}
\lref\IsKR{C. J. Isham, Y. Kubyshin and P. Renteln, \CQG{7} (1990) 1053.}
\lref\WhA{J. A. Wheeler, in {\it Relativity, Groups and Topology}, eds. C.
DeWitt and B. S. DeWitt, (Gordon and Breach, New York, 1963), p.\ 315.}
\lref\WhB{J. A. Wheeler, in {\it Batelles Rencontres}, eds.\ C. DeWitt and J.
A. Wheeler, (Benjamin, New York, 1968), p.\ 242.}
\lref\DeW{B. S. DeWitt, \PR{160} (1967) 1113.}
\lref\MisA{C. W. Misner, \PR{186} (1969) 1319; in {\it Relativity}, eds.
M. Carmeli, Finkler and L. Witten, (Plenum, New York, 1970), p.\ 55.}
\lref\MisB{C. W. Misner, in {\it Magic Without Magic: John Archibald Wheeler},
ed. J. Klauder, (W.H. Freeman, San Francisco, 1972), p.\ 441.}
\lref\HawA{S. W. Hawking, in {\it Astrophysical Cosmology}, eds.\ H. A.
Br\"uck, G. V. Coyne and M. S. Longair, (Pontifica Academia Scientarium {\it
Scripta Varia} {\bf48}, Vatican City, 1982) p.\ 563.}
\lref\HarHaA{J. B. Hartle and S. W. Hawking, \PR{D28} (1983) 2960.}
\lref\HawC{S. W. Hawking, \NP{B239} (1984) 257.}
\lref\VilA{A. Vilenkin, \PL{117B} (1982) 25; \PR{D27} (1983) 2848.}
\lref\VilB{A. Vilenkin, \PR{D30} (1984) 509; \NP{B252} (1985) 141.}
\lref\Vilc{A. Vilenkin, \PR{D33} (1986) 3560.}
\lref\VilC{A. Vilenkin, \PR{D37} (1988) 888.}
\lref\VilD{A. Vilenkin, \PR{D50} (1994) 2581.}
\lref\VilE{A. Vilenkin, \PR{D39} (1989) 1116.}
\lref\VilF{A. Vilenkin, \PRL{74} (1995) 846.}
\lref\VilG{A. Vilenkin, in {\it Proceedings of the
International School of Astrophysics `D. Chalonge', Eriche, 1995}, ed.
N. Sanchez, (Kluwer, Dordrecht), to appear.}
\lref\Lin{A. D. Linde, Zh.\ Eksp.\ Teor.\ Fiz.\ {\bf87} (1984) 369 [Sov.\
Phys.\ JETP {\bf60} (1984) 211]; Lett.\ \NC{39} (1984) 401; Rep.\ Prog.\ Phys.
{\bf47} (1984) 925.}
\lref\HarA{J. B. Hartle, in {\it Quantum Cosmology and Baby Universes}, eds.
S. Coleman, J. B. Hartle T. Piran and S. Weinberg, (World Scientific,
Singapore, 1991), p.\ 65.}
\lref\HaPMZu{J. J. Halliwell, J. Perez-Mercader, and W. H. Zurek (eds.), {\it
The Physical Origins of Time Asymmetry}, (Cambridge University Press, 1994).}
\lref\St{A. Strominger, in {\it Quantum Cosmology and Baby Universes}, eds.
S. Coleman, J. B. Hartle, T. Piran and S. Weinberg, (World Scientific,
Singapore, 1991), p.\ 269.}
\lref\CsGr{A. Csord\'as and R. Graham, \PRL{74} (1995) 4129.}
\lref\Wald{R. M. Wald, {\it General Relativity}, (Chicago University Press,
1984).}
\lref\Yor{J. W. York, \PRL{28} (1972) 1082.}
\lref\DirA{P. A. M. Dirac, \PRS{A246} (1958) 326; 333.}
\lref\HaRT{A. Hanson, T. Regge and C. Teitelboim, {\it Constrained Hamiltonian
Systems}, (Accademia Nationale dei Lincei, Rome, 1976).}
\lref\Higgs{P. W. Higgs, \PRL{1} (1958) 373.}
\lref\HaPaA{S. W. Hawking and D. N. Page, \NP{B264} (1986) 185.}
\lref\LouA{J. Louko, \AP{N.Y.}{181} (1988) 318.}
\lref\GiHaA{G. W. Gibbons and S. W. Hawking, \PR{D15} (1977) 2752.}
\lref\HawB{S. W. Hawking, in {\it General Relativity: An Einstein Centenary
Survey} (Cambridge University Press, 1979), p.\ 746.}
\lref\HarHaB{J. B. Hartle and S. W. Hawking, \PR{D13} (1976) 2188.}
\lref\GiHaB{G. W. Gibbons and S. W. Hawking, \PR{D15} (1977) 2738.}
\lref\GiPeA{G. W. Gibbons and M. J. Perry, \PRS{A358} (1978) 467.}
\lref\EgHa{T. Eguchi and A. J. Hanson, \PL{74B} (1978) 249.}
\lref\GiHaC{G. W. Gibbons and S. W. Hawking, \PL{78B} (1978) 430; \CMP{66}
(1979) 291.}
\lref\GiPo{G. W. Gibbons and C. N. Pope, \CMP{66} (1979) 267.}
\lref\GibA{G. W. Gibbons, in {\it Fields and Geometry}, ed.\ A. Jadczyk,
(World Scientific, Singapore, 1986).}
\lref\GaSt{D. Garfinkle and A. Strominger, \PL{256B} (1991) 146.}
\lref\DGGH{H. F. Dowker, J. P. Gauntlett, S. B. Giddings and G. T. Horowitz,
\PR{D50} (1994) 2662.}
\lref\GiHaPe{G. W. Gibbons, S. W. Hawking and M. J. Perry, \NP{B138} (1978)
141.}
\lref\Schl{K. Schleich, \PR{D36} (1987) 2342.}
\lref\HalHarA{J. J. Halliwell and J. B. Hartle, \PR{D43} (1991) 1170.}
\lref\KoCh{A. S. Kompaneets and A. S. Chernov, Zh.\ Eksp.\ Teor.\ Fiz.\ {\bf47}
(1964) 1939 [Sov.\ Phys.\ JETP {\bf20} (1965) 1303].}
\lref\KaSa{R. Kantowski and R. K. Sachs, \JMP{7} (1966) 443.}
\lref\KuRy{K. V. Kucha\v{r} and M. P. Ryan, \PR{D40} (1989) 3982.}
\lref\HuSi{S. Sinha and B. L. Hu, \PR{D44} (1991) 1028.}
\lref\Mazz{F. D. Mazzitelli, \PR{D46} (1992) 4758.}
\lref\IsIs{A. Ishikawa and T. Isse, \MPL{A8} (1993) 3413.}
\lref\MacC{M. A. H. MacCallum, in {\it General Relativity: An Einstein
Centenary Survey} eds. S. W. Hawking and W. Israel, (Cambridge University
Press, 1979), p.\ 533.}
\lref\Bian{L. Bianchi, Mem.\ di Mat.\ Soc.\ Ital.\ Sci.\ {\bf11} (1897) 267.}
\lref\MisC{C. W. Misner, \PRL{22} (1969) 1071.}
\lref\BeLiKh{V. A. Belinskii, E. M. Lifschitz and I. M. Khalatnikov, Usp.\
Fiz.\ Nauk.\ {\bf102} (1970) 463 [Sov.\ Phys.\ Usp.\ {\bf13} (1970) 745].}
\lref\LaMuCo{A. Latifi, M. Musette and R. Conte, \PL{194A} (1994) 83.}
\lref\HalB{J. J. Halliwell, \PR{D38} (1988) 2468.}
\lref\HalC{J. J. Halliwell, \PR{D36} (1987) 3626.}
\lref\HalD{J. J. Halliwell, \PR{D39} (1989) 2912.}
\lref\HaLa{S. Habib and R. Laflamme, \PR{D42} (1990) 4056.}
\lref\CaGo{E. Calzetta and J. J. Gonzalez, \PR{D51} (1995) 6821.}
\lref\Cal{E. Calzetta, \PR{D43} (1991) 2498.}
\lref\HuPS{B. L. Hu, J. P. Paz and S. Sinha, in {\it Directions in General
Relativity, Vol.\ I}, eds. B. L. Hu, M. P. Ryan and C. V. Vishveshwara,
(Cambridge University Press, 1993), p.\ 145.}
\lref\Hu{B. L. Hu, in [\HaPMZu], p.\ 475.}
\lref\MosA{I. G. Moss, \AIHP{49} (1988) 341.}
\lref\PagB{D. N. Page, \JMP{32} (1991) 3427.}
\lref\Ban{T. Banks, \NP{B249} (1985) 332.}
\lref\Kim{S. P. Kim, \PR{D52} (1995) 3382.}
\lref\KoTu{E. W. Kolb and M. S. Turner, {\it The Early Universe},
(Addison-Wesley, Reading, Mass., 1990).}
\lref\Wilt{D. L. Wiltshire, \PR{D36} (1987) 1634.}
\lref\Hen{M. Henneaux, Lett. \NC{38} (1983) 609.}
\lref\GHSt{G. W. Gibbons, S. W. Hawking and J. M. Stewart, \NP{B281} (1987)
736.}
\lref\KaVi{D. J. Kaup and A. P. Vitello, \PR{D9} (1974) 1648.}
\lref\HaPaB{S. W. Hawking and D. N. Page, \NP{B298} (1988) 789.}
\lref\HaPaC{S. W. Hawking and D. N. Page, \PR{D42} (1990) 2655.}
\lref\SuYo{W. M. Suen and K. Young, \PR{D39} (1989) 2201.}
\lref\CoZe{H. D. Conradi and H. D. Zeh, \PL{154A} (1991) 321.}
\lref\Con{H. D. Conradi, \PR{D46} (1992) 612; \CQG{12} (1995) 2423.}
\lref\HalLoA{J. J. Halliwell and J. Louko, \PR{D39} (1989) 2206.}
\lref\HalLoB{J. J. Halliwell and J. Louko, \PR{D40} (1989) 1868.}
\lref\HalLoC{J. J. Halliwell and J. Louko, \PR{D42} (1990) 3997.}
\lref\HalHarB{J. J. Halliwell and J. B. Hartle, \PR{D41} (1990) 1815.}
\lref\Kuch{K. V. Kucha\v{r}, \JMP{22} (1981) 2640.}
\lref\LoVa{J. Louko and T. Vachaspati, \PL{223B} (1989) 21.}
\lref\Ryan{M. P. Ryan, {\it Hamiltonian Cosmology} (Springer {\it Lecture Notes
in Physics} {\bf 13}, Berlin, 1972).}
\lref\Teit{C. Teitelboim, \PRL{38} (1977) 1106.}
\lref\TabTe{R. Tabensky and C. Teitelboim, \PL{69B} (1977) 453.}
\lref\MOR{A. Mac\'{\i}as, O. Obreg\'on and M. P. Ryan, \CQG{4} (1987) 1477.}
\lref\DeHu{P. D. D'Eath and D. I. Hughes, \PL{214B} (1988) 498; \NP{B378}
(1992) 381.}
\lref\Grah{R. Graham, \PRL{67} (1991) 1381; \PL{277B} (1992) 393.}
\lref\DeHO{P. D. D'Eath, S. W. Hawking and O. Obreg\'on, \PL{300B} (1993) 44.}
\lref\PagC{D. N. Page, \PR{D34} (1986) 2267.}
\lref\PagD{D. N. Page, in {\it Quantum Concepts in Space and Time}, eds.
R. Penrose and C. J. Isham, (Oxford University Press, 1986) p.\ 274.}
\lref\Lyo{G. W. Lyons, \PR{D46} (1992) 1546.}
\lref\GiGr{G. W. Gibbons and L. P. Grishchuk, \NP{B313} (1989) 736.}
\lref\PagE{D. N. Page, \CQG{7} (1990) 1841.}
\lref\HarB{J. B. Hartle, \PR{D37} (1988) 2818, {\bf D43} (1991) 1434.}
\lref\HarC{J. B. Hartle, \PR{D38} (1988) 2985.}
\lref\HaMy{J. J. Halliwell and R. C. Myers, \PR{D40} (1989) 4011.}
\lref\KSuB{I. Klebanov, L. Susskind, and T. Banks, \NP{B317} (1989) 665.}
\lref\BaKa{A. O. Barvinsky and A. Yu. Kamenshchik, \CQG{7} (1990) L181.}
\lref\Barv{A. O. Barvinsky, Phys.\ Rep.\ {\bf230} (1993) 237.}
\lref\GrRoA{L. P. Grishchuk and L. V. Rozhansky, \PL{234B} (1990) 9.}
\lref\GrRoB{L. P. Grishchuk and L. V. Rozhansky, \PL{208B} (1988) 369.}
\lref\Luk{A. Lukas, \PL{347B} (1995) 13.}
\lref\HalHaw{J. J. Halliwell and S. W. Hawking, \PR{D31} (1985) 1777.}
\lref\LafA{R. Laflamme, \PL{198B} (1987) 156.}
\lref\Vach{T. Vachaspati, \PL{217B} (1989) 228.}
\lref\VaVi{T. Vachaspati and A. Vilenkin, \PR{D37} (1988) 898.} 
\lref\BuDa{T. S. Bunch and P. C. W. Davies, \PRS{A360} (1978) 117.}
\lref\Alle{B. Allen, \PR{D32} (1985) 3136.}
\lref\WadA{S. Wada, \NP{B276} (1986) 729.}
\lref\HaDE{J. J. Halliwell and P. D'Eath, \PR{D35} (1985) 1100.}
\lref\WadB{S. Wada, \PRL{59} (1987) 2375.}
\lref\HawD{S. W. Hawking, \PL{115B} (1982) 295.}
\lref\HawE{S. W. Hawking, \PR{D32} (1985) 2489.}
\lref\GuPi{A. H. Guth and S. Y. Pi, \PRL{49} (1982) 1110; \PR{D32} (1985)
1899.}
\lref\COBE{K. M. Gorski, G. Hinshaw, A. J. Banday, C. L. Bennett, E. L. Wright,
A. Kogut, G. F. Smoot, and P. Lubin, Astrophys.\ J.\ {\bf 430} (1994) L89.}
\lref\GolA{T. Gold (ed.), {\it The Nature of Time}, (Cornell University Press,
Ithaca, 1967).}
\lref\GolB{T. Gold, in {\it La Structure at l'\'Evolution de l'Univers, 11th
Solvay Conference}, (Edition Stoops, Brussels, 1958), p.\ 81; Am.\ J.\ Phys.\
{\bf30} (1962) 403.}
\lref\BeFe{S. C. Beluardi and R. Ferraro, \PR{D52} (1995) 1963.}
\lref\HiWa{A. Higuchi and R. M. Wald, \PR{D51} (1995) 544.}
\lref\Mar{D. Marolf, \CQG{12} (1995) 2469.}
\lref\HaWu{S. W. Hawking and Z. C. Wu, \PL{151B} (1985) 15.}
\lref\HawF{S. W. Hawking, Vist.\ Astr.\ {\bf 37} (1993) 559; and in [\HaPMZu],
p.\ 346.}
\lref\PagF{D. N. Page, \PR{D32} (1985) 2496.}
\lref\PagG{D. N. Page, \PRL{70} (1993) 4034.}
\lref\LaSh{R. Laflamme and E. P. S. Shellard, \PR{D35} (1987) 2315.}
\lref\LafB{R. Laflamme, {\it Time and Quantum Cosmology}, (Ph.D.\ thesis,
University of Cambridge, 1988); in [\HaPMZu], p.\ 358.}
\lref\HaLL{S. W. Hawking, R. Laflamme and G. W. Lyons, \PR{D47} (1993) 5342.}
\lref\KiZe{C. Kiefer and H. D. Zeh, \PR{D51} (1995) 4145.}
\lref\Dea{P. D'Eath, \PR{D48} (1993) 713.}
\lref\HawG{S. W. Hawking, \PR{D37} (1988) 904.}
\lref\GrLu{R. Graham and H. Luckock, \PR{D49} (1994) R4981.}
\lref\dWNM{B. de Wit, H. Nicolai and H. J. Matschull, \PL{318B} (1994) 115.}
\lref\CFOP{S. M. Carroll, D. Z. Freedman, M. E. Ortiz and D. N. Page, \NP{B423}
(1994) 661.}

\centerline{\bf AN INTRODUCTION TO QUANTUM COSMOLOGY}\vskip5\baselineskip
\centerline{DAVID L. WILTSHIRE}
\centerline{\it Department of Physics and Mathematical Physics,}
\centerline{\it University of Adelaide, S.A. 5005, Australia.}
\vskip3\baselineskip \centerline{CONTENTS}\bigskip\leftskip=36pt\rightskip=36pt
\Tenpoint \noindent \vbox{\halign{#\hfil\ &#\hfil\cr 1.&Introduction\cr
\bomb1.1\ Quantum cosmology and quantum gravity\cr \bomb1.2\ A brief history
of quantum cosmology\cr 2.&Hamiltonian Formulation of General Relativity\cr
\bomb2.1\ The $3+1$ decomposition\cr\bomb2.2\ The action\cr
3.&Quantisation\cr \bomb3.1\ Superspace\cr \bomb3.2 Canonical quantisation\cr
\bomb3.3\ Path integral quantisation\cr \bomb3.4\ Minisuperspace\cr
\bomb3.5\ The WKB approximation\cr \bomb3.6\ Probability measures\cr
\bomb3.7\ Minisuperspace for the Friedmann universe with massive scalar
field\cr 4.&Boundary Conditions\cr
\bomb4.1\ The no-boundary proposal\cr \bomb4.2\ The tunneling proposal\cr
5.&The Predictions of Quantum Cosmology\cr 
\bomb5.1 The period of inflation\cr \bomb5.2 The origin of density
perturbations\cr \bomb5.3 The arrow of time\cr
6.&Conclusion\cr}}
\par\nobreak \leftskip=0pt\rightskip=0pt\normalpoint

\newsec{Introduction}

These lectures present an introduction to quantum cosmology for an audience
consisting for a large part of astronomers, and also a number of particle
physicists. As such, the material covered will invariably overlap with that of
similar introductory reviews$^{\HalA,\PagA}$, although I hope to emphasise,
where possible, those aspects of quantum cosmology which are of most interest
to astronomers.

This is not an easy task -- quantum cosmology is not often discussed at
Summer Schools such as this where there is a large emphasis on astrophysical
phenomenology, for the very good reason that the ideas involved are at present
rather tentative, and quantitative predictions are thus much more
difficult to arrive at. Nevertheless we should remember that not too long ago
the whole enterprise of cosmology was viewed as the realm of wild speculation.
Vladimir Lukash remarked that when he started out in research the advice he
was given was that ``cosmology was all right for someone like Zeldovich to
potter around with after he had already established himself by producing
a body of serious work, but it was inappropriate for a young physicist
embarking on a career''. Happily the situation is quite different today -- I
am sure the earlier lectures in this School will have convinced you that
modern cosmology is a hardcore quantitative science, and that with the new
technology and techniques now being developed we can expect to accurately
measure all the important cosmological parameters within the next decade and
thus enter into a ``Golden Era'' for cosmological research.

Quantum cosmology, however, still enjoys the sort of status that all of
cosmology had until not so very long ago: essentially it is a dangerous field
to work in if you hope to get a permanent job.
I hope to convince you nevertheless that quantum cosmology represents a vitally
important frontier of research, and that although it is by nature somewhat
speculative, such speculations are vital if we are to understand the entire
history of the universe.

On the face of it the very words ``quantum'' and ``cosmology'' do appear to
some physicists to be inherently incompatible. We usually think of cosmology
in terms of the very large scale structure of the universe, and of quantum
phenomena in terms of the very small. However, if the hot big bang is the
correct description of the universe -- which we can safely assume given the
overwhelming evidence described in the earlier lectures -- then the universe
did start out incredibly small, and there must have been an epoch when quantum
mechanics applied to the universe as a whole.

There are people who would take issue with this. In the standard Copenhagen
interpretation of quantum mechanics one always has a classical world in which
the quantum one is embedded. We have observers who make measurements -- the
observers themselves are well described by classical physics. If the whole
universe is to be treated as a quantum system one does not have such a luxury,
and some would argue that our conventional ideas about quantum physics cease
to make sense. Yet if quantum mechanics is a universal theory then it is
inevitable that some form of ``quantum cosmology'' was important at the
earliest of conceivable times, namely the Planck time, $t\ns{Planck} =\left(G
\hbar\right)^{1/2}/c^{5/2}=5.4\times10^{-44}$ sec, (equivalent to $10^{19}$
GeV as an energy, or $1.6\times10^{-35}$ m as a length). At such
scales, where the Compton wavelength of a particle is roughly equal to its
gravitational (Schwarzschild) radius, irreducible quantum fluctuations render
the classical concept of spacetime meaningless. Whether or not our current
efforts at constructing a theory of quantum cosmology are physically valid is
therefore really a question of whether our current understanding of
quantum physics is adequate for considering the description of processes at
the very beginning of the universe, or whether quantum mechanics itself has to
be revised at some level. Such a question can really only be answered
by extensive work on the problem.

Setting aside the question of the fundamentals of quantum mechanics, let us
briefly review the problems which are left unanswered in the standard hot big
bang scenario. These are:

\noindent 1.\ The {\it value of} $\et=N\Ns{baryons}/N\Ns{photons}$: the exact
value of this parameter is unexplained in the hot big bang but is crucial in
determining abundances of light elements through primordial nucleosynthesis.\pr
2.\ The {\it horizon problem}: the isotropy of the cosmic microwave background
radiation indicates that all regions of the sky must have been in thermal
contact at some time in the past. However, in the standard
Friedmann-Robertson-Walker (FRW) models regions separated by more than a
couple of degrees have non-intersecting particle horizons -- i.e., they cannot
have been in causal (and hence thermal) contact.\pr
3.\ The {\it flatness problem}: given the range of possible values of the
ratio of the density of the universe to the critical density at the present
epoch, $\OM\Z0$, then the FRW models predict that $\OM(t)$ must have been
incredibly close to the spatially flat case $\OM(t)\simeq1$ at early times;
e.g., assuming $\OM\Z0\gsim0.1$ we find $|\OM-1|\lsim10^{-26}$ at the lepton
era, and $|\OM-1|\lsim10^{-53}$ at the GUT era.\pr
4.\ The {\it unwanted relic problem}: models of the early universe which
involve phase transitions often produce copious amounts of topological defects,
such as monopoles produced at the GUT scale. If one puts the numbers in one
finds that the density of such relics is so great that they would exceed the
critical density by such a margin in the standard FRW models that the
universe should have ended long ago!\pr
5.\ The {\it origin of density perturbations} is unexplained.\pr
6.\ The {\it arrow of time} is a physical mystery. On the one hand the laws
of physics are CPT-invariant, and on the other there is a {\it thermodynamic
arrow of time}, as prescribed by the Second Law of Thermodynamics, and it
appears to match the {\it cosmological arrow of time}, as prescribed by the
expansion of the universe.\pr
7.\ The {\it initial conditions} of the universe must be
put in by hand, rather than being physically prescribed.

The first four problems on the list are ones that can be explained without
appealing to quantum cosmology. The value of $\et$ (problem 1) is
predicted by models of baryogenesis, which typically take place at
the GUT scale. Problems 2--4 are solved by the inflationary universe scenario:
an early phase of exponential expansion of the universe drastically changes
the past light cone, thereby removing the horizon problem, while
also driving $\OM$ close to unity, and diluting unwanted relics to such very
low densities that they are close to unobservable.

Problems 5--7, on the other hand, are of a nature which is beyond the scope
of the inflationary universe scenario to satisfactorily explain. Inflation
provides a mechanism whereby initial small ``quantum'' perturbations are
inflated to all length scales to form a scale-free Harrison-Zeldovich
spectrum, but it does not address the question as to exactly how these
perturbations arise. Furthermore, a typical model of the very early universe
might possess both inflationary and non-inflationary solutions, so that the
precise initial conditions of the universe can be crucial for determining
whether the universe undergoes a period of inflation {\it sufficiently long} to
be consistent with observation. The length of the period of inflation is
precisely the sort of quantitative result that we might hope quantum cosmology
should provide. Questions such as the origin of the arrow of time might appear
to be of a more philosophical nature -- however, quantum cosmology
should provide a calculational framework in which such questions can begin
to be addressed.

\subsec{Quantum cosmology and quantum gravity}

Quantum cosmology is perhaps most properly viewed as one attempt among many to
grapple with the question of finding a quantum theory of gravity. As a field
theory general relativity is not perturbatively renormalisable, and
attempts to reconcile general relativity with quantum physics have not yet
succeeded despite the attentions of at least one generation of physicists. It
is perhaps not surprising that the problem is such a difficult one since
general relativity is a theory about the large scale structure of spacetime,
and to quantise it we have to quantise spacetime itself rather than simply
quantising fields that live in spacetime.

Many ideas have been considered in the quest for a fundamental quantum theory
of gravity -- whether or not these ideas have brought us closer to that goal
is difficult to say without the benefit of hindsight. However, such ideas
have certainly profoundly increased our knowledge about the nature of
possible physical theories. Some important areas of research have included:\pr
1.\ {\it Supergravity}$\,^{\vaNi\hS{\West}\SaSa}$. Using supersymmetry, a
symmetry between fermions and bosons, one can enlarge the gravitational degrees
of freedom to include one or more spin--$3\over2$ gravitinos, $\ps_\mu$, in
addition to the spin--$2$ graviton, $g_{\mu\nu}$. Such a symmetry can cure some
but not all of the divergences of perturbative quantum general
relativity\foot{*}{For details of the application of
perturbative techniques to quantum cosmology, both with and without
supersymmetry, see [\Espo].}. In
particular, while pure Einstein gravity is perturbatively non-renormalisable at
two loops$^{\GoSa}$, or at one loop if interacting with matter$^{\tHVe\hS{\DeNi
}\DeNT}$, in the case of supergravity renormalisability fails only at the
3-loop level$^{\DeKS}$\foot{\blacktriangleright}{The status of the result concerning the 3-loop
divergence of supergravity is not quite as rigorous as the other examples
mentioned, as the complete 3-loop calculation has not been done. However, it is
known that a 3-loop counterterm exists for all extended supergravities and
there is no reason to expect the coefficient of the counterterm to be zero,
making 3-loop finiteness extremely unlikely. For a review of the ultraviolet
properties of supersymmetric field theories see [\HoSt].}.\pr
2.\ {\it Superstring theory}$\,^{\GSW}$. Progress can be made if in addition to
using supersymmetry, one constructs a theory in which the fundamental objects
have an extension rather than being point-like: a theory of strings rather than
particles. Much interest in string theory was generated in the mid 1980s with
the discovery that certain string theories appear to be finite {\it at each
order} of perturbation theory. In some sense stringy physics ``smears out'' the
problems associated with pointlike interactions. The entire sum of all terms
in the perturbation expansion diverges in the case of the bosonic string$^{
\GrPe}$, however, and it is believed that similar results should apply to the
superstring\foot{\circledast}{The question of finiteness of superstring
perturbation theory is a difficult technical question, which has still to be
resolved -- see, e.g., [\Da].}. Furthermore, despite what many see as the
mathematical beauty of string theory, there has unfortunately as yet been no
definitive success in deriving concrete phenomenological predictions.\pr
3.\ {\it Non-perturbative canonical quantum gravity}$\,^{\AshA,\Rov}$. The fact
that general relativity is not perturbatively renormalisable might simply be
a failure of flat spacetime quantum field theory techniques to deal with such
an inherently non-linear theory, rather than reflecting an inherent
incompatibility between general relativity and quantum physics. Given the
divergence of the string perturbation series mentioned above, a
non-perturbative formulation of string theory would also be desirable. As a
starting point a systematic investigation of a non-perturbative canonical
formalism based on general relativity could provide deep insights into quantum
gravity. Such a programme has been investigated by Ashtekar and coworkers,
mainly from the mid 1980s onwards. One principal difference from the
canonical formalism which I will describe in section 2 is that instead of
taking the metric to be the fundamental object to describe the quantum
dynamics, one bases such a dynamics on a {\it connection}, in this case an
$SL(2,\Cop)$ spin connection$^{\Sen,\AshB}$. If one integrates this connection
around a closed loop one arrives at ``loop variables'', which might be
considered to be analogous to the magnetic flux, $\mathchar"0B08$, obtained by
integrating
the electromagnetic gauge potential, $A_\mu$, around a closed loop. In the
``loop representation'' quantum states are represented by functionals of such
loops on a 3-manifold$^{\RoSm,\Blen}$, rather than by functionals of classical
fields. Although a number of technical difficulties remain, considerable
progress has been made with the Ashtekar formulation, and the reformulation
of quantum cosmology in the Ashtekar framework$^{\Kod\hS{\SoCh\SmSo}\LouB}$
could be an area for some interesting future work.
\pr
4.\ {\it Alternative models of spacetime}. All the above approaches assume that
the basic quantum variables, be they a metric or a connection, are defined on
differentiable manifolds. Given that it is highly possible that ``something
strange'' happens at the Planck length it is plausible that one might have to
abandon this assumption in order to effectively describe the quantum dynamics
of gravity. A number of ideas have been considered on these lines. One
framework which has been widely used both for numerical relativity and studies
of quantum gravity is that of {\it Regge calculus}$\,^{\Re}$\foot{\vartriangle}
{For a brief review and extensive bibliography see [\WiTu].} in which one
replaces smooth manifolds by spaces consisting of piecewise linear simplicial
blocks. Naturally, other possibilities for discretised spacetime structure also
exist -- {\it causal sets}$\,^{\BLMS}$ being another example which has not yet
been so widely explored. A further possibility is {\it topological
quantisation} whereby one replaces a manifold by a set and quantises all
topologies on that set$^{\Ish,\IsKR}$.\pr

Many of the above alternatives fall into the category of being attempts to
construct a fundamental theory of quantum gravity. The current quantum
cosmology programme is not quite as ambitious. One begins by making the
assumption that whatever the exact nature of the fundamental theory of quantum
gravity is, in its semiclassical limit it should agree with the semiclassical
limit of a canonical quantum formalism based on general relativity alone. Thus
we study the semiclassical properties of quantum gravity based solely on
Einstein's theory, or some suitable modification of it.

Clearly, any predictions made from such a foundation must be treated with
caution. In particular, a new fundamental theory of quantum gravity might
introduce radically new physical processes at an energy scale relevant for
cosmology. String theory, for example, introduces its own fundamental scale
which expressed as a temperature (the {\it Hagedorn temperature}) is given by:
$$T\Ns{Hagedorn}={\sqrt{\hbar c}\over4\pi k\Z B\sqrt{\al'}}\,,\eqn\THagedorn$$
the constant $\al'$ being the Regge slope parameter, which is inversely
proportional to the string tension. It is not known exactly what the value of
$\al'$ is; however, if $T\Ns{Hagedorn}$ is comparable to or significantly lower
than the Planck scale, then it is clear that fundamentally ``stringy''
processes will be very important in the very early universe, if string theory
is indeed the ``ultimate'' theory.

Although new fundamental physics could drastically change the predictions of
quantum cosmology, I believe nevertheless that studying quantum cosmology
based even just on Einstein's theory is an important activity. General
relativity is a remarkably successful theory; in seeking to replace it by
something better it is important that we study processes at the limit of its
applicability, thereby challenging our understanding. General relativity is
limited by the Planck scale -- the physical arenas in which this scale is
approached include: (i) very small black holes; (ii) the very early universe.
Since it seems unlikely that we will ever be able to create energies of order
$10^{19}\w{GeV}$ in the laboratory, the consequences of quantum gravity for
the physics of the very early universe will remain the one way of indirectly
``testing'' it, at least for the foreseeable future.

It is thus important that we consider quantum gravity in a cosmological
context. Even if our current attempts do not fully reach the mark, in that we
do not yet have a fully-fledged quantum theory of gravity, they nonetheless
constitute a vital part of the process of trying to find such a theory.

\subsec{A brief history of quantum cosmology}

The quantum cosmology programme which I will describe in these lectures has
gone through three main identifiable phases to date:

\noindent 1.\ {\it Defining the problem}. The canonical formalism,
including the definition of the wavefunction of the universe, $\PS$, its
configuration space -- superspace -- and its evolution according to the
Wheeler-DeWitt equation, was set up in the late 1960s$^{\WhA\hS{\WhB,\DeW,\MisA
}\MisB}$.

\noindent 2.\ {\it Boundary conditions}. Quantum cosmology research went into
something of a lull during the 1970s but was revived in the mid 1980s when the
question of putting appropriate boundary conditions on the wavefunction of the
universe was treated seriously. The idea is that such boundary conditions
should describe the ``creation of the universe from nothing''$^{\HawA,\VilA}$,
where {\it nothing} means the absence of space and time. A number of proposals
for such boundary conditions emerged -- two major contenders being the {\it
``no-boundary''} proposal of Hartle and Hawking$^{\HarHaA,\HawC}$ and the
{\it``tunneling''} proposal advocated by Vilenkin$^{\Vilc\hS{\VilC}\VilD}$.

\noindent 3.\ {\it Quantum decoherence}. The mechanism of the transition from
quantum physics to classical physics (``quantum decoherence'') becomes vitally
important when quantum physics is applied to the universe as a whole. The
issues involved have begun to occupy many researchers in the early 1990s\foot
{\blacktriangledown}{See [\HarA] for a review and [\HaPMZu] for a collection of
recent papers on the subject.}.

\noindent Two other important areas of quantum cosmology (or related) research
have been: (i) quantum wormholes and ``baby universes''; (ii) supersymmetric
quantum cosmology.

Quantum wormholes\foot{\divideontimes}{See [\St] for a review.} were extremely
fashionable in the particle physics community in the years 1988--1990. Such
states arise when one considers topology change in the path integral
formulation of quantum gravity: {\it quantum wormholes} are instanton solutions
which play an important role in the Euclidean path integral. One deals directly
with a ``third quantised'' formalism, (i.e., quantum field theory over
superspace), which includes operators that create and destroy universes --
so-called {\it baby universes}. Much of the excitement in the late 1980s was
associated with the idea that such processes could fix the fundamental
constants of nature -- in particular, driving the cosmological constant to
zero.

Supersymmetric quantum cosmology has emerged recently as one of the most
active areas of current research\foot{\pitchfork}{See [\CsGr] and references
therein.}. In considering the quantum creation of the universe we are of
course dealing with the very earliest epochs of the universe's existence, at
which time it is believed that supersymmetry would not yet be broken. The
inclusion of supersymmetry could therefore be vital from the point of view of
physical consistency. 

Since the focus of this School is on cosmology, my intention in these lectures
is to cover topics 1 and 2 above, and then to proceed to discuss the
predictions of quantum cosmology. The third topic of quantum decoherence raises
questions which have not been resolved even in ordinary quantum mechanics,
since the question of decoherence really amounts to understanding what happens
during the ``collapse of the wavefunction''. Although this is a fascinating
issue it has more to do with the fundamentals of quantum mechanics than
directly with cosmology. Likewise, I will only briefly touch upon quantum
wormholes and supersymmetric quantum cosmology, as these areas are still in
their infancy, and one is still at the point of trying to resolve basic
questions concerning quantum gravity. I hope the reader will not be
disappointed by this -- however, given the vast scope of quantum gravity and
quantum cosmology one must necessarily be rather selective.

\newsec{Hamiltonian Formulation of General Relativity}
\subsec{The $3+1$ decomposition}

In order to discuss quantum cosmology a fair amount of technical machinery is
required.
In the canonical formulation we begin by making a $3+1$-split of the
4-dimensional spacetime manifold, $\Mi{}$, which will describe the universe,
foliating it into spatial hypersurfaces, $\SI_t$, labeled by a global time
function, $t$. Thus we take the 4-dimensional metric to be given by\foot
{\blacklozenge}{We use a $(-+++)$ Lorentzian metric signature, and natural
units in which $c=\hbar=1$.}
$$\dd s^2=g_{\mu\nu}\dd x^\mu\dd x^\nu=-\Bom^0\otimes\Bom^0+h_{ij}\Bom^i
\otimes\Bom^j,\eqn\coordsA$$
where
$$\eqalign{\Bom^0&=\NN\dd t\cr \Bom^i&=\dd x^i+\NN^i\dd t.\cr}\eqn\coordsB$$
Such a decomposition is possible in general if the manifold
$\Mi{}$ is {\it globally hyperbolic}$\,^{\Wald}$.
The quantity $\NN(t,x^k)$ is called the {\it lapse function} -- it measures
the difference between the coordinate time, $t$, and proper time, $\ta$, on
curves normal to the hypersurfaces $\SI_t$, the normal being $n_\al=(-\NN,0,0,0
)$ in the above coordinates. The quantity $\NN^i(t,x^k)$ is called the
{\it shift vector} -- it measures the difference between a spatial point,
$p$, and the point one would reach if instead of following $p$ from one
hypersurface to the next one followed a curve tangent to the
normal $\B n$. That is to say, the spatial coordinates are ``comoving'' if $\NN
^i=0$. Finally, $h_{ij}(t,x^k)$ is the {\it intrinsic metric} (or {\it first
fundamental form}) induced on the spatial hypersurfaces by the full
4-dimensional metric, $g_{\mu\nu}$. In components we have
$$\left(g_{\mu\nu}\right)=\left[\matrix{-\NN^2+\NN^k\NN_k&\NN_j\cr\NN_i&h_{ij}
\cr}\right]\,,\eqn\metric$$
with inverse
$$\left(g^{\mu\nu}\right)=\left[\matrix{{-1\over\;\,\NN^2}&{\NN^j\over\NN^2}\cr
{\NN^i\over\NN^2}&h^{ij}-{\NN^i\NN^j\over\NN^2}\cr}\right]\,,\eqn\invmetric$$
where $h^{ij}$ is the inverse to $h_{ij}$, and the intrinsic metric is used to
lower and raise spatial indices: $\NN^k\NN_k\equiv h^{jk}\NN_j\NN_k=h_{jk}\NN^j
\NN^k$ etc. \Ifig\SPLIT{The $3+1$ decomposition of the manifold, with lapse
function, $N$, and shift vector, $N^i$.}{\epsfxsize=10cm \epsfbox{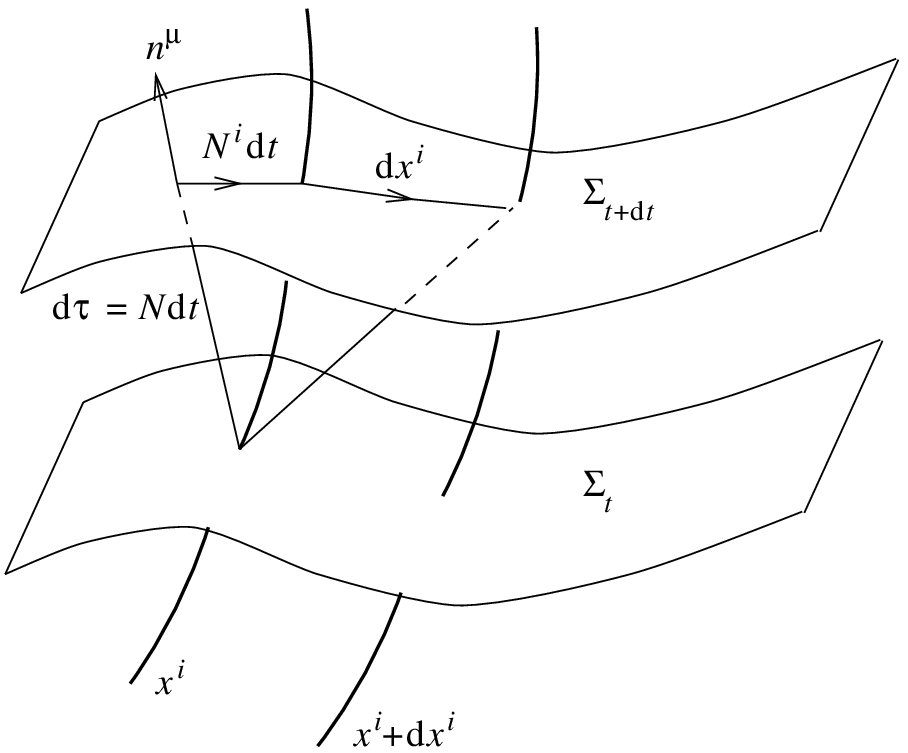}}

One can construct an {\it intrinsic curvature} tensor $^3\!R^i_{\ \;jkl}(h)$
from the intrinsic metric alone -- this of course describes the curvature
intrinsic to the hypersurfaces $\SI_t$. One can also define an {\it extrinsic
curvature}, (or {\it second fundamental form}), which describes how the spatial
hypersurfaces $\SI_t$ curve with respect to the 4-dimensional spacetime
manifold within which they are embedded. This is given by
$$\eqalign{K_{ij}\equiv&-n_{i;j}=-\GA\Y0_{\ \;ij}n\Z0\cr &={1\over2\NN}\left(
\NN_{i|j}+\NN_{j|i}-{\pt h_{ij}\over\pt t}\right),\cr}\eqn\extrinsic$$
where a semicolon denotes covariant differentiation with respect to the
4-metric, $g_{\mu\nu}$, and a vertical bar denotes covariant differentiation
with respect to the 3-metric, $h_{ij}$: $\NN_{i|j}\equiv\NN_i,_j-\GA^k_{\ \;ij}
\NN_k$ etc.

For a given foliation of $\Mi{}$ by spatial hypersurfaces, $\SI_t$, it is
always possible to choose {\it Gaussian normal coordinates}, in which
$$\dd s^2=-\dd t^2+h_{ij}\dd x^i\dd x^j.\eqn\GNC$$
These are comoving coordinates ($\NN^i=0$) with the additional property that $t
$ is the proper time parameter ($\NN=1$). This is the standard ``gauge choice''
that is made in classical cosmology, and in such coordinates $K_{ij}=-\dot h_
{ij}$, where dot denotes differentiation with respect to $t$. In making the
$3+1$ decomposition, however, we are only free to make a specific choice of
coordinates such as \GNC\ {\it after} variation of the action if we want to be
sure to obtain Einstein's equations, and thus we must retain the lapse and
shift function for the time being. 

\subsec{The action}

A relevant action for use in quantum cosmology is that of Einstein
gravity plus a possible cosmological term, $\LA$, and matter, given by
$$S={1\over4\KA}\left[\;\int\limits_{\Mi{}}\dd^4x\sqrt{-g}\left(
\ric4-2\LA\right)+2\int\limits_{\pt\Mi{}}\dd^3x\sqrt{h}\,K\right]+S\Ns{matter},
\eqn\actionA$$
where $\KA=4\pi G=4\pi m^{-2}\ns{Planck}$, $K\equiv K^i_{\ \;i}$ is the trace
of the extrinsic curvature, and for many simple models the matter is specified
by a single scalar field, $\PH$, with potential, $\V(\PH)$,
$$S\Ns{matter}=\int\limits_{\Mi{}}\dd^4x\sqrt{-g}\left(
-{1\over2}g^{\mu\nu}\pt_\mu\PH\pt_\nu\PH-\V(\PH)\right).\eqn\actionB$$
The boundary term$^{\Yor}$ in \actionA\ does not of course contribute to the
classical field equations, and this term is usually omitted in a first course
on general relativity. However, in quantum physics we are often interested in
phenomena which occur when the classical field equations do not apply, that is
``off-shell'', and thus it is vitally important to retain the surface term. The
simple matter action \actionB\ given here should simply be seen as being
representative of the type of matter action one might consider. Although the
example given by \actionB\ is sufficient for studying many inflationary
universe models, many other alternatives might also be of interest, such as
extra matter from a supergravity multiplet, or matter corresponding to the
low-energy limit of string theory. In the latter case, if one works in the
``string conformal frame'' it is also necessary to alter the gravitational part
of the action, as one characteristic of string theory is the presence of the
scalar dilaton, $\PH$, which couples universally to matter (at least
perturbatively). In that case \actionA, \actionB\ would be replaced by
$$\eqalign{S={1\over4\KA}\Biggl[\int\limits_{\Mi{}}\dd^4x\sqrt{-g}
\e^{-2\PH}&\left(\ric4+4g^{\mu\nu}\pt_\mu\PH\pt_\nu\PH-8\V(\PH)+\dots\right)\cr
&\hbox to22true mm{\hfil}+2\int\limits_{\pt\Mi{}}\dd^3x\sqrt{h}\e^{-2\PH}K
\Biggr],\cr}\eqn\actionC$$
where the ellipsis denotes any additional matter degrees of freedom, and we
have allowed for the possibility of the generation of a dilaton potential,
$\V(\PH)$, via some non-perturbative symmetry breaking mechanism. However, for
simplicity the specific examples we will deal with here will be confined to
models of the type \actionA, \actionB. 

We now wish to express \actionA, \actionB\ in terms of the variables of the
$3+1$ split. One can show that
$$\ric4=\ric3-2R_{nn}+K^2-K^{ij}K_{ij},\eqn\RicS$$
where $\ric3$ is the Ricci scalar of the intrinsic 3-geometry, and
$$R_{nn}\equiv R_{\al\be}n^\al n^\be=-K^{ij}K_{ij}+K^2+\left(n^\al K+a^\al
\right)_{;\al},\eqn\RicZ$$
with $a^{\al}\equiv n^\be n\UD\al{;\be}$.
Combining \actionA, \RicS\ and \RicZ, and noting that the boundary integral
involving $a^\al$ vanishes identically since $n^\al a_\al=0$, we obtain
$$S\equiv\int\dd t\;L={1\over4\KA}\int\dd t\,\dd^3x\NN\sqrt{h}\left(K_{ij}K^
{ij}-K^2+\ric3-2\LA\right)+S\Ns{matter}.\eqn\actionD$$
As in the Hamiltonian formulation of field theory we define canonical momenta
in the standard fashion
$$\eqalignno{\PI^{ij}&\equiv{\de L\over\de\dot h_{ij}}=-{\sqrt{h}\over4\KA}
\left(K^{ij}-h^{ij}K\right),&\Eqn\Pih\cr
\PIPH&\equiv{\de L\over\de\dot\PH}={\sqrt{h}\over\NN}\left(\dot\PH-\NN^i\PH,_i
\right),&\Eqn\PiPH\cr
\PIz&\equiv{\de L\over\de\dot\NN}=0,&\Eqn\PiNo\cr
\PI^i&\equiv{\de L\over\de\dot\NN_i}=0.&\Eqn\PiN\cr}$$
The fact that the momenta conjugate to $\NN$ and $\NN_i$ vanish means that we
are dealing with {\it primary constraints} in Dirac's terminology$^{\DirA,\HaRT
}$.

If we use $\NN$, $\NN_i$, $h_{ij}$, $\PH$ and their conjugate momenta as the
basic variables we obtain\foot{\diamondsuit}{Details of all missing steps in
this section will be provided upon request in a plain brown envelope.} a
Hamiltonian
$$\eqalign{H&\equiv\int\dd^3x\left(\PIz\dot\NN+\PI^i\dot\NN_i+\PI^{ij}\dot h_
{ij}+\PIPH\dot\PH\right)-L\cr&=\int\dd^3x\left(\PIz\dot\NN+\PI^i\dot\NN_i+\NN
\HH+\NN_i\HH^i\right),\cr}\eqn\Ham$$
where
$$\eqalignno{\HH&={\sqrt{h}\over2\KA}\left(G\UD\hz\hz-2\KA T\UD\hz\hz\right)\cr
&=4\KA\GG\Z{ijkl}\PI^{ij}\PI^{kl}-{\sqrt{h}\over4\KA}\left(\ric3-2\LA\right)+
\half\sqrt{h}\left[{\PIPH^{\ 2}\over h}+h^{ij}\PH,_i\PH,_j+2\V\right]\!,&\Eqn
\Enden\cr
\HH^i&={\sqrt{h}\over2\KA}\left(G^{\hz\hi}-2\KA T^{\hz\hi}\right)\cr&
=-2\PI\UD{ij}{|j}+h^{ij}\PH,_j\PIPH\,,&\Eqn\Momden\cr}$$
and
$$\GG_{ijkl}=\half h^{-1/2}\left(h_{ik}h_{jl}+h_{il}h_{jk}-h_{ij}h_{kl}\right)
\eqn\DeWitt$$
is the {\it DeWitt metric}$^{\DeW}$. The hats in \Enden\ and \Momden\ denote
orthonormal frame components of the Einstein and energy-momentum tensors.
In terms of these variables the action \actionD\
then becomes
$$S=\int\dd t\,\dd^3x\left(\PIz\dot\NN+\PI^i\dot\NN_i-\NN\HH-\NN_i\HH^i\right)
.\eqn\actionHam$$
If we vary \actionHam\ with respect to $\PI^{ij}$ and $\PIPH$ we obtain their
defining relations \Pih\ and \PiPH. The lapse and shift functions now act as
Lagrange multipliers; variation of \actionHam\ with respect to the lapse
function, $\NN$, yields the {\it Hamiltonian constraint}
$$\HH=0,\eqn\HamCoN$$
while variation of \actionHam\ with respect to the shift vector, $\NN_i$,
yields the {\it momentum constraint}
$$\HH^i=0.\eqn\MomCoN$$
From \Enden\ and \Momden\ it is clear that these constraints are simply the
$(00)$ and $(0i)$ parts of Einstein's equations. In Dirac's terminology these
are {\it secondary} or {\it dynamical constraints}.

The $3+1$ decomposition of our spacetime looks to be very counterintuitive to
the usual ideas of general relativity. This is so by choice. Einstein
described the l.h.s.\ of his field equations which encodes the 4-dimensional
geometry as a ``hall of marble'', which does not encourage people to tamper
with it. However, to quantise spacetime we must do just that: we must
deconstruct spacetime and replace it by something else. Thus far we have not
done that -- we have simply split our 4-dimensional manifold into a sequence
of spatial hypersurfaces, $\SI_t$. Time is the natural variable to base this
split upon since it plays a special role in quantum mechanics -- it is a
parameter rather than an operator.

Classically the evolution of one spatial hypersurface to the next is completely
well-defined, (provided that the manifold, $\Mi{}$, is globally hyperbolic),
and given initial data, $h_{ij}$, $\PH$, on an initial hypersurface, $\SI$,
we can use the Cauchy development to stitch the hypersurfaces $\SI_t$
together to recover the 4-dimensional manifold, $\Mi{}$. To quantise the
theory, however, we want to perform a path integral over {\it all} geometries,
not just the classically allowed ones. Thus we must consider sequences of
geometries at the quantum level which {\it cannot} be stitched together in a
regular Cauchy development to form a 4-manifold which solves Einstein's
equations. (See \fig\GEOMETRY.) We must therefore abandon Einstein's ``hall of
marble'' -- spacetime is no longer a fundamental object.
\ifig\GEOMETRY{Quantum geometrodynamics: in addition to the classical Cauchy
development from $\SI$ to $\SI'$ (left), the path integral includes a sum over
all 4-manifolds which interpolate between the initial and final configurations.
The weighting by $\e^{iS}$ means that the greater the number of classically
forbidden 3-geometries (shaded slices) is in the interpolating 4-manifold, the
smaller its contribution is to the path integral.}
{\epsfxsize=15cm\epsfbox{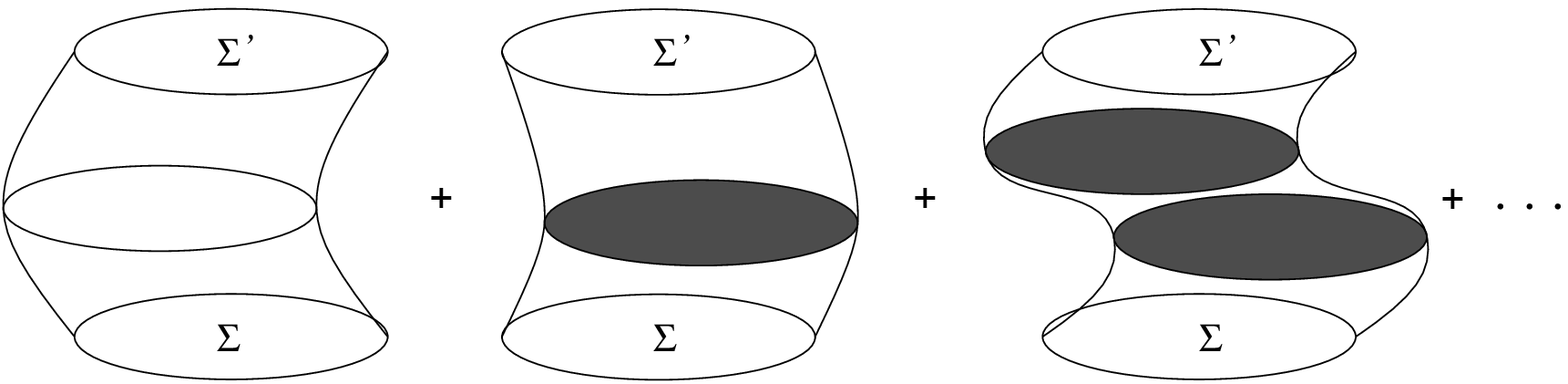}}

As was mentioned in the Introduction the ``deconstruction'' of spacetime that
we adopt here is probably the most conservative choice we could make.
Even though we abandon the notion of spacetime in discussing the quantum
dynamics of gravity, our fundamental objects are still defined on regular
3-manifolds, $\SI$. A more radical departure would be to replace these spatial
hypersurfaces by some more general set by using, for example, the ideas of
Regge calculus, causal sets or topological quantisation.

\newsec{Quantisation}
\subsec{Superspace}

As a prelude to the canonical quantisation of gravity let us first introduce
the relevant configuration space on which the quantum dynamics will be defined.

Consider the space of all Riemannian 3-metrics and matter configurations on the
spatial hypersurfaces, $\SI$,
$$\w{Riem}(\SI)\equiv\left\{h_{ij}(x),\PH(x)\;|\;x\in\SI\right\}.\eqn\Riem$$
This is an infinite-dimensional space, on account of the label $x=\{x^i\}
$, which specifies the point on the hypersurface, but there are a finite number
of degrees of freedom at each point, $x\in\SI$. In fact, we are really
interested in geometry here and configurations which can be related to each
other by a diffeomorphism, i.e., a coordinate transformation, should be
considered to be equivalent since their intrinsic geometry is the same. Thus
we factor out by diffeomorphisms on the spatial hypersurfaces and
identify {\it superspace}\foot{\star}{The use of the terminology ``superspace''
for the configuration space of quantum cosmology predates the discovery of
supersymmetry, and of course is completely different from the combined manifold
of commuting and anticommuting coordinates which is called ``superspace'' in
supersymmetric theories.} as
$${\hbox{Riem}\,(\SI)\over\hbox{Diff}\X0(\SI)}\,,$$
where the subscript zero denotes the fact that we consider only diffeomorphisms
which are connected to the identity. This infinite-dimensional space will be
the basic configuration space of quantum cosmology.

The DeWitt metric \DeWitt\ then provides a metric on superspace which we can
write as
$$\GG\X{AB}(x)\equiv\GG_{(ij)(kl)}(x),\eqn\altDeWitt$$
where the indices $A$, $B$ run over the independent components of the intrinsic
metric $h_{ij}$:
$$A,B\;\in\{h\Z{11},h\Z{12},h\Z{13},h\Z{22},h\Z{23},h\Z{33}\}.$$
The DeWitt metric has a $(-+++++)$ signature at each point $x\in\SI$,
regardless of the signature of the spacetime metric, $g_{\mu\nu}$. To
incorporate all the degrees of freedom, we also need to extend the range of
the indices $A$, $B$ to include the matter fields, by appropriately defining
$\GG\X{\PH\PH}(x)$ (and other components if more than one matter field is
present), thereby obtaining the full {\it supermetric}.

An inverse DeWitt metric, $\GG\W{AB}(x)=\GG^{(ij)(kl)}(x)$, can be defined by
the requirement
$$\GG^{(ij)(kl)}\GG_{(kl)(mn)}=\half\left(\de\UD im\de\UD jn+\de\UD in\de\UD jm
\right),\eqn\definv$$
which gives
$$\GG^{(ij)(kl)}=\half\sqrt{h}\left(h^{ik}h^{jl}+h^{il}h^{jk}-2h^{ij}h^{kl}
\right).\eqn\invDeWitt$$

\subsec{Canonical quantisation}

We will perform canonical quantisation by taking the wavefunction of the
universe, $\PS[h_{ij},\PH]$, to be a functional on superspace. Unlike the case
of ordinary quantum mechanics, the wavefunction, $\PS$, does not depend
explicitly on time here. This is related to the fact that general relativity
is an ``already parametrised'' theory -- the original Einstein-Hilbert action
is time-reparametrisation invariant. Time is contained implicitly in the
dynamical variables, $h_{ij}$ and $\PH$.

According to Dirac's quantisation procedure$^{\DirA}$ we make the following
replacements for the canonical momenta
$$\PI^{ij}\rarr-i\Der\de{h_{ij}},\quad\PIPH\rarr-i\Der\de\PH,\quad
\PIz\rarr-i\Der\de\NN,\quad\PI^i\rarr-i\Der\de{\NN_i}\;.\eqn\quantise$$
and then demand that the wavefunction is annihilated by the operator version of
the constraints. For the primary constraints we have
$$\eqalign{\hat\PI\PS&=-i\Deriv\de\NN\PS=0,\cr \hat\PI^i\PS&=-i\Deriv\de{\NN_i}\PS=
0,\cr}\eqn\primCoN$$
which implies that $\PS$ is independent of $\NN$ and $\NN^i$. The momentum
constraint yields
$$\hat\HH^i\PS=0\qquad\hence\qquad i\left[\Deriv\de{h_{ij}}\PS\right]_{|j}=2\KA
\hat T^{\hz\hi}\PS.\eqn\momCoN$$
Using \momCoN\ one can show$^{\Higgs}$ that $\PS$ is the same for
configurations $\left\{h_{ij}(x),\PH(x)\right\}$ which are related by a
coordinate transformation in the spatial hypersurface, $\SI$, which accords
with the rationale for factoring out by diffeomorphisms in our definition of
superspace. Finally, the Hamiltonian constraint yields
$$\hat\HH\PS=\left[-4\ka^2\GG_{ijkl}{\de^2\hphantom{h_{ij}h_kl}\over\de h_{ij}
\de h _{kl}}+{\sqrt{h}\over4\ka^2}\left(-\ric3+2\LA+4\ka^2\hat T^{\hz\hz}\right
)\right]\PS=0,\eqn\WdW$$
where for our scalar field example
$$\hat T^{\hz\hz}={-1\over2h}{\de^2\ \over\de\PH^2}+\half h^{ij}\PH,_i\PH,_j+
\V(\PH).\eqn\Tnndef$$

Eq.\ \WdW\ is known as the {\it Wheeler-DeWitt equation}$^{\WhB,\DeW}$. In
fact, it is not a single equation but is actually one equation at each point, $
x\in\SI$. It is a second-order hyperbolic functional differential equation on
superspace. On account of factor-ordering ambiguities it is not completely
well-defined, although there exist ``natural'' choices$^{\DeW,\HaPaA}$ of
ordering for which the derivative pieces become a Laplacian in the supermetric.

\subsec{Path integral quantisation}

An alternative to canonical quantisation which perhaps better accommodates an
intuitive understanding (c.f.\ \GEOMETRY) of the quantisation procedure is the
path integral approach. Path integral techniques in quantum gravity were
pioneered in the late 1970s$^{\GiHaA,\HawB}$. The starting point for this
is Feynman's idea that the amplitude to go from one state with an intrinsic
metric, $h_{ij}$, and matter configuration, $\PH$, on an initial hypersurface,
$\SI$, to another state with metric, $h'_{ij}$ and matter configuration $\PH'$
on an final hypersurface, $\SI'$, is given by a functional integral of $\e^{iS}
$ over all 4-geometries, $g_{\mu\nu}$, and matter configurations, $\pH$, which
interpolate between the initial and final configurations
$$\langle h'_{ij},\PH',\SI'|h_{ij},\PH,\SI\rangle=\sum_{\Mi{}}\int\DD{\B g}\,
\DD\pH\,\e^{iS[g_{\mu\nu},\pH]}.\eqn\amplitude$$

In ordinary quantum field theory in flat spacetime the path integral suffers
from the difficulty that since the action $S[g_{\mu\nu},\pH]$ is finite the
integral oscillates and therefore fails to converge. Furthermore, the solution
which extremises the action is given by solving a hyperbolic equation between
initial and final boundary surfaces -- which is not a mathematically well-posed
problem, and may have either no solutions or an infinite number of solutions.
To deal with this problem one performs a ``Wick rotation'' to ``imaginary
time'' $t\rarr-i\ta$ and considers a path integral formulated in terms of the
Euclidean action, $I=-iS$. The action is then positive-definite, so that the
path integral is exponentially damped and should converge. Also the problem of
finding the extremum becomes that of solving an elliptic equation with
boundary conditions, and this is well-posed.

One may attempt to apply a similar approach to quantum gravity, replacing $S$
in \amplitude\ by the Euclidean action\foot{\boxplus}{Strictly speaking one
should call this the Riemannian action, since ``Euclidean'' spaces are those
which are {\it flat}, whereas {\it curved} manifolds with $(++++)$ signature
are known as {\it Riemannian spaces}. However, the terminology ``Euclidean''
action which has been taken over from flat space quantum field theory seems to
have stuck, despite the fact that we are of course no longer dealing with $\Rop
^4$.} $I[g_{\mu\nu},\pH]=-iS[g_{\mu\nu},\pH]$, and taking the sum in
\amplitude\ to be over all metrics with signature $(++++)$, which induce the
appropriate 3-metrics $h_{ij}$ and $h'_{ij}$ on the past and future
hypersurfaces. This approach to quantum gravity has had some important
successes -- most notably, it provides: (i) an elegant way of deriving the
thermodynamic properties of black holes$^{\HarHaB\hS{\GiHaB}\GiPeA}$; and (ii)
a natural means for discussing the effects of gravitational instantons$^{\EgHa
\hS{\GiHaC}\GiPo}$. Gravitational instantons have been found to provide the
dominant contribution to the path integral in processes such as pair creation
of charged black holes in a magnetic field$^{\GibA\hS{\GaSt}\DGGH}$, and
therefore provide a means of calculating the rates of such processes
semiclassically.

The problems associated with the Euclidean approach to quantum gravity are
considerable, however. Firstly, unlike ordinary field theories\foot{\bullet}
{The action for fermi fields in ordinary quantum field theory is not
positive-definite, but that is not a problem since one can treat them as
anticommuting quantities so that the path integral over them converges.} such
as Yang-Mills theory the gravitational action is not positive-definite$^
{\GiHaPe}$, and thus the path integral does not converge if one restricts the
sum to real Euclidean-signature metrics. To make the path integral converge it
is necessary to include complex metrics in the sum$^{\GiHaPe,\Schl}$. However,
there is no unique contour to integrate along in superspace$^{\HalLoA\Hs
{\HalLoB}\HalLoC}$ and the result one obtains for the path integral may depend
crucially on the contour that is chosen$^{\HalLoA}$. Furthermore, the measure
in \amplitude\ is ill-defined. It is really only in the last ten years that
mathematicians have succeeded in making path integration in ordinary quantum
field theory rigorously defined. Clearly, we may have to wait some time before
the same can be said of path integrals in quantum gravity. 

The physicists' approach is to set aside the issues involved in making the
formalism rigorous and to see what can be learned nevertheless. We thus take
the wavefunction $\PS$ of the universe on a hypersurface, $\SI$, with intrinsic
3-metric, $h_{ij}$, and matter configuration, $\PH$, to be defined$^{\HarHaA,
\HawC}$ by the functional integral
$$\PS\left[h_{ij},\PH,\SI\right]=\sum_{\Mi{}}\int\DD{\B g}\,\DD\pH\,\e^{-I[g_
{\mu\nu},\pH]}.\eqn\wavefn$$
where the sum is over a class of 4-metrics, $g_{\mu\nu}$, and matter
configurations, $\pH$, which take values $h_{ij}$ and $\PH$ on the boundary $
\SI$. Alternative definitions of the wavefunction have been proposed. In
particular, Linde$^{\Lin}$ has argued that one should Wick rotate with the
opposite sign, i.e., $t\to+i\ta$ instead of $t\to-i\ta$ as above, leading to a
factor $\e^{+I}$ instead of $\e^{-I}$ in \wavefn. Alternatively, one could
stick with a Lorentzian path integral$^{\VilB}$, with $\e^{iS}$ instead of
$\e^{-I}$ in \wavefn. In any case, in order to achieve convergence of the path
integral\foot{\circleddash}{Linde's suggested modification [\Lin] to \wavefn\
yields a convergent path integral for the scale factor, which is all that one
needs in the simplest minisuperspace models, but does not lead to convergence
if one includes matter or inhomogeneous modes of the metric.} it is
necessary to include complex manifolds in the sum, which somewhat obscures the
distinctions between these alternative proposed definitions of $\PS$. 
The real distinction between the alternative definitions
arises when it comes to imposing boundary conditions, thereby restricting the
4-manifolds included in the sum in \wavefn. For example, one could view the
Euclidean sector as being the appropriate sector of the quantum theory in which
an ``initial'' boundary condition on $\PS$ should be imposed, which would make
\wavefn\ the natural starting point, as is the case for the no-boundary
proposal$^{\HarHaA,\HawC}$. Alternative boundary conditions would favour the
Lorentzian path integral$^{\VilB}$.

The path integral definition of the wavefunction
\wavefn\ is consistent with the earlier definition based on canonical
quantisation to the extent that wavefunctions defined according to \wavefn\ can
be shown to satisfy the Wheeler-DeWitt equation \WdW\ provided that the action,
the measure and the class of paths summed over are invariant under
diffeomorphisms$^{\HalHarA}$. 

In the canonical quantisation formalism any particular solution to the
Wheeler-DeWitt equation will depend upon the specification of boundary
conditions on the wavefunction. In the path integral formulation the particular
solution of the Wheeler-DeWitt equation that one obtains will similarly depend
on the contour of integration chosen in superspace, and the class of 4-metrics
one sums over in \wavefn. Unfortunately it is not known how the choice of
contour and class of paths prescribes the boundary conditions on the
wavefunction in the general case, although it can be answered for specific
models. The question of boundary conditions is naturally of prime importance
for cosmology, and we shall return to this question in \S4.

\subsec{Minisuperspace}

In practice to work with the infinite dimensions of the full superspace is
not possible, at least with the techniques that have been developed to date.
One useful approximation therefore is to truncate the infinite degrees of
freedom to a finite number, thereby obtaining some particular {\it
minisuperspace} model. An easy way to achieve this is by considering
homogeneous metrics, since as was observed earlier for each point $x\in\SI$
there are a finite number of degrees of freedom in superspace.

The results we shall obtain by this approach will be somewhat satisfying in
that they do appear to have some predictive power. However, the truncation to
minisuperspace has not as yet been made part of a rigorous approximation scheme
to full superspace quantum cosmology. As they are currently formulated
minisuperspace models should therefore be viewed as toy models, which we
nonetheless hope will capture some of the essence of quantum cosmology. Since
we are simultaneously setting most of the field modes and their conjugate
momenta to zero, which violates the uncertainty principle, this approach
might seem rather {\it ad hoc}. However, in classical cosmology homogeneity and
isotropy are important simplifying assumptions which do have a sound
observational basis. Therefore it is not entirely unreasonable to hope that a
consistent truncation to particular minisuperspace models with particular
symmetries might be found in future\foot{\clubsuit}{For some discussions of
the validity of the minisuperspace approximation see [$\KuRy\hS{\HuSi,\Mazz}
\IsIs$].}.

Let us thus consider {\it homogeneous} cosmologies for simplicity. Instead of
having a separate Wheeler-DeWitt equation for each point of the spatial
hypersurface, $\SI$, we then simply have a single Wheeler-DeWitt equation for
all of $\SI$. The standard FRW metrics, with
$$h_{ij}\dd x^i\dd x^j=a^2(t)\left[{\dd r^2\over1-kr^2}+r^2\dOM\right],\qquad k
=-1,0,1,\eqn\FRW$$
are of course one special example. In that case
our model with a single scalar field simply has two minisuperspace coordinates,
$\{a,\PH\}$, the cosmic scale factor and the scalar field. Many more general
homogeneous (but anisotropic) models can also be considered. Indeed all such
models can be classified\foot{\vartriangleleft}{See [\MacC] for a review.} and
apart from the FRW models the other cases of interest are: (i) the {\it
Kantowski-Sachs models}$^{\KoCh,\KaSa}$, which have a 3-metric
$$h_{ij}\dd x^i\dd x^j=a^2(t)\dd r^2+b^2(t)\dOM,\eqn\KantSachs$$
and thus three minisuperspace coordinates, $\{a,b,\PH\}$;
and (ii) the Bianchi models.

The {\it Bianchi models} are the most general homogeneous cosmologies with a
3-dimensional group of isometries. These groups are in a one--to--one
correspondence with 3-dimensional Lie algebras, which were classified long ago
by Bianchi$^{\Bian}$. There are nine distinct 3-dimensional Lie algebras, and
consequently nine types of Bianchi cosmology. The 3-metric of each of these
models can be written in the form
$$h_{ij}\dd x^i\dd x^j=h_{ij}(t)\Bom^i\otimes\Bom^j,\eqn\Bianchigen$$
where $\Bom^i$ are the invariant 1-forms associated associated with the
isometry group. The simplest example is Bianchi I, which corresponds to the
Lie Group $\Rop^3$. In that case we can choose $\Bom^1=\dd x$, $\Bom^2=\dd y$,
and $\Bom^3=\dd z$, so that
$$h_{ij}\dd x^i\dd x^j=a^2(t)\dd x^2+b^2(t)\dd y^2+c^2(t)\dd z^2,\eqn
\Bianchione$$
and the minisuperspace coordinates are $\{a,b,c,\PH\}$. If we take the spatial
directions to be compact such a universe will have toroidal topology. In the
special case that $a(t)=b(t)=c(t)$ we retrieve the spatially flat ($k=0$) FRW
universe.

The most complicated, and possibly the most interesting, Bianchi model is
Bianchi IX, associated to the Lie group $SO(3,\Rop)$. In this case the
invariant 1-forms may be written as
$$\eqalign{\Bom^1=&-\sin\ps\,\dd\th+\sin\th\cos\ps\,\dd\varphi,\cr
\Bom^2=&\phantom{-}\cos\ps\,\dd\th+\sin\th\sin\ps\,\dd\varphi,\cr
\Bom^3=&\phantom{-}\cos\th\,\dd\varphi+\dd\ps,\cr}\eqn\Bianchiforms$$
in terms of the Euler angles, $(\ps,\th,\varphi)$, on the 3-sphere, $S^3$. The
spatial sections of the geometry resulting from \Bianchigen, \Bianchiforms\
have the topology of $S^3$, but are only spherically symmetric in the special
case that $h\Z{11}(t)=h\Z{12}(t)=\dots h\Z{33}(t)$, which corresponds to the
$k=+1$ FRW universe. Geometrically the spatial hypersurfaces can thus
be thought of as squashed, twisted 3-spheres [see \fig\BIANCHI]. Bianchi IX has
played an important role in classical cosmological studies of the initial
singularity -- it is the basis of the so-called ``mixmaster universe''$^{\MisC,
\BeLiKh}$. As a classical dynamical system Bianchi IX is extremely
interesting because it appears to be chaotic, but only just on the verge of
being so. Over the years there has been much debate as to whether Bianchi IX
is or is not chaotic, and this seems to have been recently resolved by an
explicit demonstration that it is not integrable\foot{\blacksquare}{Technically
speaking, what has been proved is the failure of integrability in the
Painlev\'e sense. While this does not guarantee the existence of chaotic
regimes, it does make their existence extremely probable.}$^{\LaMuCo}$. The
corresponding minisuperspace model will have six independent coordinates in
addition to the scalar field coordinate.
\ifig\BIANCHI{Schematic geometry of spatial hypersurfaces in the Bianchi IX
universe.}{\epsfysize=5cm\epsfbox{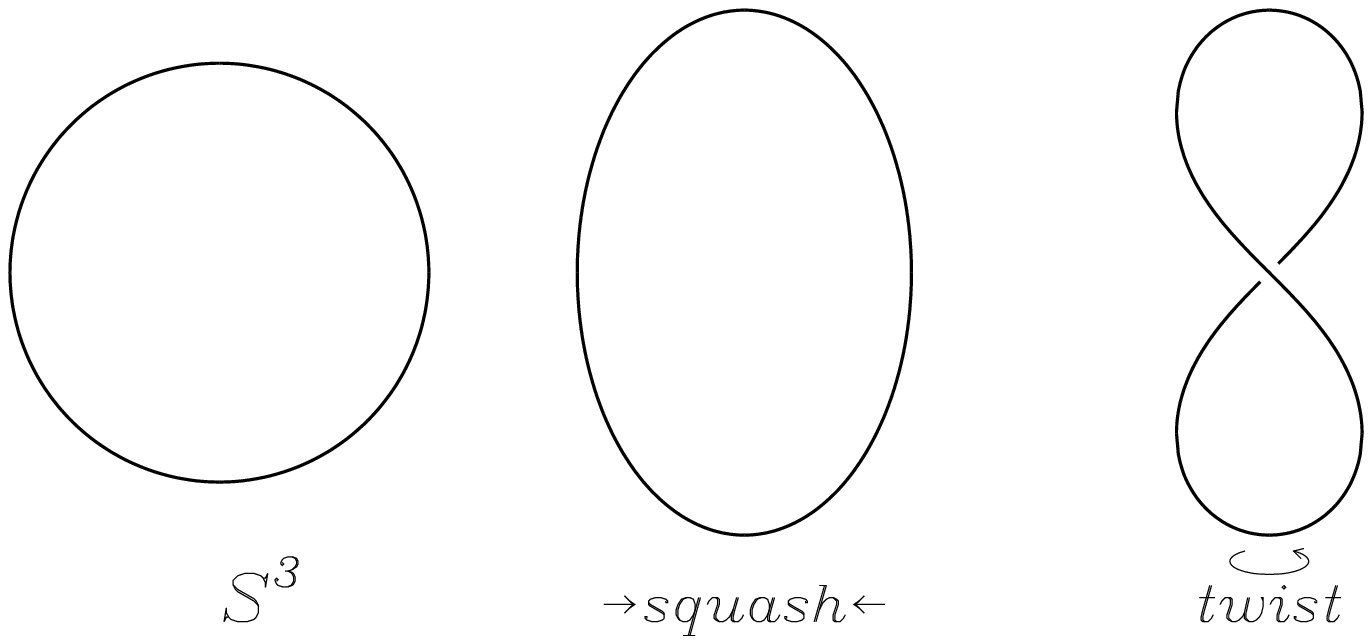}}

Let us now consider the minisuperspace corresponding to an arbitrary
homogeneous cosmology. We will assume that the minisuperspace is of dimension
$n$, and will denote the minisuperspace coordinates by $\{\qA\}$. Since $\NN^i=
0$ by assumption, it follows from the definitions \extrinsic\ and \invDeWitt\
that
$$\GG^{ijkl}\dot h_{ij}\dot h_{kl}=4\sqrt{h}\NN^2\left(K_{ij}K^{ij}-K^2\right),
\eqn\stepA$$
and consequently the Lorentzian action \actionD\ now takes the form
$$\eqalignno{S=&\int\dd t\left[{1\over2\NN}\GG\X{AB}(q)\dqA\dqB-\NN\UU(q)\right
],&\Eqn\actionmini\cr
\noalign{\hbox{where}}
\GG\X{AB}\dd\qA\dd\qB=&\int\dd^3x\left[{1\over8\KA}\GG^{ijkl}\de h_{ij}\de h_
{kl}+\sqrt{h}\de\PH\de\PH\right],&\Eqn\DeWittmini\cr}$$
is the {\it minisupermetric}, which is now of finite dimension, $n$, and
$$\UU=\int\dd^3x\sqrt{h}\left[{1\over4\KA}\left(-\ric3+2\LA\right)+\V(\PH)
\right].\eqn\Upot$$
The action \actionmini\ is simply equivalent to that for a ``point particle''
moving in a ``potential'' $\UU$. Variation of \actionmini\ with respect to
$\qA$ thus yields a geodesic equation with force term,
$${1\over\NN}\Der dt\left(\dqA\over\NN\right)+{1\over\NN^2}\gABC\dot q
\W B\dot q\W C=-\GG\W{AB}\Deriv\pt{\qB}\UU,\eqn\geodesic$$
where $\gABC$ are the Christoffel symbols determined from the minisupermetric,
while variation of \actionmini\ with respect to $\NN$ yields the Hamiltonian
constraint
$${1\over2\NN^2}\GG\X{AB}(q)\dqA\dqB+\UU(q)=0.\eqn\miniHamcon$$
The general solution to \geodesic, \miniHamcon\ will have $2n-1$ independent
parameters, one of which is always trivial in the sense that it corresponds to
a choice of origin of the time parameter. In studying any particular
minisuperspace model we must take care to check that what we have done above
is consistent, as it does not always follow that substituting a particular
ansatz into an action before varying it will yield the same result as
substituting the same ansatz into the field equations obtained from variation
of the original action. Eqs.\ \geodesic\ and \miniHamcon\ should correspond
respectively to the $(ij)$ and $(00)$ components of the original Einstein
equations, while the $(0i)$ equation is trivially satisfied in the present
case.

Quantisation is greatly simplified because now that our configuration
space is finite-dimensional we are effectively dealing with the quantum
{\it mechanics} of a constrained system. The canonical momenta and Hamiltonian
are respectively given by
$$\eqalignno{\PI\X A=&\Deriv\pt{\dqA}L={\GG\X{AB}\dqB\over\NN}\,,&\Eqn\PIqA\cr
H=\PI\X A\dqA-L=&\NN\left[\half\GG\W{AB}\PI\X A\PI\X B+\UU(q)\right]\equiv
\NN\hh.&\Eqn\HamqA\cr}$$
The $\PI\X A$ are related to the canonical momenta \Pih--\PiN\ defined
earlier by integration over the 3-volume of the hypersurfaces of homogeneity
in \DeWittmini. In terms of the new variables the action \actionmini\ and
Hamiltonian constraint \miniHamcon\ are respectively
$$\eqalignno{S&=\int\dd t\left[\PI\X A\dqA-\NN\hh\right],&\Eqn\actionqA\cr\half
&\GG\W{AB}\PI\X A\PI\X B+\UU(q)=0.&\Eqn\miniHamconqA\cr}$$
Under canonical quantisation \miniHamconqA\ yields the Wheeler-DeWitt equation
$$\eqalignno{\hat\hh\PS=&\left[-\half\Lap^2+\UU(q)\right]\PS=0,&\Eqn\miniWdW\cr
\noalign{\hbox{where}} \Lap\equiv&{1\over\sqrt{-\GG}}\pt\X A\left[\sqrt{-\GG}\,
\GG\W{AB}\pt\X B\right]&\Eqn\miniLap\cr}$$
is the Laplacian operator of the minisupermetric. In arriving at \miniWdW\ we
have made an explicit ``natural choice'' of factor ordering$^{\DeW,\HaPaA}$ in
order to accommodate the factor-ordering ambiguity. This choice is favoured by
independent minisuperspace calculations of the prefactor, using zeta function
regularisation and a scale invariant measure, which can then be related to
factor ordering dependent terms through a semiclassical expansion of the
Wheeler-DeWitt equation$^{\LouA}$.

An alternative ``natural choice''$^{\MisB,\HalB\hS{\MosA}\PagB}$ of factor
ordering would yield a {\it conformally invariant} Wheeler-DeWitt equation,
$$\hat\hh\PS=\left[-\half\Lap^2+{n-2\over8(n-1)}{\cal R}+\UU(q)\right]\PS=0,
\eqn\altminiWdW$$
where $\cal R$ is the scalar curvature obtained from the minisupermetric.

\subsec{The WKB approximation}

In view of the difficulties associated with solving the Wheeler-DeWitt equation
in general, the best we can realistically hope for in many minisuperspace
models is to look for appropriate approximate solutions in the semiclassical
limit\foot{\eqcirc}{The semiclassical limit of the full superspace
Wheeler-DeWitt equation has been treated more formally by a number of authors.
See, e.g., [\Ban], [\Kim] and references therein.}, in which
$$\PS\simeq\sum_n\PS_n\equiv\sum_n{\cal A}_n\e^{-I_n},\eqn\semiclass$$
where the sum is over saddle points of the path integral, the ${\cal A}_n$
being appropriate (possibly complex) prefactors. In general we might
expect to find regions in which the wavefunction is exponential, $\PS\simeq\e^
{-I}$, and regions in which it is oscillatory, $\PS\simeq\e^{iS}$. The latter
could be viewed as the wavefunction of a universe in the classical
``Lorentzian'' or ``oscillatory'' region, while the former would correspond to
a universe in a classically inaccessible ``Euclidean'' or ``tunneling'' region.
As has already been mentioned, the sum in
\semiclass\ will in general contain a number of saddle points with an action,
$I_n$, which is neither purely real nor purely imaginary.

Our own universe is of course Lorentzian at late times, and therefore the only
minisuperspace models which can be of direct physical relevance are those for
which the Wheeler-DeWitt equation does possess approximate solutions of the
oscillatory type. Approximate solutions of this type can be obtained by
performing a WKB expansion, for which purpose it is necessary to restore $\hbar
$ in the minisuperspace Wheeler-DeWitt equation \miniWdW. If we assume that
each component $\PSn$ satisfies \miniWdW\ separately, then
$$\eqalign{0&=\hat\hh\PS_n=\left[-\half\hbar^2\Lap^2+\UU\right]\An\e^{-I_n/
\hbar}\cr &=\e^{-I_n/\hbar}\left\{\left[-\half\left(\LAP I_n\right)^2+\UU
\right]\An+\hbar\left[\LAP I_n\cdot\LAP\An+\half\An\Lap^2I_n\right]+\OO(\hbar^
2)\right\},\cr}\eqn\WdWwkb$$
where the dot implies contraction with the minisupermetric $\GG\W{AB}$. The $
\OO(\hbar^0)$ and $\OO(\hbar)$ terms give two equations for $I_n$ and $\An$.
If we decompose $I_n$ into real and imaginary parts according to $I_n=\Rn-i\Sn$
then the real and imaginary parts of the $\OO(\hbar^0)$ term in \WdWwkb\ give
$$\eqalignno{-\half\left(\LAP\Rn\right)&^2+\half\left(\LAP\Sn\right)^2+\UU=0,&
\Eqn\WdWwkbA\cr &\LAP\Rn\cdot\LAP\Sn=0.&\Eqn\WdWwkbB\cr}$$
Provided that the imaginary part of the action varies much more rapidly than
the real part, i.e., $(\LAP\Rn)^2\ll(\LAP\Sn)^2$, then \WdWwkbA\ is the
Lorentzian Hamilton-Jacobi equation for $\Sn$:
$$\half\GG\W{AB}\Deriv\pt\qA\Sn\Deriv\pt\qB\Sn+\UU(q)=0.\eqn\wkbHJ$$
Comparison of \wkbHJ\ with \miniWdW\ suggests a strong correlation between
coordinates and momenta, and invites the identification
$$\PI\X A=\Deriv\pt\qA\Sn.\eqn\wkbmom$$
If we differentiate \wkbHJ\ w.r.t.\ $\qC$ we obtain
$$\half\GG\W{AB},\X{C}\Deriv\pt\qA\Sn\Deriv\pt\qB\Sn+\GG\W{AB}\Deriv\pt\qA\Sn
{\pt^2\Sn\hphantom{\pt}\over\pt\qB\pt\qC}+\Deriv\pt\qC\UU=0.\eqn\foo$$
If we define a minisuperspace vector field
$$\Der\dd s\equiv\GG\W{AB}\Deriv\pt\qA\Sn\Der\pt\qB,\eqn\minivector$$
then combining \wkbmom, \foo\ and \minivector\ we obtain
$$\Deriv\dd s{\PI\X C}+\half\GG\W{AB},\X C\PI\X A\PI\X B+\Deriv\pt\qC\UU=0,\eqn
\wkbgeo$$
which after raising indices is the same geodesic equation \geodesic\
obtained earlier provided we identify the parameter, $s$, with the proper time
on the geodesics.

We can now solve the equation given by the $\OO(\hbar)$ term of \WdWwkb.
Since $|\LAP\Rn|\ll|\LAP\Sn|$ it follows that the terms involving $\Rn$ can be
neglected, and thus
$$\LAP\Sn\cdot\LAP\An\equiv\GG\W{AB}\Deriv\pt\qA\Sn\Deriv\pt\qB\An\equiv\Deriv
\dd s\An=-\half\An\Lap^2\Sn,\eqn\WdWwkbC$$
which may be readily integrated. We thus obtain a first-order WKB wavefunction
$$\PSn=\CC_n\exp\left(i\Sn-\half\int\dd s\,\Lap^2\Sn\right),\eqn\wkbwavefn$$
where $\CC_n$ is an arbitrary (complex) constant to be appropriately
normalised, and we have reverted to natural units in which $\hbar=1$.

The wavefunction \wkbwavefn\ could be considered to be the analogue of the
wavefunction for coherent states in ordinary quantum mechanics,
$$\pS_n(\xx,t)=\cn\e^{i\mathchar"0B70_n\xx}\exp\left(-(\xx-\bar\xx_
n(t))^2\over\mathchar"0B1B^2\right)\,,\eqn\coherent$$
which describes a wave packet which is ``peaked'' about a classical particle
trajectory, $\bar\xx_n(t)$, and which thus roughly ``predicts'' classical
behaviour. This becomes problematic, however, if we consider a superposition,
$\pS=\sum_n\pS_n$, of such states since interference between different wave
packets will in general destroy the classical behaviour. In order to interpret
the total wavefunction as saying that the particle follows a roughly classical
trajectory, $\bar\xx_n(t)$, with probability $|\cn|^2$, it is necessary that a
{\it decoherence} mechanism should exist which renders this quantum mechanical
interference negligible$^{\HalD}$.

The issue of quantum decoherence is clearly also of great importance in
quantum cosmology, since in order to interpret $\PS$ in \semiclass\ in a
similar fashion a similar mechanism must exist.
It has been recently argued, furthermore, that decoherence is a necessary
feature of the WKB interpretation of quantum cosmology, since without
decoherence the existence of chaotic cosmological solutions would lead to a
breakdown of the WKB approximation$^{\CaGo}$. This is analogous to similar
problems with the commutativity of the limits $t\to\infty$ and $\hbar\to0$
in ordinary quantum mechanics when applied to chaotic systems.

The issues involved in decoherence pose complex conceptual questions for the
fundamentals of quantum mechanics itself, quite apart from the problems
specific to quantum cosmology\foot{\boxtimes}{For further details see, e.g.,
[\HaLa]$\hS{[\Cal\HuPS}$[\Hu] and references therein.}. Here we
will merely assume that such a mechanism exists, and we will take the view
that $\PS$ can be considered to ``predict'' a classical spacetime if there
exist WKB-type solutions \wkbwavefn, which yield a strong correlation
between $\PI\W A$ and $\qA$ according to \wkbmom. The sense in which the
minisuperspace positions and momenta are ``strongly correlated'' can be made
more precise through the use of quantum distribution functions, such as the
Wigner function$^{\HaLa,\HalC}$. By use of the Wigner function one may show$^
{\HalC}$ that wavefunctions of the oscillatory type, $\PS\goesas\e^{i\SS}$,
predict a strong correlation between coordinates and momenta, whereas
wavefunctions of the type $\PS\goesas\e^{-\mathchar"0B49}$, which are also
typical minisuperspace solutions, do not. Such exponential wavefunctions can
thus be considered as describing universes in a purely quantum ``tunneling''
regime, before the quantum to classical transition. We will interpret
wavefunctions, $\PS\goesas\e^{i\SS}$, as corresponding to classical spacetime,
or rather a set of classical spacetimes as $\SS$ is a first integral of the
equations of motion.

\subsec{Probability measures}

Given a solution, $\PS$, to the Wheeler-DeWitt equation it is necessary to
construct a probability measure in order to make predictions. One central
question in quantum cosmology is how one should construct such a measure.

The minisuperspace Wheeler-DeWitt equation \miniWdW\ is a second-order
equation very much like the Klein-Gordon equation in ordinary field theory, and
it readily follows from \miniWdW\ that the current$^{\DeW}$
$$\JJ=-\half i\left(\PSB\LAP\PS-\PS\LAP\PSB\right)\eqn\current$$
is conserved: $\LAP\cdot\JJ=0$. The similarity to the Klein-Gordon current
extends to the fact that the natural inner product$^{\DeW}$ constructed from
$\JJ$ is not positive-definite and so gives rise to negative probabilities. In
quantum field theory this is not a problem since one
can split the wavefunction up into positive and negative frequency components
which correspond to particles and anti-particles. However, as has already been
mentioned there is no well-defined notion of positive frequencies in superspace
on account of its lack of symmetries$^{\Kuch}$. A further problem is that many
natural wavefunctions would have zero norm with this definition. For example,
the no-boundary wavefunction is real and gives $\JJ=0$.

The similarity of $\PS$ to the Klein-Gordon field has suggested to many people
that one should turn $\PS$ into an operator, $\hat\PS$, thereby introducing
quantum field theory on superspace, or ``third quantisation''. One then arrives
at operators which create and annihilate universes. However, as we do not
perform measurements over a statistical ensemble of universes it is not clear
how we can arrive at sensible probabilities using such a formulation.

The difficulties with the Klein-Gordon current of course led Dirac to introduce
the Dirac equation, and it is worth mentioning that a similar resolution of
the problem is available in supersymmetric quantum cosmology. In particular,
one can go to a theory which includes fermionic variables by considering
quantum cosmology based on supergravity\foot{\blacktriangleleft}{In fact, a
na\"{\i}ve first-order Hamiltonian formulation for the minisuperspaces of
homogeneous cosmologies was found early on [\Ryan], but until the development
of supergravity there was no natural interpretation of the Dirac-type
constraint equation obtained.} rather than the purely bosonic Einstein theory.
The constraints of supergravity, which may be viewed as the Dirac square root
of the constraints of general relativity$^{\Teit,\TabTe}$, are reducible to
first-order equations. Furthermore, this also translates into simplifications
in homogeneous minisuperspace models -- the appropriate constraint equation
which determines the quantum evolution of the wavefunction can be considered to
be the Dirac square root of the Wheeler-DeWitt equation$^{\MOR\Hs{\DeHu\Grah}
\DeHO}$. As a result it is possible to construct$^{\MOR}$ a Dirac-type
probability density which is conserved by the equation ${\cal Q}\PS=0$, where
$\cal Q$ is the supercharge.

Another alternative to the question of the probability measure is to use $|\PS|
^2$ directly as a probability measure$^{\HaPaA,\VilE}$, by defining the
probability of the universe being in a region, $\euA$, of superspace by
$$\PP(\euA)\propto\int_\euA|\PS|^2\One\eqn\ProbOne$$
where $\One$ is the volume-element on superspace, $*$ being the Hodge dual in
the supermetric. This definition of a
necessarily positive-definite probability density works very well for
homogeneous minisuperspaces, for which the volume form $\One$ is independent
of $x\in\SI$. This is perhaps not surprising since as was observed in \S3.4
the assumption of homogeneity reduces the problem to one of quantum mechanics,
and $|\PS|^2$ is of course the probability density in conventional quantum
mechanics.

Problems with the definition \ProbOne\ do arise since even in some
simple examples the wavefunction is not normalisable, but instead $\left\langle
\PS|\PS\right\rangle=\infty$.
One further problem is that whereas in ordinary quantum mechanics $|\PS|^2$
describes a probability density in configuration space -- i.e., the space of
particle positions -- in quantum cosmology the configuration space is
(mini)superspace and time is implicitly contained in the (mini)superspace
coordinates. These coordinates cannot therefore be thought of as the mere
analogues of particle positions. As a result the recovery of the conservation
of probability and the standard interpretation of the quantum mechanics for
small subsystems is not necessarily straightforward in the approach based on
\ProbOne, and may involve understanding some subtle questions about the role of
time in quantum gravity.

Ultimately a formulation such as \ProbOne\ which is based on absolute
probabilities may not be required since it is impossible to measure
statistical ensembles of universes and thus all we can really test are {\it
conditional probabilities} rather than absolute probabilities. For example, a
relevant testable probability might be the probability, $\PP(\euA|\euB)$, of
finding $\PS$ in a region $\euA$ of superspace given that $\PS$ started in
another region $\euB$ of superspace. Page$^{\PagA,\PagC}$ has explored the
construction of conditional probabilities in quantum cosmology without the use
of absolute probabilities.

Clearly the issues surrounding the choice of probability measure involve some
deep conceptual problems which may perhaps get to the heart of the broader
conceptual basis of quantum gravity. Such issues have been discussed by a
number of authors in the context of quantum cosmology$^{\PagA,\HarA,\VilE\hS
{\HarB}\PagG}$ and I will not address them here in any detail.

For the purposes of examining
how we might hope to make predictions from the proposed boundary conditions
of \S4.1,4.2 we shall merely consider quantum cosmology in the WKB limit. In
this limit it follows from \wkbwavefn\ that each of the components $\PSn$ in
\semiclass\ has a conserved Klein-Gordon-type current \current\ given by
$$\Jn\simeq|\An|^2\LAP\Sn,\eqn\wkbcurrent$$
which flows very nearly along the direction of the classical trajectories.
The current conservation law $\LAP\!\!\X A\,\Jn\W{\ \,A}=0$ implies that
$$\dd\PP=\Jn\W{\ \,A}\dd\HYP\!\X A,\eqn\wkbprob$$
is a conserved probability measure on the set of trajectories with tangent
$\LAP\Sn$, where $\dd\HYP\!\X A$ is the element of a hypersurface, $\HYP$, in
minisuperspace which cuts across the flow and intersects each curve in the
congruence once and only once. One finds that for a pencil of trajectories near
the classical trajectory the probability density \wkbprob\ is
positive-definite. Vilenkin has argued$^{\VilE}$ that positive-definiteness of
the probability measure is really only required in the semiclassical limit, as
this is the only limit in which we obtain a universe accessible to observation
where the conventional laws of physics apply. Therefore given that the current
\current\ works in the WKB limit, this is all that is needed if we are
content that ``probability'' and ``unitarity'' are only approximate concepts
in quantum gravity. One can also show$^{\PagA,\HaPaA}$ that the manifestly
positive definition \ProbOne\ yields essentially the same result as \wkbprob\
in the WKB limit.

\subsec{Minisuperspace for the Friedmann universe with massive scalar field}

Let us now apply our results to the particularly simple case of a homogeneous,
isotropic universe with a single scalar field, with a potential which allows
for inflationary behaviour. A quadratic potential is possibly the simplest
example with this property, and thus has been much studied in quantum
cosmology.

For convenience we will introduce a numerical normalisation factor $\si^2=\KA/(
3\Vv)$ into the metric, where $\Vv$ is the 3-volume of the unit hypersurface --
e.g., $\Vv=2\pi^2$ for the 3-sphere, $k=+1$. We are considering {\it closed
universes only}, which requires making topological identifications for the $k=0
,-1$ cases, so that $\Vv$ remains finite. 
In place of \coordsA, \coordsB\ and \FRW\ we then have a metric
$$\dd s^2=\si^2\left\{-\NN^2\dd t^2+a^2(t)\left[{\dd r^2\over1-kr^2}+r^2
\dOM\right]\right\}.\eqn\FRWalt$$
We also take the scalar field action to be normalised by
$$S\Ns{matter}={3\over\KA}\int\limits_{\Mi{}}\dd^4x\sqrt{-g}\left(
-{1\over2}g^{\mu\nu}\pt_\mu\ph\pt_\nu\ph-{V(\ph)\over2\si^2}\right).
\eqn\actionE$$
Thus $\NN\to\si\NN$, $a\to\si a$, $\PH=\sqrt{3}\ph/\ka$ and $\V=3V/(2\KA\si^2)$
relative to our earlier definitions, and $a$, $\ph$ and $V$ are now
dimensionless.
As for the general homogeneous minisuperspace discussed in the previous
section, we use a gauge in which $\NN_i=0$. From \FRWalt\ it follows that
$\ric3={6k\over\si^2a^2}$ and $K_{ij}=-{\dot a\over\si\NN a}h_{ij}$,
and thus the action takes the form \actionmini\ with a minisupermetric
$$\GG\X{AB}\dd\qA\dd\qB=-a\dd a^2+a^3\dd\ph^2,\eqn\miniFRWmetric$$
and potential
$$\UU=\half\left(a^3V(\ph)-ka\right).\eqn\miniFRWpot$$
Alternatively, it is sometimes useful to
express \miniFRWmetric\ in terms of a conformal gauge
$$\eqalignno{\GG\X{AB}\dd\qA\dd\qB&=\e^{3\al}\left(-\dd\al^2+\dd\ph^2\right),&
\Eqn\miniFRWconformal\cr&=-(4uv)^{-1/4}\dd u\dd v,&\Eqn\miniFRWnull\cr}$$
where $\al=\ln a$, or alternatively in null coordinates
$$\eqalign{u=&\half\e^{2(\al-\ph)}=\half a^2\e^{-2\ph},\cr v=&\half\e^{2(\al+
\ph)}=\half a^2\e^{2\ph}.\cr}\eqn\nullcoord$$
The canonical momenta are given by
$$\PIz=0,\qquad\PI_a={-a\dot a\over\NN},\qquad\PI_\ph={a^3\dot\ph\over\NN},
\eqn\FRWcanmom$$
and the classical equations of motion yield
$$\eqalignno{&\hh=\half\left(-{a\dot a^2\over\NN^2}+{a^3\dot
\ph^2\over\NN^2}-ka+a^3V\right)=0,&\Eqn\mFRWa\cr
&{1\over\NN}\Der\dd t\left(\dot a\over\NN\right)+{2a\dot\ph^2\over
\NN^2}-aV=0,&\Eqn\mFRWb\cr &{1\over\NN}\Der\dd t\left(\dot\ph\over\NN\right
)+3{\dot a\dot\ph\over a\NN^2}+\half\Deriv\dd\ph V=0,&\Eqn\mFRWc\cr}$$
which are equivalent to the geodesic equation \geodesic\ with metric
\miniFRWmetric\ and potential \miniFRWpot.
The lapse function is not of physical relevance classically since we can choose
an alternate proper time parameter, $\dd\ta=\NN\dd t$, or equivalently choose a
gauge $\NN=1$ in \mFRWa--\mFRWc\ so that $t$ is the proper time. One of these
equations depends on the other two by virtue of the Bianchi identity, as is
always the case in general relativity. We thus effectively have
two independent differential equations in two unknowns, one first order and one
second order, or equivalently an autonomous system of three first order
differential equations. The solution therefore depends on three free
parameters, but as mentioned above one of these amounts to a choice of the
origin of time which is of no physical importance. Thus there is a
two-parameter family of physically distinct solutions.

The classical solutions cannot be written in a simple closed form except for
certain special values of $k$ and the potential $V(\ph)$, which
unfortunately does not even include the quadratic potential\foot
{\vartriangleright}{The Einstein equations for the FRW universe with a massive
scalar field can be solved approximately [\HaPaB] in various limits, however.}
$V(\ph)=m^2\ph^2$. The qualitative property of the solutions may
nonetheless be determined by studying the 3-dimensional phase space. Instead of
choosing a particular potential, however, let us suppose that we are in a
region of the phase space for which $V$ can be approximated by a constant. Such
conditions are more or less met in the ``slow-rolling approximation''$^{\KoTu}$
of inflationary cosmology, in which $|V'/V|\ll6$ and $|V''/V|\ll9$. In this
approximation the dynamics is described by setting $\dot\ph\simeq0$ in \mFRWa\
and \mFRWb, and setting $\ddot\ph\simeq0$ in \mFRWc. In approximating the
nearly flat region of the potential $V$ by a cosmological constant we ignore
the slow change of the scalar field determined from \mFRWc. We thus obtain a
simplified model which possesses classical inflationary solutions, provided
the constant $V$ is chosen to be {\it positive}. This will serve as a useful
test model for quantum cosmology.

In the case that $V$ is constant, eq.\ \mFRWc\ integrates to
give $\dot\phi=Ca^{-3}$, where $C$ is an arbitrary constant, and it therefore
follows that the Friedmann equation \mFRWa\ can be written in terms of an
elliptic integral in $a^2(\et)$,
$$\et=\half\int_0^{a^2}{\dd z\over\sqrt{Vz^3-kz^2+C^2}}\,,\eqn\elliptic$$
where $\et$ is the conformal time parameter defined by $\dd\et=a^{-1}\dd t$. It
is thus possible to express the general solution in terms of elliptic
functions. Since the properties of such functions are not very transparent
perhaps, we can alternatively plot the 2-dimensional phase space -- e.g., in
terms of the variables $\dot\ph$ and $\dot\al$, as in \fig\PHASE\
in order to understand the qualitative features of the solutions. There are
four critical points, at $(\dot\ph,\dot\al)=\left\{\left(0,\pm1/\sqrt{V}\right)
,\left(\pm1/\sqrt{2V},0\right)\right\}$, the first two being nodes, $A_\pm$,
which are endpoints for all values of the spatial curvature, $k$, and the
latter two saddle points, $B_\pm$, for the $k=+1$ solutions.
\ifig\PHASE{The 2-dimensional phase plot of $\dot\al=\dot a/a$ versus $\dot\ph$
for the simplified model with constant potential, $V$.}{\epsfbox{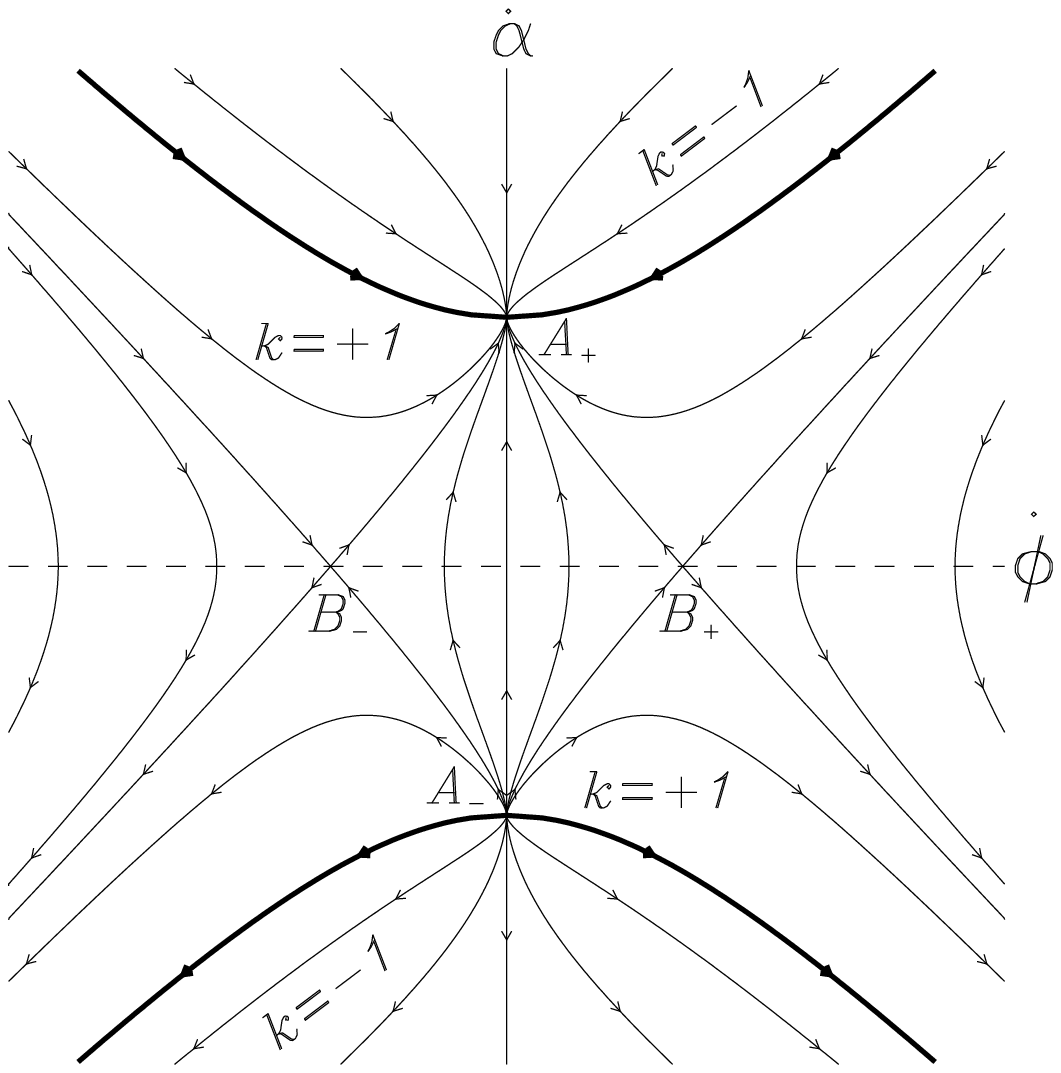}}

The general solution for the spatially flat case ($k=0$), which corresponds to
the bold separatrices shown in \PHASE, is given by$^{\Wilt}$
$$\eqalign{a&=C_a\left|\sinh\frac32\sqrt{V} t\right|^{1/3}\left|\cosh\frac32
\sqrt{V}t\right|^{1/3},\cr\ph&=\frac13\ln\left|\tanh\frac32\sqrt{V}t\right|+C_
\ph,\cr}\eqn\separatrixA$$
in closed form, where $C_a$ and $C_\ph$ are arbitrary constants. At late times
the solutions \separatrixA\ have an exponential scale factor, $a\to C_a\exp(
\sqrt{V}\,t)$ as $t\to\infty$, and constant scalar field $\ph\to C_\ph$.
Furthermore, one
can see from \PHASE\ that all the $k=-1$ solutions in the upper half-plane,
and a number of the $k=+1$ solutions are also attracted to the point $A_+$
with a similar inflationary behaviour at late times. (The corresponding point
$A_-$ on the $k=0$ separatrix corresponds to the time-reversed solution, with
an inflationary phase as $t\to-\infty$.) The simplified universe corresponding
to \PHASE\ of course is far from being the complete picture, as the model
does not allow for any exit from inflation. However, \PHASE\ illustrates the
typical situation that a given model will possess regimes with inflationary
behaviour and regimes with non-inflationary behaviour. In \PHASE\ the $k=+1$
solutions to the right and left of the separatrices that pass through $B_\pm$
fall into the latter category, for example. The situation becomes even more
involved when one considers the full 3-dimensional phase space for some
particular potential $V(\ph)$.

The case of the $k=+1$ solutions in \PHASE\ illustrates the general feature
that classical dynamics are highly dependent on initial conditions. In order to
obtain a sufficiently long inflationary epoch to overcome the problems
mentioned in the Introduction, (of order 65 $e$-folds growth in the scale
factor), the initial values of $\ph$ and $\dot\ph$ must be restricted to a
particular region of the phase space. In particular, $\dot\ph$ must be small
initially. Classically, there is no {\it a priori} reason for one choice of
initial conditions over any other choice, unless further ingredients are added.
The degree to which inflationary initial conditions are preferred relative to
other initial conditions -- i.e., how probable is inflation? -- is precisely
the sort of question that we might therefore hope quantum cosmology could
answer.

It is possible to attempt to solve this question without resorting to quantum
cosmology. To do this one must construct a measure on the set of all universes$
^{\Hen,\GHSt}$, and then compare the number of inflationary solutions with
a sufficiently long exponential phase to the number of other solutions.
Preliminary results$^{\GHSt}$ seemed to indicate that almost all models with a
massive scalar field undergo a period of inflation. However, a more careful
analysis$^{\HaPaB}$ revealed that the answer is ambiguous, as both the set of
inflationary solutions and the set of non-inflationary solutions have infinite
measure.

Alternatively, if as we expect the universe began in some sort of tunneling
process or similar transition from a quantum regime, then we could expect the
``initial'' classical parameters to be determined, at least in a
probabilistic fashion, from more fundamental quantum processes. The question
of the most probable state of the universe is then pushed back a level and
becomes: ``what is a typical wavefunction for the universe?''

In the context of the present minisuperspace model, therefore, we can proceed
by quantising the Wheeler-DeWitt equation \mFRWa, to obtain
$$\eqalign{\hat\hh\PS=\left[-\half\Lap^2+\UU\right]\PS
&=\half\left[{1\over a^3}\left(a\Der\pt a\, a\Der\pt a-\DDer\pt\ph2
\right)-ka+a^3V(\ph)\right]\PS\cr &=\half\e^{-3\al}\left[\DDer\pt\al2-\DDer\pt
\ph2-k\e^{4\al}+\e^{6\al}V(\ph)\right]\PS\cr &=(4uv)^{1/4}\left[2\Deriv\pt{u\pt
v}{^2\hphantom{uv}}-{k\over2}+(uv)^{1/2}V\right]\PS\cr&=0\cr}\eqn\mFRWd$$
in terms of the various sets of coordinates given earlier. In general boundary
conditions will have to specified in order to solve \mFRWd. However, we can
consider the approximate form of the WKB solutions without considering boundary
conditions for the time being.

We will confine ourselves to regions in which the potential can be approximated
by a cosmological constant, as in the analysis of \PHASE, so that we can drop
the term involving derivatives with respect to $\ph$ in \mFRWd, thereby
obtaining a simple 1-dimensional problem which is amenable to a standard WKB
analysis. The first order WKB wavefunction \wkbwavefn\ which solves \mFRWd\ in
this approximation is
$$\PS(a,\ph)\simeq\cases{\dsp{{\cal B}(\ph)\over a\left(a^2V(\ph)-k\right)^{1/4
}}\exp\left[{\pm i\over3V(\ph)}\left(a^2V(\ph)-k\right)^{3/2}\right],&$a^2V>k$,
\phantom{\hfill\Eqn\wkbpsO}\rlap{\hskip-23.885pt\wkbpsO}\cr
\noalign{\vskip2pt}
\dsp{\CC(\ph)\over a\left(k-a^2V(\ph)\right)^{1/4}}
\exp\left[{\pm1\over3V(\ph)}\left(k-a^2V(\ph)\right)^{3/2}\right],&$a^2V<k$.
\phantom{\hfill\Eqn\wkbpsE}\rlap{\hskip-22.54192pt\wkbpsE}\cr}
$$
If $V$ is positive, as was assumed above, then oscillatory type solutions will
thus exist for large values of the scale factor, while the exponential type
solutions will exist for small values of the scale factor if $k=+1$.
\Pfig\PENROSE{Conformal diagrams of the 2-dimensional minisuperspace. The
region where oscillatory WKB solutions exist, as given by the rough criterion
$a^2V>1$, is shaded for various potentials: (a) $V=0.25$ (\const); (b) $V=4$
(\const); (c) $V=0.25\ph^2$;\ifx\epsfbox\UnDeFiNeD{}\else\break\fi
(d) $V=25\ph^2$; (e) $V=\left[1-\e^{-\ph/f}\right]^2$ with $f=1.5$; (f)
$V=4\sinh^2\ph\exp\left[-f\e^{-2\ph}\right]$ with $f=0.1$.}
{\vbox{\offinterlineskip{\halign{&\epsfxsize=64mm\epsfbox{#}\cr 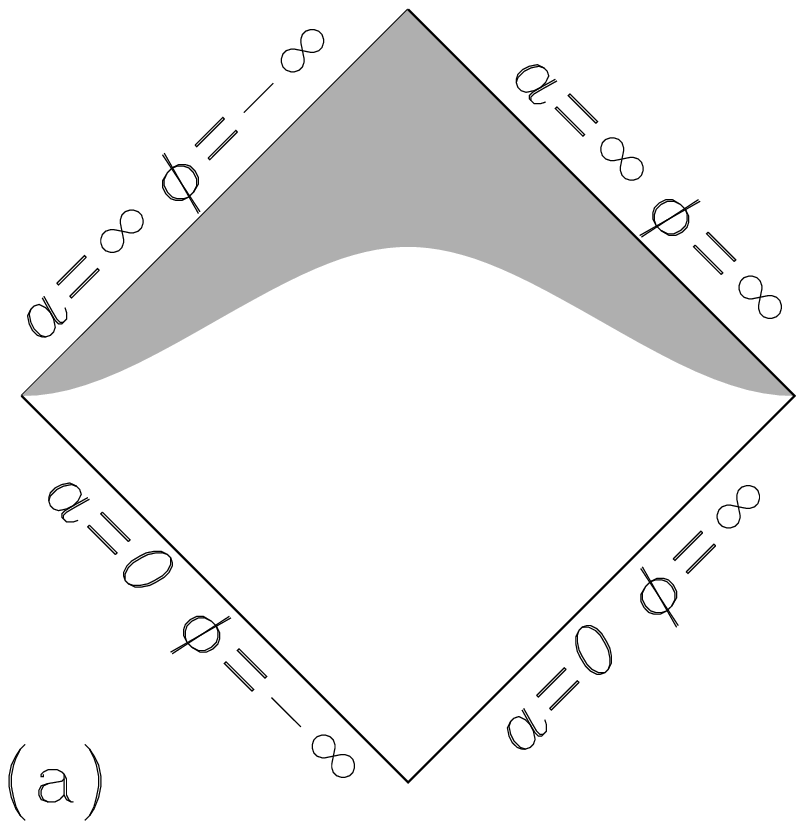&
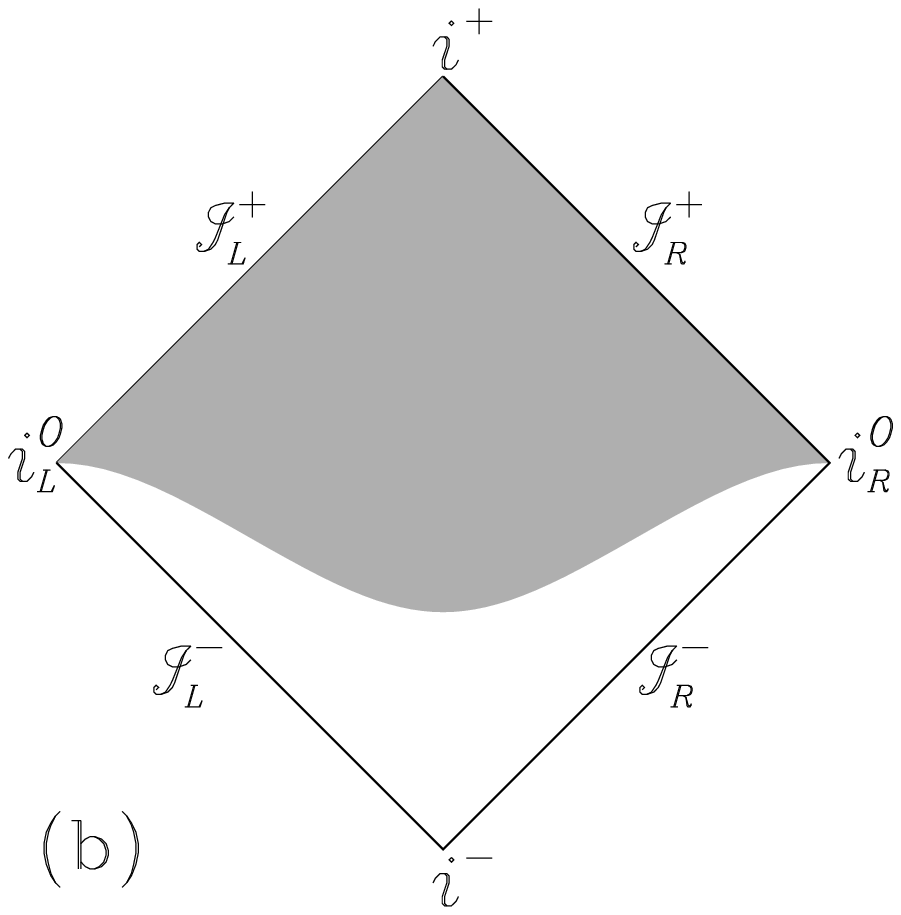\cr 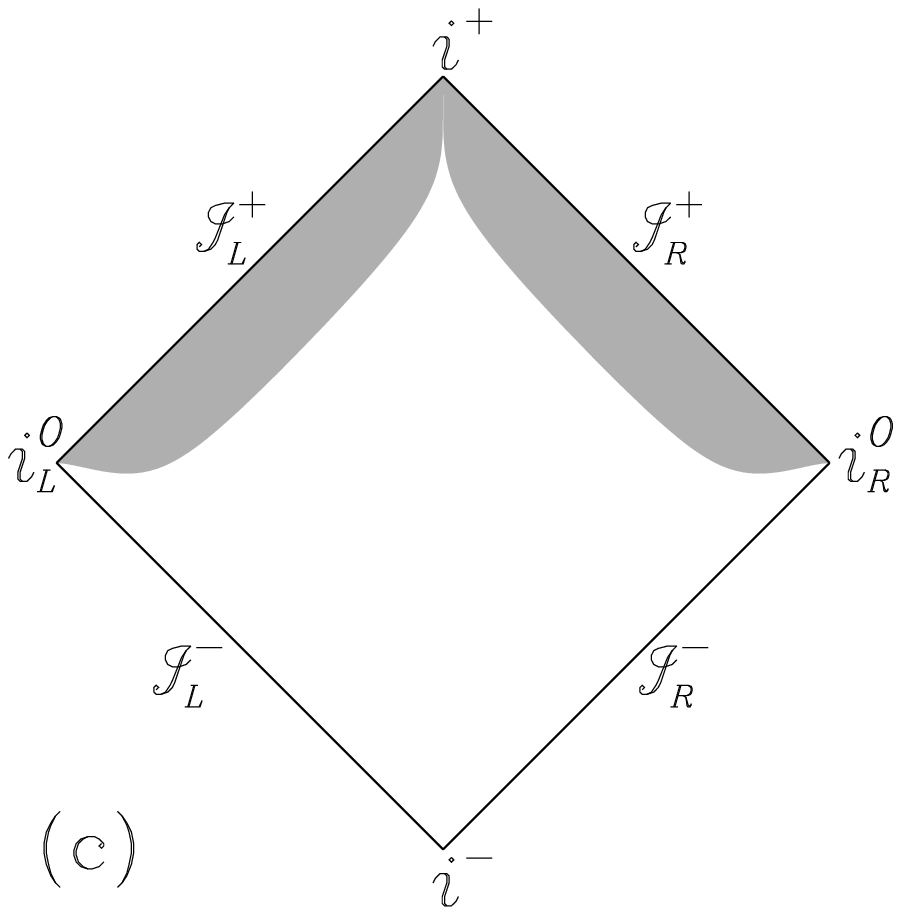&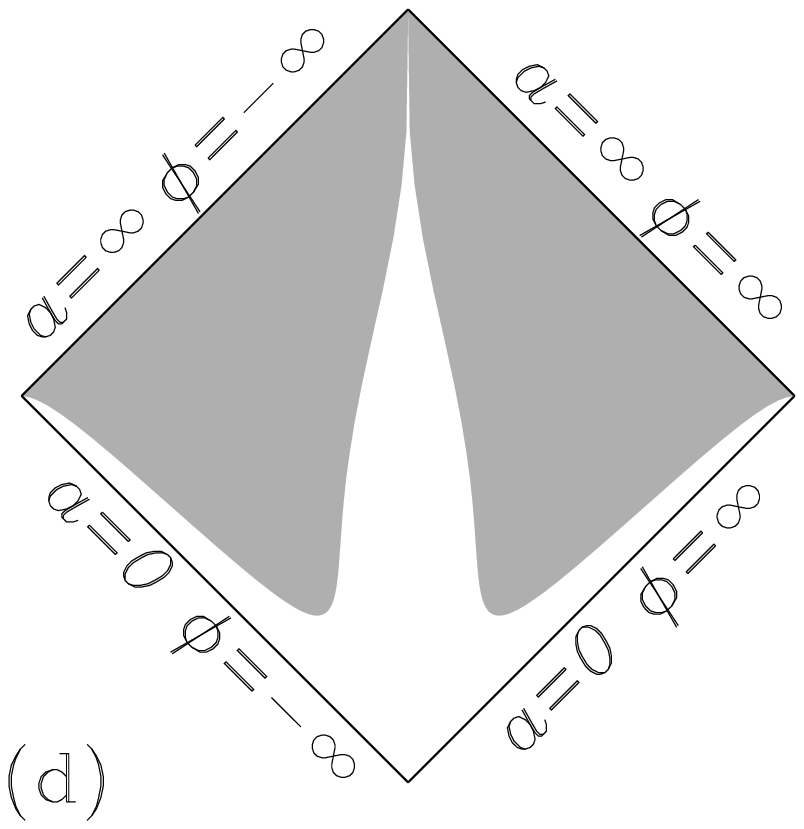\cr 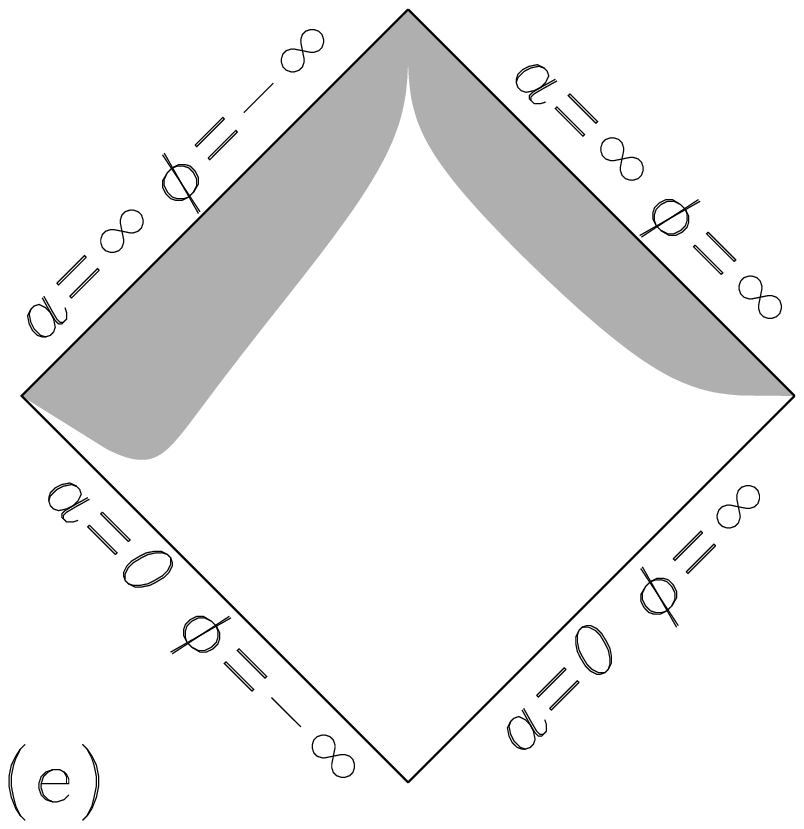&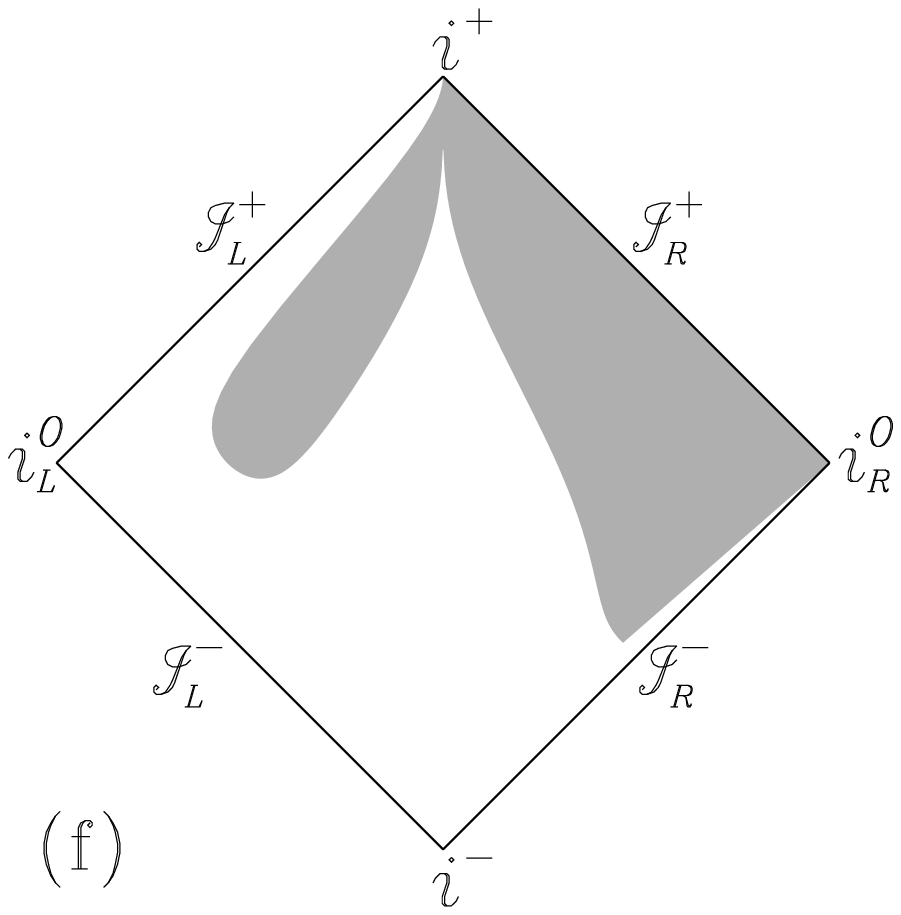\cr}}}}

The oscillatory solutions are of the form $\PS\goesas\e^{i\SS}$ (neglecting
the prefactor), where $\SS$ satisfies the Hamilton-Jacobi equation \wkbHJ.
Comparing this to the Hamiltonian constraint \miniHamconqA\ we find a strong
correlation \wkbmom\ between momenta and coordinates. For large scale factors,
$a^2V\gg|k|$, so that $\SS\simeq\pm\frac13a^3\sqrt{V}$. In this
limit \PIqA, \wkbmom\ and \miniFRWconformal\ thus yield
$$\eqalign{\PI_\al=\Deriv\pt\al\SS\quad&\hence\quad\dot\al\simeq\pm\sqrt{V},\cr
\PI_\ph=\Deriv\pt\ph\SS\quad&\hence\quad\dot\ph\simeq0,\cr}\eqn\wkbinflat$$
which correspond in fact to the inflationary points, $A_\pm$, of \PHASE. The
oscillatory wavefunction thus ``picks out'' classical inflationary universes.

Since the minisupermetric \miniFRWconformal\ is conformal to 2-dimensional
Minkowski space in the coordinates $(\al,\ph)$, it is convenient to represent
it by a Carter-Penrose conformal diagram (see \PENROSE).
In each case we plot $(p-q)$ horizontally and $(p+q)$ vertically, where $\tan
p=\al+\ph$, and $\tan q=\al-\ph$. The boundary consists
of points corresponding to past timelike infinity, $\ii^-=\left\{(a,\ph)\,
|\,a=0,\ph\w{finite}\right\}$, future timelike infinity, $\ii^+=\left\{(a,
\ph)\,|\,a=\infty,\ph\w{finite}\right\}$, left and right spacelike infinity,
$\ii^0\Z{L,R}=\left\{(a,\ph)\,|\,a=\w{finite},\ph=\pm\infty\right\}$;
and past and future null boundaries, $\scri^-\Z{L,R}=\left\{(a,\ph)\,|\,a=0,
\ph=\pm\infty\right\}$ and $\scri^+\Z{L,R}=\left\{(a,\ph)\,|\,a=0,\ph=\pm
\infty\right\}$. In each case the subscript $L$ (left) is associated with
$\ph\to-\infty$, and the subscript $R$ (right) with $\ph\to+\infty$. The
approximate region for which oscillatory WKB solutions exist is shown in
\PENROSE(a,b) for the approximate minisuperspace with a cosmological constant,
in \PENROSE(c,d) for $V(\ph)=m^2 \ph^2$, and in \PENROSE(e,f) for potentials,
$V(\ph)$, typically found in higher-derivative gravity theories and in string
theory with supersymmetry breaking.

Naturally it is of interest to know whether the inflationary WKB wavefunctions
are {\it typical} solutions to the Wheeler-DeWitt equation. To determine a
typical wavefunction for the universe, we need to make a choice of boundary
conditions for $\PS$ when solving \mFRWd.
\goodbreak
\newsec{Boundary Conditions}

The specification of boundary conditions for the Wheeler-DeWitt equation may
seem a disappointment, as it might appear that we are just replacing an
arbitrary initial choice of parameters which describe the classical evolution
of the universe by an arbitrary initial choice of parameters which describe its
quantum evolution. However, if quantum mechanics is a universal theory then it
must have applied at the earliest epochs of the existence of the universe, in
which case it is natural that the quantum dynamics precedes the classical
dynamics. This justifies a quantum boundary condition for the universe as being
more fundamental than a classical one. In any case, the only
alternative to choosing quantum boundary conditions would be that mathematical
consistency might be enough to guarantee a unique solution to the
Wheeler-DeWitt equation, as DeWitt originally hoped$^{\DeW}$. If the experience
gained from the study of minisuperspace models translates to superspace, then
this would not appear to be the case, however.

The question naturally arises as to whether there should be some natural
boundary condition, which once and for all determines the quantum evolution
of the universe at early times, or alternatively whether the nature of the
quantum dynamics might be somewhat indifferent to such choices. Deep conceptual
problems are involved in trying to make headway with this question. Unlike
other situations in quantum physics, where boundary conditions are readily
specified by the symmetry of particular problems, such as spherical symmetry in
the case of the hydrogen atom, the origin of the universe poses a situation
in which all intuition must be abandoned and we can at best proceed on
aesthetic grounds alone.

Having made a choice of boundary condition, we can
of course solve the Wheeler-DeWitt equation and study the physical consequences
for the evolution of the universe. However, without some additional principle
we should by rights study many different boundary conditions before we can
begin to have any confidence about the predictions made. To arrive at a
principle which would circumvent this problem is an immense challenge: it would
more or less amount to an additional
law of physics which must be appended to the others which describe the quantum
evolution of the universe. The
situation might be considered to be the same as trying to describe the phase
transition from gas to liquid if all the physical phenomena that we knew about
related to the gaseous phase only. The universe appears to have undergone a
phase transition when it was formed, but the only experience we have
available involves the ``after'' state of the universe alone.

Progress can of course only be made by attempting to define natural boundary
conditions for the wavefunction of the universe, and examining the consequences.
This became an important activity in the 1980s. I will only discuss the two
most studied boundary condition proposals, the ``no-boundary proposal'' and
the ``tunneling proposal''. However, other proposals have been put forward,
including the ``all possible boundaries proposal'' of Suen and Young$^{\SuYo}$
and the ``symmetric initial condition'' of Conradi and Zeh$^{\CoZe,\Con}$.

\subsec{The no-boundary proposal}

The proposal of Hartle and Hawking$^{\HarHaA,\HawC}$ is that one should
restrict the sum in the definition of the wavefunction of the universe \wavefn\
to include only compact Euclidean 4-manifolds, $\Mi{}$, for which the spatial
hypersurface $\SI$ on which $\PS$ is defined forms the only boundary, and only
matter configurations which are regular on these geometries. The
universe then has no singular boundary to the past, as is the case for the
standard FRW cosmology. The sum \wavefn\ thus includes manifolds such
as those shown in \fig\NOBDY, but not those shown in \GEOMETRY. As Hawking$^
{\HawC}$ puts it: {\it the boundary conditions of the universe are that it has
no boundary}.\ifig\NOBDY{Geometries allowed by the Hartle-Hawking no-boundary
proposal.}{\epsfbox{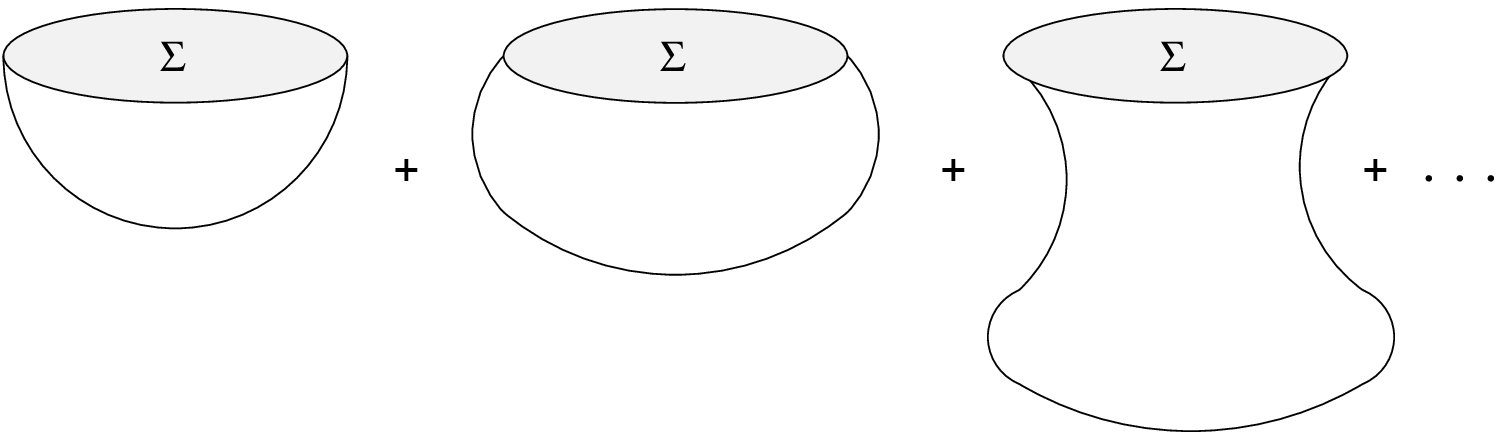}}

Intuitively, what Hartle and Hawking had in mind in formulating this proposal
was to get rid of the initial singularity by ``smoothing the geometry of the
universe off in imaginary time''. For example, whereas a surface with $\sqrt{h}
=0$ would be singular in a Lorentzian signature metric, this is not necessarily
the case if the metric is of Euclidean signature, as can be seen from the
example of $S^4$ shown in \fig\SLICE.
Ideally, the no-boundary proposal should tell us what initial
conditions to set when we take manifolds with $\sqrt{h}\to0$ or any similar
limit consistent with the proposal. If the limit is taken at an initial
time $\ta=0$, the no-boundary proposal would lead to conditions on
$h_{ij}(x,0)$, $\ph(x,0)$ and their derivatives.
In practice, quantum cosmology is rarely studied beyond the semiclassical
approximation, in which $\PS\simeq{\cal A}\e^{-I_{\rm cl}}$, where $I_{\rm cl}$
is the classical (possibly complex) action evaluated along the solution to the
Euclidean field equations. In the semiclassical approximation one therefore
works only with boundary conditions on the metric and matter fields which
correspond to the no-boundary proposal at the classical level. In particular,
we demand: (i) that the $4$-geometry is closed; and (ii) that the saddle points
of the functional integral correspond to regular solutions of the classical
field equations which match the prescribed initial data on $\SI$.
\ifig\SLICE{Slicing a 4-sphere of radius R embedded in flat
5-dimensional space:\br (a) a surface $\left|x^5\right|=\hbox{a}<\hbox{R}$
intersects the 4-sphere in a 3-sphere of non-zero radius; (b) when $\left|x^5
\right|=\hbox{R}$ the 3-sphere shrinks to zero radius but there is no
singularity of the 4-geometry.}{\epsfxsize=15cm\epsfbox{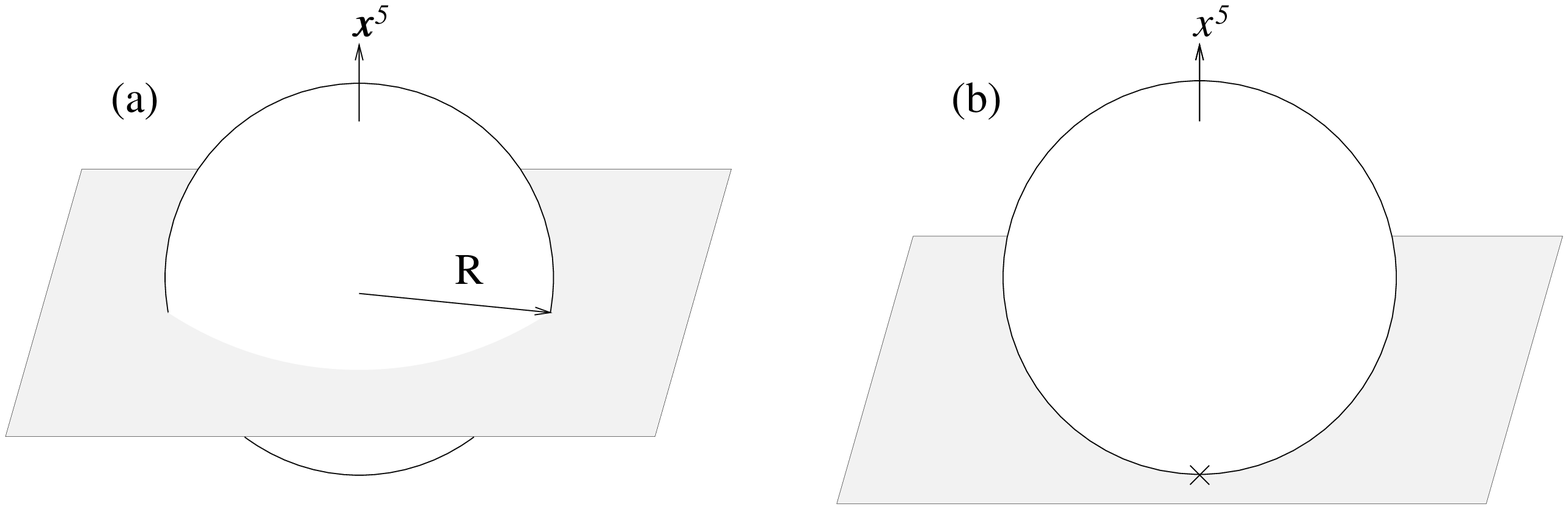}}

One question which is not explained by the no-boundary proposal is the choice
of a contour of integration for the path integral. As was mentioned earlier,
the path integral over real Euclidean metrics does not converge, and thus it is
necessary to include complex metrics to make the path integral converge. Such
metrics will generally include ones which are neither truly Euclidean
nor truly Lorentzian, and thus the na\"{\i}ve picture of a compact Euclidean
geometry sewn onto a Lorentzian one, which is suggested by ``smoothing the
geometry of the universe off in imaginary time'', is not completely accurate.
In general, one might expect a truly complex metric to interpolate the
Euclidean and Lorentzian ones, and in general the initial geometry might be
only {\it approximately Euclidean} and the final geometry only
{\it approximately Lorentzian}$^{\Lyo}$. Unfortunately the criteria for
achieving convergence of the path integral do not single out a unique contour
of integration, and the no-boundary proposal does not appear to offer any
further clues as to how the contour should be chosen$^{\HalHarB}$.
\Ifig\SFOUR{Euclidean solutions which correspond to matching a given 3-sphere
hypersurface, $\SI$, to a 4-sphere which is: (a) less than half filled;
(b) more than half filled.}{\epsfbox{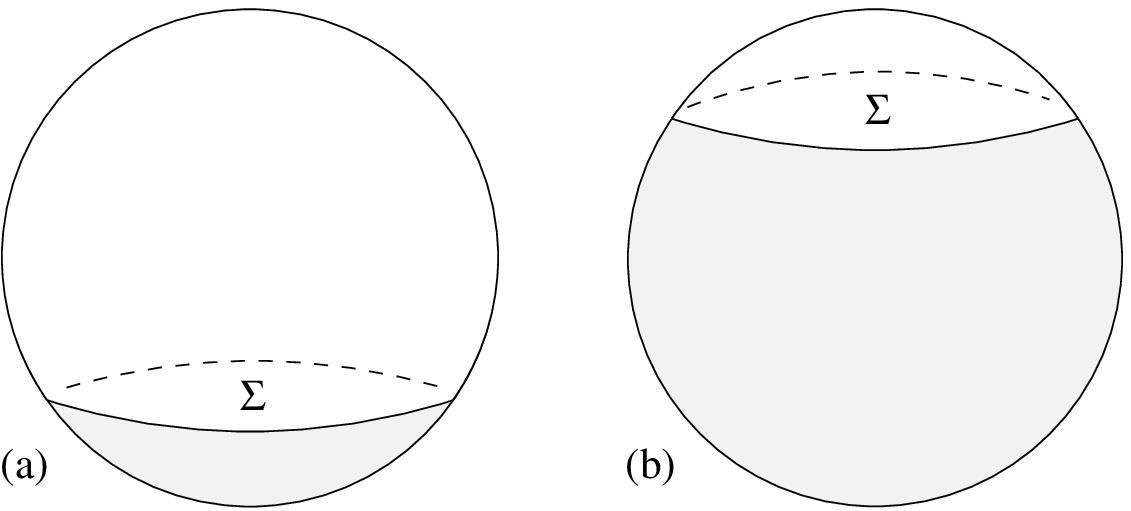}}

Non-uniqueness of the contour of integration is already a problem in the
simplest conceivable non-trivial minisuperspace model, namely the $k=+1$ FRW
universe with a cosmological term and no other matter -- the ``de Sitter
minisuperspace'' model. At the semiclassical level one can calculate
$\e^{-I_{\rm cl}}$ by the steepest descents method. Hartle and Hawking
discussed this in their original ``no boundary'' paper$^{\HarHaA}$, and argued
heuristically that one particular saddle point would yield the dominant
contribution to the path integral, namely the saddle point corresponding to the
classical Euclidean solution which matches the $S^3$ hypersurface $\SI$ to a
less than half filled 4-sphere, rather than the solution which matches $S^3$ to
a more than half filled 4-sphere (see \SFOUR). However, their argument did
not stand up to a more rigorous analysis. Halliwell and Louko$^{\HalLoA}$ found
a means of evaluating the minisuperspace path integral exactly, and thereby
explicitly determined convergent contours. They showed that this simple model
possesses inequivalent contours for which the path integral converges. These
pass through different saddle points and lead to different semiclassical
wavefunctions, $\PS$. There are thus many different no-boundary wavefunctions,
each corresponding to a different choice of contour. The problem persists in
more complicated models$^{\HalLoB,\HalLoC}$. Since different
no-boundary wavefunctions could lead to different physical predictions, the
ambiguity associated with the choice of contour would appear to be the most
significant problem with the no-boundary proposal which still needs to be
resolved.

Let us return to the minisuperspace example of \S3.7 and examine the
implications of imposing the Hartle-Hawking boundary condition. For
simplicity we will specialise to $k=+1$ models. The other values of $k$ have
also been discussed in the literature.$^{\PagA,\PagC,\GiGr}$

The minisuperspace no-boundary wavefunction for the $k=+1$ models is given
by
$$\PS\Ns{HH}[\Ta,\Tp]=\int^{(a,\ph)}\DD\a\,\DD\pH\,\DD\NN\,\e^{-I[\a,\pH,\NN\,
]},\eqn\nbps$$
where\foot{\blacktriangle}{I shall use a different font to distinguish the
minisuperspace coordinates in the functional integral, $(\a,\pH)$, from their
boundary values, $(\Ta,\Tp)$, in the 3-geometry on the hypersurface $\SI$.}
$$I=\half\int_0^{\ta_f}\dd\ta\NN\left[{-\a\over\NN^2}\left(\Deriv\dd\ta\a\right)
^2+{\a^3\over\NN^2}\left(\Deriv\dd\ta\pH\right)^2-\a+\a^3V\right].\eqn\nbI$$
The integral is taken over a class of paths which match the values
$$\a(\ta_f)=\Ta,\qquad\pH(\ta_f)=\Tp,\eqn\nbinit$$
on the final surface, and the origin of the Euclidean time coordinate, $\ta$,
has been chosen to be zero.

There are two approaches we can take to solving the Wheeler-DeWitt equation
\mFRWd: either (i) attempt to interpret the Hartle-Hawking boundary condition
directly in terms of boundary conditions of $\PS$ on minisuperspace; or (ii)
take a saddle-point approximation to the path integral \nbps. For consistency
these two approaches should agree.

Let us first consider \mFRWd\ directly. Hawking and Page$^{\HaPaA,\PagD}$ have
argued that one can approximate the Hartle-Hawking boundary condition in
minisuperspace by saying that in an appropriate measure one should have $\PS
=1$ when $a\to0$ with $\ph$ regular, and $\PS=1$ also along the past null
boundaries, in order to provide sufficient Cauchy data to solve
\mFRWd\ everywhere in the $(\al,\ph)$ plane. It is possible to
solve \mFRWd\ exactly for a massless scalar field$^{\PagB,\KaVi}$, (i.e., $V=0
$), but exact solutions are not known for the massive scalar field. Approximate
solutions can be found in various regimes$^{\HaPaA,\PagD,\HaPaC}$, however.

Firstly, since $\PS$ must be regular as $a\to0$, we see that $\PS$ must be
independent of $\ph$ in this limit, $\Deriv\pt\ph\PS\simeq0$, in order to
overcome the divergence of the $a^{-3}$ factor (using coordinates $(a,\ph)$)
in \mFRWd. Furthermore, for $a^2V\ll1$ the approximation $V\simeq0$ is a good
one\foot{{\bowtie}}{We assume that $V(\ph)$ grows less strongly than $\e^{6|\ph
|}$ as $|\ph|\to\infty$, i.e., $|V'/V|<6$, so that this approximation remains
valid for arbitrarily large $|\ph|$.}, and for $k=\pm1$ \mFRWd\ becomes a
Bessel equation in the variable $\half a^2$. For $k=+1$, which is the case of
interest to us here, we therefore obtain the solution$^{\HaPaA,\PagD}$
$$\PS\simeq\Iz(\half a^2),\eqn\BesselA$$
where $\Iz(z)\equiv\sum_{n=0}^\infty{1\over(n!)^2}\left(z\over2\right)^
{2n}$ is the zero order modified Bessel function, and the normalisation has
been fixed to satisfy the boundary condition $\PS\to1$ as $a\to0$.
For large $a$ (with $a^2V\ll1$),
$$\PS\goesas{1\over\sqrt{\pi}\,a}\exp\left(\half a^2\right)\left[1+\OO(a^{-2})
\right],\eqn\psasymA$$ i.e., the wavefunction
is of exponential type. It therefore agrees with the WKB approximation
\wkbpsE\ in the limit $a^2V\ll1$ for large $a$ provided we take the $(-)$
solution of \wkbpsE\ with normalisation
$$\CC(\ph)={1\over\sqrt{\pi}}\exp\left(+1\over3V(\ph)\right).\eqn\nbnormA$$
Let us now consider the limit $V(\ph)\gg1$ but avoid regimes in which the $
\ph$ dependence of \mFRWd\ is significant by assuming that $V$ is approximately
constant, as in the approximation of \PHASE, so that the $\ph$ derivative can
still be neglected.
The term involving the spatial curvature $k$ in \mFRWd\ is now negligible
compared to the last term and can also be neglected, yielding the solution$^
{\HaPaA,\PagD}$
$$\PS\simeq\cc(\ph)\,\Jz(\frac13a^3\sqrt{V}),\eqn\BesselB$$
where $\Jz(z)\equiv\sum_{n=0}^\infty{(-1)^n\over(n!)^2}\left(z\over2\right)^
{2n}$ is the zero order ordinary Bessel function. Since the spatial curvature
term in \mFRWd\ is always dominant for $a\to0$, the approximation which led to
\BesselB\ no longer applies in that limit, so the factor $\cc$ cannot be
normalised by the boundary condition $\PS(0,\ph)=1$. For large $a$, however,
$$\PS\goesas{\cc\over\sqrt{2\pi\SS}}\cos\left(\SS-{\pi\over4}\right),\eqn
\psasymB$$
where $\SS=\frac13a^3\sqrt{V}$, which is a superposition of the two oscillatory
WKB modes \wkbpsO\ for large $\SS$. Using the WKB connection formula to match
the $(-)$ solution of \wkbpsE\ to the oscillatory region, we find agreement
with the asymptotic limit \psasymB, provided
$$\cc=\sqrt{2\pi\over3}\,\CC=\sqrt{\frac23}\exp\left(1\over3V(\ph)\right).\eqn
\nbnormB$$
The $\ph$-dependent corrections to these wavefunctions, which result from
perturbations in directions in which $\dot \ph\not\simeq0$ have been discussed
by Page$^{\PagA,\PagD,\PagE}$.

Let us now consider the alternative method of determining $\PS$ by making a
saddle-point approximation to the path integral.
In the semiclassical approximation the wavefunction takes the form
$$\PS\goesas\exp\left(-I\ns{cl}(\Ta,\Tp)\right),\eqn\semiclassnb$$
where $I\ns{cl}$ denotes the Euclidean action \nbI\ evaluated at a classical
(possibly complex) solution to the Euclidean field equations
$$\eqalignno{&{1\over\NN^2}\left(\Deriv\dd\ta\a\right)^2-{\a^2\over\NN^2}\left(
\Deriv\dd\ta\pH\right)^2-1+\a^2V(\pH)=0,&\Eqn\nbfeA\cr
&{1\over\NN\a}\Der\dd\ta\left({1\over\NN}\Deriv\dd\ta\a\right)+{2\over\NN^2}
\left(\Deriv\dd\ta\pH\right)^2+V(\pH)=0,&\Eqn\nbfeB\cr
&{1\over\NN}\Der\dd\ta\left({1\over\NN}\Deriv\dd\ta\pH\right)+{3\over\NN\a}
\Deriv\dd\ta\a\Deriv\dd\ta\pH-\half\Deriv\dd\pH V=0,&\Eqn\nbfeC\cr}$$
which follow from \mFRWa--\mFRWc\ by replacing $t\to-i\ta$. We will henceforth
restrict ourselves to the gauge in which $\Deriv\dd\ta\NN=0$, i.e., $\NN=\const
$, in which case the functional integral over $\NN$ in \nbps\ must be
replaced by an ordinary integral.

The Hartle-Hawking boundary condition demands that
$$\a(0)=0,\qquad\left.\Deriv\dd\ta\pH\right|_0=0.\eqn\nbmini$$
To see this consider the Euclidean 4-metric which is given by
$$\dd s^2=\NN^2\dd\ta^2+\a^2(\ta)\dd\OM_3^{\ 2}.\eqn\minimet$$
The Hartle-Hawking boundary condition requires that we close this 4-geometry in
a regular fashion as $\ta\to0$. This is achieved if $\a\goesas\pm\NN\ta$ as
$\ta\to0$, since \minimet\ is then the same as the metric of the 4-sphere in
spherical polar coordinates in this limit. This suggests that we demand $\a(0)=
0$ and ${1\over\NN}\left.\Deriv\dd\ta\a\right|_0=\pm1$. However, the second
condition is guaranteed by the constraint equation \nbfeA\ if the first
condition is imposed. The second condition of \nbmini\ is obtained by noting
that \nbfeC\ will give a regular solution for $\pH$ in the limit $\ta\to0$ only
if $\left.\Deriv\dd\ta\pH\right|_0=0$, since the middle term of \nbfeC\
diverges otherwise.

If as before we make the simplifying assumption that $V$ can be approximated by
a cosmological constant, and $\Deriv\dd\ta\pH\simeq0$, then it follows that
there exist solutions satisfying the boundary conditions \nbinit and \nbmini,
which may be written
$$\a(\ta)\simeq{\Ta\sin(\NN\sqrt{V}\,\ta)\over\sin(\NN\sqrt{V}\,\ta_f)},\eqn
\nbsolA$$
where
$$\sin^2(\NN\sqrt{V}\,\ta_f)=\Ta^2V,\eqn\nbsolB$$
which follows from solving the constraint \nbfeA. If $\Ta^2V<1$ then \nbsolB\
will give an infinite number of real solutions for the constant $\NN$, which
may be conveniently parametrised
$$\NN=\NN_{n\pm}\equiv{1\over\sqrt{V}\,\ta_f}\left[(n+\half)\pi\pm\cos^{-1}(\Ta
\sqrt{V})\right],\quad n\in\Zop,\eqn\nbsolC$$
with $\cos^{-1}(\Ta\sqrt{V})$ taken in the principal range $(0,\frac\pi2)$.
Substitution of \nbsolC\ in \nbsolA\ then gives
$$\a(\ta)\simeq(-1)^nV^{-1/2}\sin(\NN\sqrt{V}\,\ta).\eqn\nbsolD$$
In addition to the real solutions \nbsolC\ of \nbsolB, there will also be
complex solutions if $\Ta^2V>1$.

Using the solution \nbsolD\ it is straightforward to evaluate the classical
action \nbI. For $n=0$, for example, we find
$$I_\pm={-1\over3V(\Tp)}\left[1\pm\left(1-\Ta^2V(\Tp)\right)^{3/2}\right].\eqn
\nbclassI$$
If we substitute \nbsolD\ into \minimet\ we obtain the metric of the 4-sphere,
and thus the classical solutions correspond to matching a given 3-sphere
to 4-sphere(s). Furthermore, for $n=0$ we find $\left.\Deriv\dd\ta\a\right|_{
\ta_f}=\mp\NN\sqrt{1-\Ta^2V}$, so that the $(-)$ solution of \nbsolC\ and 
\nbclassI\ has $\left.\Deriv
\dd\ta\a\right|_{\ta_f}>0$, which corresponds to matching the 3-geometry to a
less than half filled 4-sphere (\SFOUR(a)), while the $(+)$ solution similarly
corresponds to the more than half filled 4-sphere case (\SFOUR(b)). Values of
$n>0$ would appear to give cases in which the 4-geometry pinches off to
zero a number of times and then ``bounces back'' resulting in linear chains of
contiguous 4-spheres$^{\HaMy,\KSuB}$. However, this interpretation of the
saddle points as corresponding to universes with bounce solutions does not
appear to remain valid$^{\Lyo}$ once one considers complex
solutions to the field equations \nbfeA--\nbfeC.

As was discussed earlier the no-boundary condition does not prescribe a
unique contour of integration, and thus it is not completely clear which of
the above saddle points should be included in the semiclassical wavefunction
\semiclass. Halliwell and Hartle have shown$^{\HalHarB}$ that the points with
$n<0$, which have a negative lapse function, lead to difficulties with the
recovery of quantum field theory in curved spacetime from quantum cosmology,
and thus one might hope that these points should be avoided in the contour of
integration. However, no clear grounds present themselves for omitting the
other saddle points. For the purpose of making predictions from the no-boundary
proposal we shall therefore make a choice by assuming that the contour is such
that the solution corresponding to the less than half filled 4-sphere, with
action $I_-$, provides the dominant contribution. (This is the choice that
Hartle and Hawking originally made$^{\HarHaA}$.) Neglecting the prefactor, we
therefore obtain a no-boundary wavefunction
$$\PS\Ns{HH}(a,\ph)\propto\exp\left(1\over3V(\Tp)\right)\exp\left[{-1\over3V(
\ph)}\left(1-a^2V(\ph)\right)^{3/2}\right],\eqn\nbwavefnA$$
in the region $a^2V<1$. Using the WKB matching procedure one can show that the
corresponding solution in the region $a^2V>1$ is
$$\PS\Ns{HH}(a,\ph)\propto\exp\left(1\over3V(\Tp)\right)\cos\left[{1\over3V(\ph
)}\left(a^2V(\ph)-1\right)^{3/2}-{\pi\over4}\right],\eqn\nbwavefnB$$
which is the superposition of the two WKB components of \wkbpsO:
$$\eqalign{\PS\Ns{HH}&=\PS\X-+\PS\X+,\cr \PS_\pm&\propto\exp\left(1\over3V(\ph)
\right)\exp\left\{\pm i\left[{1\over3V(\ph)}\left(a^2V(\ph)-1\right)^{3/2}-{\pi
\over4}\right]\right\}\,.\cr}\eqn\nbwavefnC$$
One may observe that \nbwavefnA\ and \nbwavefnB\ agree with the solutions
\BesselA\ and \BesselB\ found earlier by direct examination of the
Wheeler-DeWitt equation in the appropriate limits \psasymA\ and \psasymB.
Thus although the saddle point corresponding to the less than half filled
4-sphere does not appear to be picked out in any special way by the path
integral, it is favoured by the ``approximate'' boundary condition$^{\HaPaA,
\PagD}$ that $\PS\to1$ as $a\to0$.

One must add the caveat that the approximate boundary condition should be
amended if one is to consider genuinely complex solutions of the Euclidean
field equations \nbfeA--\nbfeC. This issue has been considered by Lyons$^{\Lyo}
$. In general one must analytically continue the boundary condition \nbmini\
demanded by the no-boundary proposal. Although the simple picture of matching
a real Euclidean solution to a real Lorentzian solution at the junction is no
longer maintained, one nonetheless finds solutions which are initially
approximately Euclidean and at late times are approximately Lorentzian, with
classical inflationary behaviour$^{\Lyo}$.

\subsec{The tunneling proposal}

An alternative approach advocated by Vilenkin is that the boundary condition
for the wavefunction, $\PS$, should be such as to embody the notion that the
universe ``tunnels into existence from nothing'' without making such specific
restrictions on the ``initial'' geometry as the Hartle-Hawking proposal
does. Conceivably there are many possible ways in which such a notion could be
translated mathematically into a definition of the wavefunction, $\PS$, and
indeed many alternative formulations of the tunneling proposal have been put
forward\foot{\between}{Linde's proposal [\Lin] embodies a similar philosophy.
However, it gives a different wavefunction to Vilenkin's ``tunneling''
wavefunction in simple minisuperspace models [\VilG].}. Some early versions
were phrased in a similar
fashion to the no-boundary proposal: in particular, Vilenkin$^{\VilB}$ proposed
defining the wavefunction by a functional integral over Lorentzian metrics
which interpolate between a given matter configuration, $\PH$, and 3-geometry,
$h_{ij}$, and a vanishing 3-geometry, $\varnothing$, lying to its past
$$\PS\left[h_{ij},\PH,\SI\right]=\sum_{\Mi{}}\int^{(h,\PH)}_\varnothing\DD{\B
g}\,\DD\pH\,\e^{iS[g_{\mu\nu},\pH]}.\eqn\wavefnalt$$

Vilenkin has also given an alternative formulation of the tunneling proposal
in terms of a boundary condition on superspace$^{\Vilc,\VilC}$ rather than a
restriction on manifolds included in the path integral. In order to formulate
boundary conditions on superspace it is necessary to consider its boundary,
which can be thought of as consisting of 3-metric and matter configurations for
which the 3-curvature is infinite, or $|\PH|\to\infty$ etc. As we
have already seen from the example of $S^4$ (\SLICE), not all singular
3-geometries will correspond to singular 4-geometries, as it is possible to
obtain a singular 3-geometry by a degenerate slicing of the 4-geometry.
Therefore we should distinguish points on the boundary of superspace which
correspond to genuine singularities of the 4-geometry from those that
correspond to degenerate slicings\foot{\bigstar}{This distinction can be made
more precise using Morse functions [\VilD].}. We call the former the {\it
singular boundary} of superspace, and the latter the {\it non-singular
boundary}.

The tunneling proposal of Vilenkin$^{\VilC}$ is that {\it the wavefunction, $
\PS$, should be everywhere bounded, and at singular boundaries of superspace $
\PS$ includes only outgoing modes, i.e., those that carry a flux out of
superspace}. Thus ingoing modes can only enter at the nonsingular boundary.
This definition is somewhat vague as there is no obvious rigorous
definition of positive and negative frequency modes in superspace due to the
fact that it possesses no Killing vectors$^{\Kuch}$, and thus there is no clear
notion which modes are ``ingoing'' and which are ``outgoing''. Furthermore, the
structure of superspace is not completely understood, and it has not been
rigorously shown that its boundary can be split into singular and non-singular
parts. 

The tunneling proposal has in fact been formulated with the minisuperspace WKB
approximation in mind, in which case the notion of the boundary of
minisuperspace and the notions of ingoing and outgoing modes are more clearly
defined. Since each oscillatory WKB mode $\PS\goesas\e^{i\Sn}$ has a current
\wkbcurrent, we can classify the modes as ingoing or outgoing according to the
direction of $\LAP\Sn$ on the surface in question. Heuristically, the idea
underlying the Vilenkin boundary condition is that the ensemble of universes
described by $\PS$ should not include any universes contracting down from
infinite size$^{\Vilc}$, but only those that correspond to ``tunneling from
nothing''. As we shall see, the ``outgoing flux'' condition$^{\VilC}$ accords
with this notion at least in the case of simple minisuperspace models.

The outgoing-flux version of the tunneling proposal agrees with the path
integral formulation \wavefnalt\ in the case of the simplest de Sitter
minisuperspace$^{\VilD,\HalLoA}$, but the two versions would not appear to be
equivalent in general$^{\HalLoC}$. Whereas the no-boundary wavefunction fixes
the initial data but leaves the contour of path integration ambiguous, the
tunneling proposal fixes the contour of integration but leaves some ambiguity
in the specification of the initial data, even when the outgoing-flux
condition is imposed$^{\HalLoC,\LoVa}$.

Consider the minisuperspace model of \S3.7. In the diagrams of \PENROSE\ all
surfaces $\scri^{\pm}\Z{L,R}$ and the points $\ii^0\Z{L,R}$ and $\ii^+$ will
be part of the singular boundary, whereas the point $\ii^-$, which corresponds
to $a\to0$ ($\al\to-\infty$) with $\ph$ finite, is the only point of the
nonsingular boundary. Of course, the condition that $a\to0$ is not enough
to guarantee regularity of the 4-geometry \minimet, and in general one might
not expect boundary points to cleanly fall into the category of the ``singular
boundary'' or the ``nonsingular boundary''. However, as we have already
observed in the last section, in the present minisuperspace model the
additional requirement for regularity that ${1\over\NN}\left.\Deriv\dd\ta a
\right|_0=\pm1$ is guaranteed by the constraint equation \nbfeA\ if $a(0)=0$
is imposed.

The oscillatory WKB region, as shaded in the conformal diagrams of \PENROSE\
and \fig\TUNNEL, is always bounded by $\ii^0\Z{R}$, $\scri^+\Z R$
and $\ii^+$, and in all cases except that of \PENROSE(f)\foot{\triangledown}
{Unlike the other cases the potential of \PENROSE(f), which corresponds to
a potential in which supersymmetry is broken through gaugino condensation in
string theory, vanishes as $\ph\to-\infty$. This limit is the ``weak coupling
limit'' of string theory.} the oscillatory region is bounded by $\ii^0\Z L$ and
$\scri^-\Z L$ also. The wavefunction in this region is given by a
superposition of terms $\e^{i\Sn}$, where $\Sn$ is a solution to the
Hamilton-Jacobi equation \wkbHJ, which in terms of the variables $(\al,\ph)$
may be written
$$-\left(\Deriv\pt\al\SS\right)^2+\left(\Deriv\pt\ph\SS\right)^2+\uu(\al,\ph)=0
,\eqn\mwkbHJ$$
where $\uu(\al,\ph)\equiv2\e^{3\al}\UU(\al,\ph)=\e^{4\al}\left(\e^{2\al}V(\ph)-1\right)$, and for convenience in what follows we have suppressed the index $n
$. The characteristics of \mwkbHJ\ satisfy
$${\dd\al\over2\SS,_\al}={\dd\ph\over-2\SS,_\ph}={\dd\SS\over2\uu}=
{\dd(\SS,_\al)\over\uu,_\al}={\dd(\SS,_\ph)\over\uu,_\ph}\,.\eqn\pdechar$$
Thus each $\SS(\al,\ph)$ describes a congruence of classical paths with
$$\Deriv\dd\al\ph=-\,{\SS,_\ph\over\SS,_\al}={\mp\SS,_\ph\over\sqrt{\SS,\DU\ph2
+\uu}}.\eqn\congruence$$
\ifig\TUNNEL{Some schematic probability flows
consistent with the tunneling proposal. An indicative oscillatory WKB region is
shown (shaded area) for the potential $V=\ph^2$.}{\epsfbox{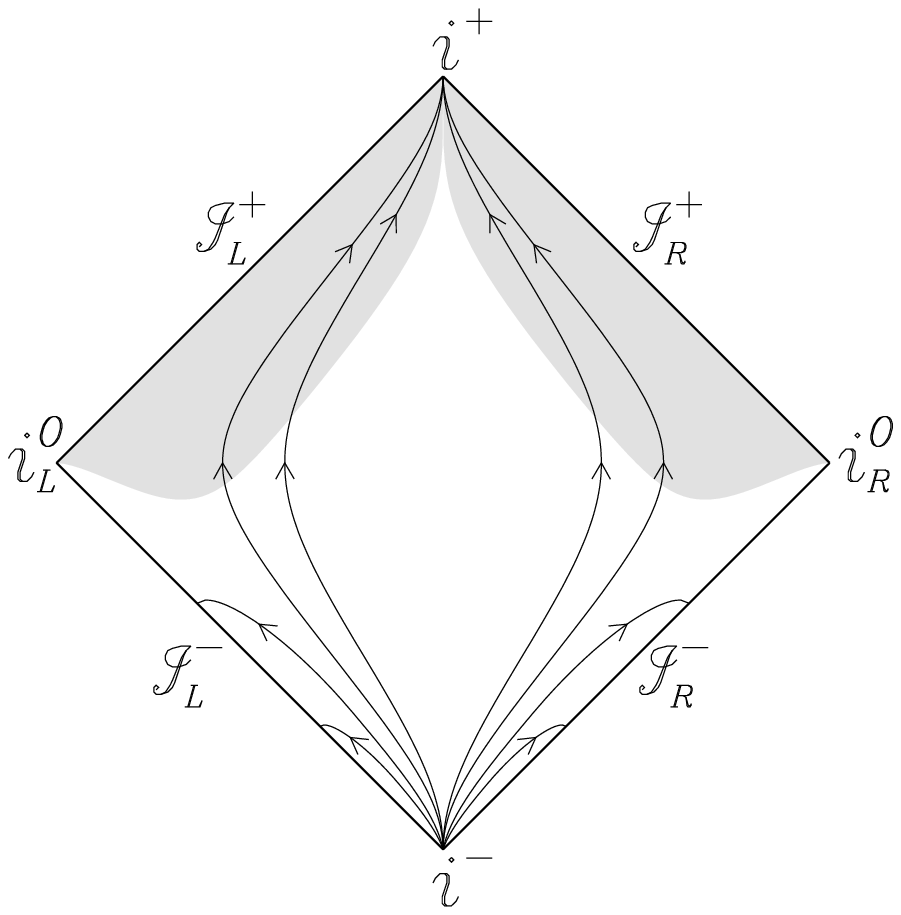}}\noindent
Since $\uu>0$ in the oscillatory region (assuming $V(\ph)>0$), it follows that
these integral curves satisfy $\left|\Deriv\dd\al\ph\right|<1$, i.e., they are
``timelike'' in the minisuperspace coordinates and have an endpoint at $\ii^+$.
Since $\PI_\al\equiv-{\e^{3\al}\dot\al\over\NN}=\SS,_\al$ it follows that
the WKB modes correspond to expanding universes ($\dot\al>0$) with $\PI_\al=\SS
,_\al<0$, or contracting universes with $\PI_\al=\SS,_\al>0$, if we assume\foot
{{\oslash}}{This choice is a matter of convention and the opposite choice would
reverse the roles of the ``ingoing'' and ``outgoing'' modes. Equivalently, the
coordinate $t$ is an arbitrary label from the point of view of general
relativity, and it is a matter of convention whether we choose $t$ to increase
or decrease towards the ``future''; the ``future'' being defined by the
expansion of the universe (the cosmological arrow) or the increase of entropy
(the thermodynamic arrow). In terms of \PHASE\ it amounts to an arbitrary
choice between $A_+$ and $A_-$ as representing the ``late-time behaviour'' of
the inflationary solutions.} $\NN>0$. Since all paths
originate at finite values of $\al$ the latter solutions will extend from $\ii^
+$ into the interior of the minisuperspace, i.e., they are ``ingoing modes''
and are to be excluded on account of the tunneling boundary condition. This
boundary condition only allows the expanding solutions which are ``outgoing''
at $\ii^+$. In the approximation in which \psasymA\ applies it follows that the
tunneling proposal demands that only the modes with $\PS\goesas\exp\left[{-i
\over3V(\ph)}\left(e^{2\al}V(\ph)-1\right)^{3/2}\right]$ are admitted in the
oscillatory region. The tunneling wavefunction is thus complex in the
oscillatory region, in contrast to the no-boundary wavefunction
\nbwavefnB\ which is real.

To extend the tunneling wavefunction into the ``tunneling region'' (the
unshaded regions of \PENROSE\ and \TUNNEL), we can use the WKB matching
procedure. Of course, there will also be additional solutions that remain
entirely in the tunneling region and never cross into the oscillatory region.
In the case of the null boundaries, $\scri^-\Z{L,R}$, which border this region
we observe that as $\al\to-\infty$ \mFRWd\ becomes the wave equation in the $(
\al,\ph)$ coordinates, and thus the solutions are asymptotically null which
leads to a notion of ``ingoing'' and ``outgoing'' modes, so that the tunneling
condition can be imposed. Some possible probability flows consistent with the
tunneling proposal are shown in the conformal diagram of \TUNNEL.

\newsec{The Predictions of Quantum Cosmology}

Given the many problems and uncertainties in the quantum cosmology programme
that have been discussed above, one could easily form the opinion that it is
premature to talk about the predictions that quantum cosmology makes.
Nevertheless, it is important to investigate the types of predictions we
might expect quantum cosmology to make about the universe, as well as
realistically evaluating the limitations of the these predictions. I will
concentrate on three key areas in this section. Other topics, most notably
the variability of the constants of nature, have been discussed recently by
Vilenkin$^{\VilG,\VilF}$.

\subsec{The period of inflation}

The construction of a suitable probability measure should allow us to answer
questions such as whether inflation is a feature of a typical universe. As we
saw in \S3.6, the construction of a general probability measure is problematic.
Thus we shall limit our discussion to the oscillatory WKB limit in which case
the Klein-Gordon-type current \current\ is adequate.

The issue of the duration of the period of inflation has been a point of some
debate between proponents of the no-boundary wavefunction and proponents of the
tunneling wavefunction. Consider the canonical minisuperspace model of \S3.7.
We have seen that in the WKB limit the Hartle-Hawking boundary condition gives
rise to a wavefunction \nbwavefnB\ in the oscillatory region, which is strongly
peaked about the set of classical solutions \wkbinflat\ that correspond to an
inflating universe: $a(t)\propto\e^{\sqrt{V}\,t}$, $\ph(t)\simeq\ph\Z0=\const$.

Vilenkin$^{\Vilc\hs\VilD,\VilB}$ has also studied the minisuperspace model of
\S3.7, but with a different choice of factor ordering in the Wheeler-DeWitt
equation \mFRWd. He concludes that similarly to \nbwavefnB\ the tunneling
boundary condition leads to a WKB wavefunction
$$\PS\Ns{V}(a,\ph)\propto\exp\left(-1\over3V(\Tp)\right)\exp\left[{-i\over3V(
\ph)}\left(a^2V(\ph)-1\right)^{3/2}+{i\pi\over4}\right]\eqn\twavefnB$$
in the oscillatory region, which is also peaked about the classical
trajectories of the inflationary universes \wkbinflat.

Both the no-boundary and tunneling wavefunctions thus predict inflation, at
least in the context of this simple minisuperspace model. The important
question is {\it how much} inflation do the models predict? This will be
largely decided by the value, $\ph\Z0$, of the scalar field with which the
universe ``nucleates'' in the semiclassical regime.

In order to study the probability flux arising from \current\ in the WKB
limit it is necessary to focus on one particular WKB component. The
no-boundary wavefunction \nbwavefnB\ is of course real, and the resulting
current \current\ is identically zero. However, $\PS\Ns{HH}$ is the
superposition \nbwavefnC\ of two WKB components, which will
correspond to contracting and expanding universes. As discussed in \S3.5\ it
is assumed that a decoherence mechanism exists so that the interference
between the two components is negligible$^{\HalD}$, and we can therefore assume
that the universe is peaked about one or other WKB component for the purposes
of determining the probability measure. The difference between the no-boundary
and tunneling wavefunctions may therefore not seem great once decoherence to a
classical universe is assumed. In particular, if we take the ``outgoing''
WKB component in the no-boundary case then the only significant difference
between $\PS\Ns{HH}$ and $\PS\Ns{T}$ is the $\ph$-dependent part of the
prefactor, $\exp\left(\pm1\over3V(\ph)\right)$,
which is obtained from boundary conditions set in the tunneling region.
This difference will have ramifications for the probability flux, however.

One question which we might hope to answer in our minisuperspace model would
be: given that a Lorentzian universe nucleates, what is the probability that it
inflates by a sufficient amount ($\goesas65$ e-folds) to solve the problems of
the standard cosmology mentioned in \S1? The answer to this would involve
integrating the probability flux \wkbprob\ on the surface separating the
tunneling and oscillatory regions, which is roughly given by $a^2V(\ph)
=1$. However, our discussion here is limited by the fact that our WKB
approximation applies to trajectories with $\dot\ph\simeq0$. For such
trajectories the probability current, $\JJ$, points chiefly in the direction of
the $a$-coordinate in minisuperspace. We can therefore attempt to answer the
question approximately by evaluating the probability current \wkbcurrent\ on a
surface $\HYP$ in minisuperspace with $a=\const$. For sufficiently large values
of $a$ such surfaces will lie almost entirely within the oscillatory regime
(see \fig \PROBSURF). On these surfaces we obtain a probability flux
$$\dd\PP=\JJ\cdot\dd\HYP\propto\cases{\exp\left(+2\over3V(\ph)\right)&
no-boundary wavefunction, $\PS\Ns{HH}$,\cr \noalign{\smallskip} \exp\left(-2
\over3V(\ph)\right)&tunneling wavefunction, $\PS\Ns{V}$.\cr}\eqn\wkbflux$$
Although the integral of \wkbflux\ may diverge for particular potentials $V(\ph
)$ of interest, this should not be viewed as a problem since, as was emphasised
in \S3.6, questions in quantum cosmology can only refer to conditional rather
than absolute probabilities. 

In the present case let us assume that inflation occurs for large values of
the scalar field, as is the case for potentials of the ``chaotic'' type, $V=\la
\ph^{2p}$. There will then be a minimum value of the scalar field, $\psuf$,
for which sufficient inflation is obtained. For $V=\la\ph^{2p}$ a universe
with $\ph=\ph\Z0$ initially will undergo
$$N_e\simeq{3\over2p}\left[\ph\DU02-\frac29p(2p-1)\right]\eqn\efold$$
e-folds of inflation\foot
{\boxdot}{In the slow-rolling approximation [\KoTu] it follows from \mFRWa\
and \mFRWc\ (with $\NN=1$) that the number of e-folds is $N_e=6\int_{\ph\X e}^
{\ph\X0}\dd\ph\,V/V'$, where $\ph\Z0$ is the initial value and $\ph\Z e$ the
final value of the scalar field at the end of the inflationary epoch. For $V=
\la\ph^{2p}$ we have $N_e={3\over2p}\left(\ph_0^{\ 2}-\ph_e^{\ 2}\right).$ The
value of $\ph_e$ can be estimated from the limit set by $|V'/V|\ll6$ and $|V''
/V|\ll9$.}, so that in the case of the quadratic potential ($p=1$) we find
$\psuf\gsim6.6$, for example. The relevant conditional probability for
sufficient inflation on an $a=\const$ surface is then given by
$$\PP\left(\ph\Z0>\psuf\;|\;\ph\Z1<\ph\Z0<\ph\Z2\right)={\dsp\int^{\ph\X2}_{\ph
_{\lower2pt\hbox{\fiverm suff}}}\;\dd\ph\Z0\exp\left(\pm2\over3V(\ph\Z0)\right)
\over\dsp\int^{\ph\X2}_{\ph\X1}\;\dd\ph\Z0\exp\left(\pm2\over3V(\ph\Z0)\right)}
\,,\eqn\probinflat$$
where the $(+)$ case refers to the no-boundary wavefunction and the $(-)$ case
to the tunneling wavefunction, and the values $\ph\Z1$ and $\ph\Z2$ are
respectively lower and upper cutoffs on the allowed values of $\ph$, which
must be determined by physical criteria. A minimum cut-off might be expected
for a variety of physical reasons, such as avoiding classical universes
which rapidly recollapse.

In their original investigations of the question of the duration of the period
of inflation for the quadratic scalar potential, Hawking and Page$^{\HaPaA}$
took $\ph\Z2=\infty$, in which case both integrals in \probinflat\ are
dominated by large values of $\ph$, leading to a probability $\PP\simeq1$, and
thus a ``prediction'' of inflation. However, this result has been
criticised by Vilenkin$^{\VilC}$ since for $k=+1$ we have $\V=\frac9{16}m^4\ns
{Planck}V$ so that values of $m\ph\gsim4/3$ are in excess of the Planck scale,
and the semiclassical approximation will no longer apply. Vilenkin suggested
that an upper cutoff, $\ph\Z2$, should be introduced at the Planck scale. If
this is the case then provided the lower cutoff $\ph\Z1$ is sufficiently close
to zero, we would find that the integral in the denominator of \probinflat\
becomes very large in the case of the no-boundary wavefunction ($+$ sign),
leading to $\PP\ll1$, whereas this would not be the case for the tunneling
wavefunction ($-$ sign), and thus the latter would predict more inflation.

Introducing a cutoff at the Planck scale might be deemed a rather arbitrary
procedure, since without any knowledge of
Planck scale physics it is impossible to be sure whether the ``real'' answer
is better approximated by the introduction of a cutoff or not. However, it has
been argued on the basis of investigations at the 1-loop level that the
wavefunction is damped at large values of $\ph$ by quantum
corrections$^{\BaKa,\Barv}$. This renders the wavefunction normalisable and
would justify a cutoff, $\ph\Z2$, near the Planck scale.
\ifig\PROBSURF{Conformal diagram for $V=0.04\ph^2
$. The oscillatory region, given roughly by $a^2V>1$, is lightly shaded. Lines
$a=\const$ are superimposed. For very large values of $\ph$ these lie almost
entirely in the oscillatory region. The region of $\ph$-values excluded by a
Planck scale cutoff is darkly shaded.}{\epsfbox{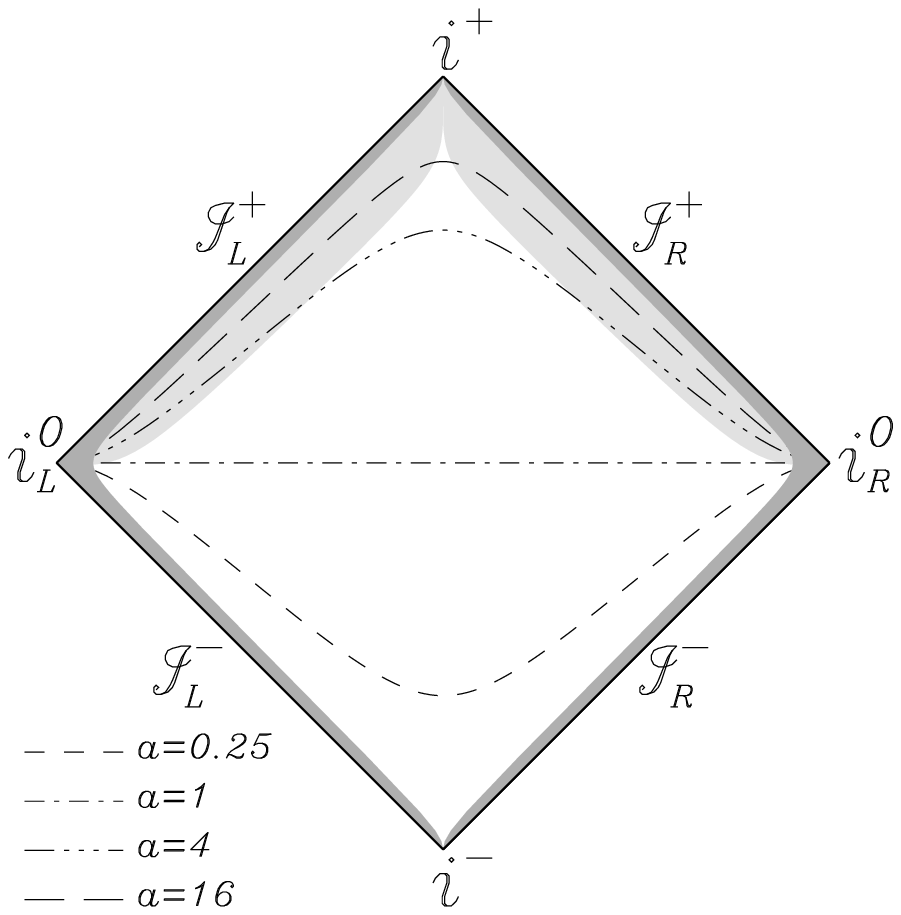}}

Given that the no-boundary wavefunction apparently yields a wavefunction peaked
around the lower cutoff, $\ph\Z1$, it is important to determine what a
reasonable value of this cutoff should be. This issue has been considered by
Grishchuk and Rozhansky$^{\GrRoA}$, and also more recently by Lukas$^{\Luk}$,
who have conducted numerical investigations to analyse the behaviour of the
caustic$^{\GrRoB}$ in the $(a,\ph)$ plane which separates the Euclidean and
Lorentzian solutions. This is illustrated in \fig\CAUSTIC, where classical
solutions to the Euclidean field equations \mFRWa--\mFRWc\ (with $\NN=1$) are
shown. The solutions beginning at $a=0$ initially follow lines with $\dot\ph
\simeq0$, but the approximation eventually breaks down when the trajectories
curl back and recollapse with $a\to0$ and $|\ph|\propto-\ln a\to\infty$. They
thus represent a flow from $\ii^-$ to $\scri^-$ in the conformal diagrams. The
solutions cross each other on the caustic, which for large values
of $\ph$ corresponds to $a^2V(\ph)=1$, in accordance with our expectation from
the WKB approximation.\ifig\CAUSTIC{Classical Euclidean trajectories for the
potential $V=\ph^4$. The approximate caustic $a^2V=1$ is indicated by a dashed
line, and the improved caustic by a dotted line. (From [\Luk].)}
{\psfig{figure=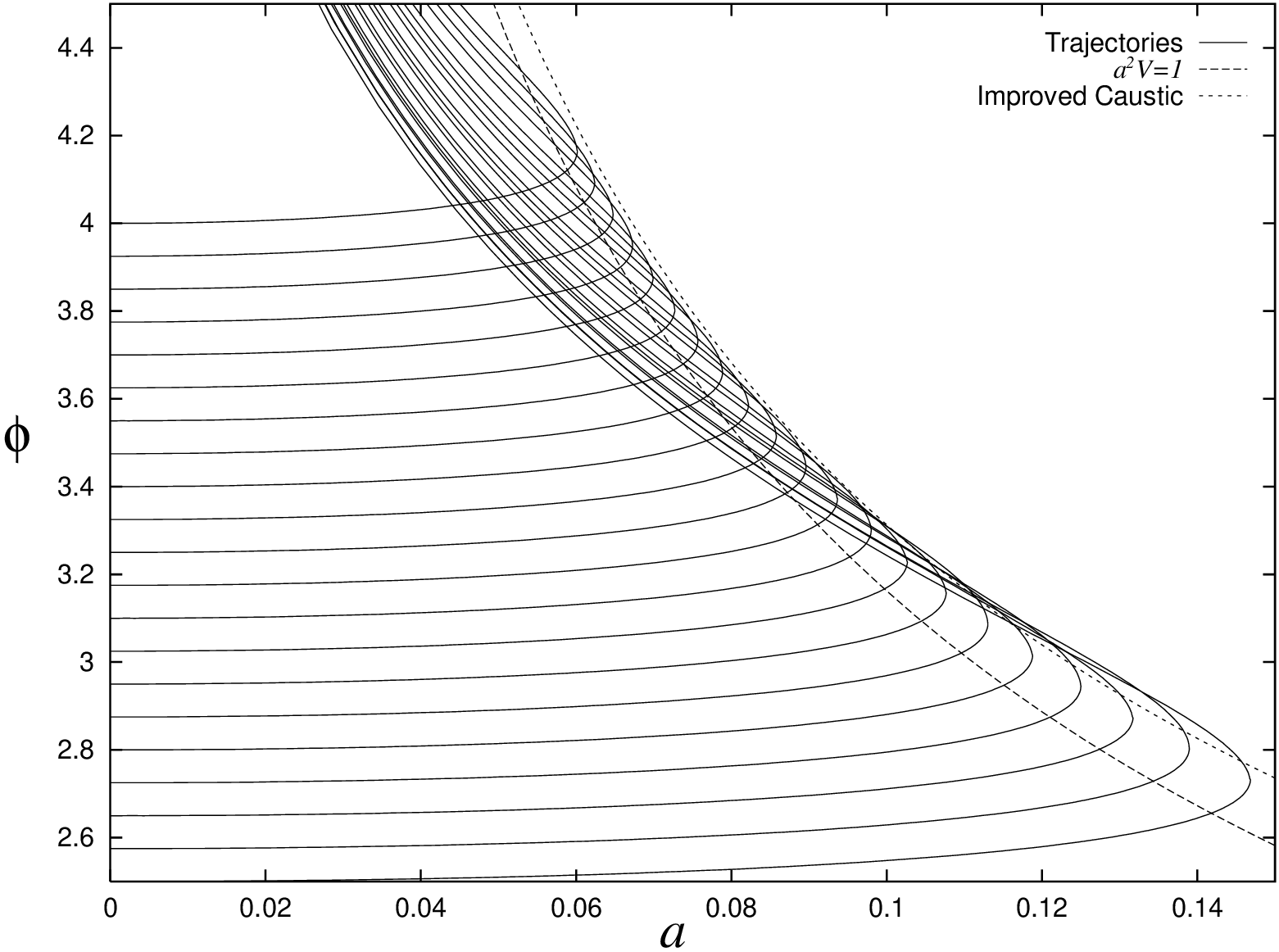,height=13.2cm,width=11cm,rheight=9.35cm,rwidth=9.2cm}}

To the right of the Euclidean solutions in \CAUSTIC\ we would find Lorentzian
solutions with $\dot\ph\simeq0$ sufficiently far away from the caustic$^{\GrRoB
}$. These solutions are not depicted here. The ``nucleation of a universe''
would thus correspond to a solution of the Wheeler-DeWitt equation which was
initially peaked about a classical Euclidean trajectory with $\dot\ph\simeq0$,
and which then crossed over the caustic to be peaked about a corresponding
trajectory with $\dot\ph\simeq0$ in the Lorentzian region. While crossing the
caustic, the wavefunction would be peaked about a complex solution which was
neither truly Euclidean nor truly Lorentzian.

It is evident from \CAUSTIC\ that for small values of $\ph$ the caustic
begins to deviate from the curve $a^2V=1$. In fact, one finds$^{\GrRoA,\Luk}$
that for sufficiently small values of $\ph$ Lorentzian universes never nucleate
at all. Accordingly, a portion of the shaded region in the conformal diagrams
should be excised about the $\ph=0$ axis in the vicinity of $\ii^+$. The
cutoff, $\ph\Z*$, at which this occurs has been estimated by Lukas$^{\Luk}$ as
being
$$\ph\Z*\simeq p+\left\{2p\left[1-(1+\e^{-1})^{-1/2}\right]\right\}^{1/2},\eqn
\cutLukas$$
in the case of the chaotic potentials, $V=\la\ph^{2p}$. Provided that the
interpretation of \probinflat\ remains valid\foot{\spadesuit}{As a cautionary
note, one should observe that since the approximation of the caustic by the
curve $a^2V=1$ breaks down for small $\ph$, the use of the surfaces $a=\const$
which was assumed in \probinflat\ may not be appropriate for small $\ph$, and
ideally one should determine the probability flux across the caustic itself.
Of course, in a more careful analysis one would consider complex solutions to
the field equations in line with [\Lyo], rather than real Euclidean and real
Lorentzian solutions with a junction condition. One would hope that this would
not alter the conclusions of the analysis much.}, we should therefore set $\ph
\Z1=\ph\Z*$ as the lower bound in the integrals. In the case of the quadratic
potential, $V=m^2\ph^2$, we then have $\ph\Z1\simeq1.5$, and using the values
of $\psuf$ and $\ph\Z2$ found earlier it is straightforward to check
that for typical values of $m\ll1$, \probinflat\ gives $\PP\ll1$ for
$\PS\Ns{HH}$ and $\PP\goesas1$ for $\PS\Ns{V}$ in accordance with the earlier
discussion.

The above analysis would appear to indicate that the no-boundary wavefunction
effectively ``predicts'' a value of $\ph\simeq\ph\Z*$, which in the case of
the quadratic potential is unfortunately less than $\psuf$. However, the
calculation is model-dependent, and if one could find a potential for which
$\ph\Z*\gsim\psuf$ then both $\PS\Ns{HH}$ and $\PS\Ns{V}$ would yield $\PP
\goesas1$. If we compare \efold\ and \cutLukas\ (for $N_e\simeq65$) we see
that in the case of the chaotic potentials, $V=\la\ph^{2p}$, this requirement
is equivalent to $p\gsim62$. However, it requires an enormously small value of
$\la$ to keep such a $V$ below the Planck scale and this is not promising.
Lukas$^{\Luk}$ has also estimated the value of $\ph\Z*$ for some other
potentials but did not find any candidates with $\ph\Z*\gsim\psuf\;$. However,
given the model-dependence of the calculations it cannot be ruled out that some
other potential, or the coupled effects of two or more scalar fields, might
give $\ph\Z*\gsim\psuf$ and thereby make inflation a ``prediction'' of the
no-boundary wavefunction.

\subsec{The origin of density perturbations}

It is expected that the anisotropies in the cosmic microwave background
radiation, which were first definitively observed in 1992, have their origin in
quantum fluctuations in the very early universe. Such fluctuations can be
described using the formalism of quantum field theory in curved spacetime.
Since the era of quantum cosmology is in a sense prior to that in which quantum
field theory in curved spacetime is applicable, one would hope to trace the
origin of the primordial perturbations back to quantum cosmology. Indeed,
this can be done$^{\HalHaw\hS{\WadA}\HaDE}$ and I will very briefly outline the
main results.

With homogeneous minisuperspaces as a starting point, one can add small
inhomogeneous perturbations to the metric and matter fields:
$$\eqalignno{h_{ij}(x,t)&=a^2(t)\left(\OM_{ij}+\ee_{ij}\right),&\Eqn
\pertA\cr \PH(x,t)&=\PH\Z0(t)+\de\PH(x,t),&\Eqn\pertB\cr \NN(x,t)&=\NN\Z0(t)+
\de\NN(x,t),&\Eqn\pertC\cr
\NN\Z i(x,t)&=0+\de\NN\Z i(x,t).&\Eqn\pertD\cr}$$
Here we have restricted attention to the $k=+1$ FRW minisuperspace model, the
subscript zero denotes the unperturbed quantities, and $\OM_{ij}$ is the
unperturbed standard round metric on the 3-sphere. The perturbations can be
expanded in terms of spherical harmonics on the 3-sphere$^{\HalHaw}$. If one
substitutes the ansatz \pertA--\pertD\ into the classical action \actionD\ and
expands to quadratic order one obtains an action which can be split into the
original minisuperspace action, $S\Z0$, and an additional action, $S\Z2$,
quadratic in the perturbations,
$$S=S\Z0[\qA,\NN\Z0]+S\Z2[\qA,\NN\Z0,\ee_{ij},\de\PH,\de\NN,\NN_i],\eqn\actionP
$$
and a corresponding Hamiltonian
$$H=\NN\Z0\left(\hh\Z0+\int\dd^3x\,\HH\Z2\right)
+\int\dd^3x\,\de\NN\,\HH\Z1+\int\dd^3x\,\de\NN_i\,\HH^i\,,\eqn\Pertham$$
where $\hh\Z0$ is the unperturbed Hamiltonian \mFRWa, and the terms $\HH\Z1$
and $\HH\Z2$ are linear and quadratic in the perturbations respectively. There
is now a non-trivial momentum constraint at each point $x\in\SI$, $\HH^i(x)=0$,
while the Hamiltonian constraint splits into a piece linear in the
perturbations, $\HH\Z1(x)=0$, plus a homogeneous piece
$$\hh\Z0+\hh\Z2\equiv\hh\Z0+\int\dd^3x\,\HH\Z2=0.\eqn\hamhom$$

We may quantise \hamhom\ in the standard fashion to obtain a modified
Wheeler-Dewitt equation
$$\hat\hh\PS=\left[-\half\Lap^2+\UU(q)+\hat\hh\Z2\right]\PS=0,\eqn\pertWdW$$
in place of \miniWdW, where the Laplacian is still defined in terms of the
original minisuperspace coordinates according to \miniLap, and $\hat\hh\Z2$ is
a second order differential operator which results from quantisation of the
homogeneous part of the perturbations. In the case of pure scalar field modes,
for example, the perturbations may be decomposed as
$$\de\PH(x,t)={1\over\si}\sum\limits_{nml}\ff_{nlm}(t)Q\UD n{lm}(x),\eqn\pertE
$$
where the $Q\UD n{lm}$ satisfy the 3-dimensional Laplace equation on $S^3$,
$$\!\W{(3)}\!\DE Q\UD n{lm}=-(n^2-1)Q\UD n{lm},\eqn\pertF$$
and one finds$^{\HalA,\HalD,\HalHaw}$
$$\hat\hh\Z2=\half\sum\limits_{nml}\left\{{-1\over\;a^3}\DDer\pt{\ff_{nlm}
}2+\left[(n^2-1)a+m^2a^3\right]\ff_{nlm}^2\right\}\eqn\hohum$$
in the case of the quadratic scalar potential.

It is possible to find solutions to the Wheeler-DeWitt equation \pertWdW\ in
which the minisuperspace coordinates, $\qA$, are treated semiclassically in
the WKB approximation, while the perturbations are treated quantum
mechanically. One can show$^{\HalA,\HalD,\HalHaw,\WadA}$ that the solutions
take the form
$$\PS=\CC(q,\PH)\,\e^{i\SS\X0(q,\PH)}\,\tilde\pS,\eqn\pspert$$
where
$\SS\Z0$ satisfies the unperturbed Hamilton-Jacobi equation \wkbHJ, the
prefactor $\CC$ depends only on the unperturbed minisuperspace coordinates, and
the functions $\tilde\pS$ satisfy the functional Schr\"odinger equation
$$i\Deriv\pt t{\tilde\pS}=\hat\hh\Z2\tilde\pS.\eqn\pertschrod$$
The different modes of the scalar perturbations \pertE\ do not interact, for
example, and in this case
$$\tilde\pS=\prod\limits_{nml}\pS_{nlm}(t,\ff_{nlm}),\eqn\pertprod$$
where each mode $\pS_{nlm}$ separately satisfies \pspert\ with $\hat\hh\Z2$
given by \hohum.

The wavefunction \pspert\ is thus peaked about classical
trajectories, with corrections, $\tilde\pS$, which satisfy the functional
Schr\"odinger equation along these trajectories. This is in fact precisely the
starting point for the treatment of matter modes in a curved spacetime
background using the formalism of quantum field theory in curved spacetime.
Thus the above result is important in that it demonstrates that quantum
cosmology is consistent with the standard approach to the quantum treatment of
cosmological perturbations. As an added bonus the imposition of a boundary
condition on $\PS$, such as the no-boundary or tunneling condition, will
result in the choice of particular solutions of the functional Schr\"odinger
equation, and consequently a particular vacuum state for the matter modes.
Both the no-boundary condition and the tunneling condition pick out$^{\VilC,
\LafA\hS{\WadB,\Vach}\VaVi}$ a de-Sitter invariant state known as the
``Euclidean'' or Bunch-Davies vacuum$^{\BuDa,\Alle}$, which is the state that
is often assumed in cosmological calculations of density perturbations. In
fact, this state is picked out by many boundary conditions$^{\WadB,\Vach}$, and
thus could be regarded as a natural quantum state for matter in quantum
cosmology.

One may solve the functional Schr\"odinger equation and study the growth$^{
\HalHaw}$ of the modes. As discussed above, the results obtained are the same
as those which are found if one begins with quantum matter fields in de-Sitter
invariant vacuum states in background inflationary spacetimes$^{\HawD,\GuPi}$.
In particular, one obtains a scale-free spectrum of density perturbations which
act as seeds for the formation of galaxies and other structures in the
universe. Such a spectrum accords well with the spectrum deduced from the
COBE measurements of cosmic microwave anisotropies$^{\COBE}$.

\subsec{The arrow of time}

The question of the origin of the arrow of time in the face of the
CPT-invariance of the laws of physics is one of the deepest unresolved
conceptual issues in theoretical physics, and it has been the subject of more
than one major conference$^{\HaPMZu,\GolA}$. The question of the nature of time
assumes prime importance in quantum gravity and quantum
cosmology\foot{\heartsuit}{The conceptual issues surrounding the nature of time
in quantum cosmology would entail a series of lectures in themselves. For
further discussion see, e.g., [\SmSo], [\HiWa]$\hS{\BeFe}$[\Mar] and references
therein.}. One might argue that if the basic laws of physics are time
symmetric, then the arrow of time could have its origin in a boundary condition
for the wavefunction of the universe.

There are many different arrows of time observed in the universe. Some of these
such as the psychological arrow of time (we remember only the past), and the
electromagnetic arrow of time (the choice of retarded as opposed to advanced
solutions of Maxwell's equations), could well be argued to be consequences of
the thermodynamic arrow of time, which arises from the second law of
thermodynamics. However, the expansion of the universe provides an alternate
cosmological arrow of time which is not obviously directly related to the
increase of entropy. Whether such a relationship does exist is a question which
might potentially be resolved by quantum cosmology.

In 1985 Hawking$^{\HawE}$ proposed that the thermodynamic arrow and
cosmological arrows of time were correlated: that is to say if the universe
were spatially closed then entropy would decrease in the contracting
phase\foot{\circledcirc}{This idea was first suggested by Gold [\GolB], and
has resurfaced a number of times in different contexts.}.
In arriving at this proposal he was influenced by the fact that the
no-boundary wavefunction is CPT-invariant, and also by some early studies of
simple minisuperspace models which possessed quasi-periodic solutions$^{\HaWu}$
which ``bounced'' instead of recontracting to a singularity as $a\to0$.
However, Hawking soon changed his mind on the issue, which he terms his
``greatest mistake''$^{\HawF}$. This change of mind was brought about by a
number of factors. Firstly, Page pointed out that CPT-invariance of the
no-boundary wavefunction does not imply CPT-invariance for an individual WKB
component of the wavefunction, which would correspond to the history of a
classical universe$^{\PagF}$. Furthermore, minisuperspace models which were
subsequently studied, such as that of the Kantowski-Sachs universe, were found
not to admit bounce solutions but always possessed singularities to the
future$^{\LaSh,\LafB}$.
Finally, as was mentioned in \S4.1, the bounce solutions do not feature even in
simple minisuperspace models once one considers the contribution of complex
metrics$^{\Lyo}$. In general, the approximate minisuperspace boundary condition
that $\PS\to1$ as $a\to0$ must be altered to allow for approximately Euclidean
metrics which also contribute a rapidly oscillating component to the
wavefunction$^{\Lyo,\HaLL}$.

Hawking, Laflamme and Lyons$^{\HaLL}$ have recently argued that a thermodynamic
arrow of time results from the imposition of the Hartle-Hawking boundary
condition. More precisely, they have considered the evolution of primordial
fluctuations as outlined in \S5.2, but accounting for the ``approximately
Euclidean'' geometries$^{\Lyo}$ which appear to be required if one considers
complex metrics. They find$^{\HaLL}$ that the gravitational wave perturbations
have an amplitude that remains in the linear regime and is roughly
time-symmetric about the time of maximum expansion. Such perturbations cannot
be said to give rise to an arrow of time. Density perturbations behave
differently, however. They start out small but grow large and become non-linear
as the universe expands, and moreover this growth continues during the
contracting phase of the spatially closed universe. This growth of
inhomogeneity therefore provides an arrow of time which could be considered to
be an essentially thermodynamic arrow. Since it does not match the cosmological
arrow in the contracting phase, the only reason for the coincidence of the two
arrows in the present epoch would appear to be an anthropic one. In particular,
the conditions that would prevail in a contracting phase would appear to
preclude the existence of life$^{\HawF,\HaLL}$. Thus the fact that we are
around to make observations means we must find ourselves in a cosmological
epoch in which the two arrows coincide.

It should be added that the debate about whether the cosmological and
thermodynamic arrows of time coincide has not yet been closed, however, and it
is still maintained by Kiefer and Zeh$^{\KiZe}$ that the boundary condition on
the Wheeler-DeWitt equation must be such that the thermodynamic arrow would
reverse in a recontracting universe.
\goodbreak
\newsec{Conclusion}

I hope to have shown you that although research in quantum
cosmology is still rather speculative and open-ended, its framework nonetheless
has the potential not only to provide answers to questions surrounding the
origin and early evolution of the universe, but also to help us unravel the
mysteries of quantum gravity. Quantum cosmology is a field in which a great
deal remains to be done, and the results of \S5 should be regarded as a
hopeful indication of the types of predictions we might hope to make. It is
too early to draw definitive conclusions about the relative merits of the
various boundary condition proposals. The results of \S5.1 ostensibly favour
the ``tunneling proposal''. However, this conclusion is only based on a few
simple models, and therefore some caution must be exercised.

As a final note, it is worth mentioning that recent results show that
supersymmetry provides a means of restricting possible boundary conditions
for the Wheeler-DeWitt equation, or the corresponding Dirac square-root
equation. The supergravity constraint equations for various homogeneous
minisuperspace models appear to be so restrictive that they only pick out the
most symmetric quantum states$^{\Grah,\DeHO,\Dea}$. In particular, simple
analytic solutions for $\PS$ in the supersymmetric Bianchi-IX
minisuperspace have been found$^{\Grah}$ in the empty and filled fermion
sectors, which have a natural interpretation as$^{\Dea}$ wormhole states$^
{\HaPaC,\HawG}$, or as$^{\GrLu}$ Hartle-Hawking no-boundary states. Some doubts
were initially cast on the relevance of these solutions, as the states appear
to have no counterpart in 4-dimensional supergravity$^{\dWNM,\CFOP}$. However,
more recent work$^{\CsGr}$ on supersymmetric minisuperspaces corresponding to
Bianchi ``class A''$^{\MacC}$ models indicates that infinitely many physical
states with finite (even) fermion number can be found, and these are direct
analogues of physical states in full supergravity. This result shows
that such minisuperspace models are likely to be physically very important for
quantum cosmology.

\bigskip\noindent{\bf Acknowledgement}\quad I would like to thank A. Lukas for
giving me permission to use \CAUSTIC, which originally appeared in [\Luk].
\par\bigbreak\bigskip
\vskip0pt plus\refminspace\penalty-400\vskip0pt plus-\refminspace
\footatend\immediate\closeout\rfile\writestoppt \baselineskip=14pt
\leftline{{\bf References}}\bigskip\message{References}{\frenchspacing%
\refskip\escapechar=` \input refs.tmp\par}
\nonfrenchspacing \leftskip=0pt
\listfigs
\bye

%% file: symbols.tex
\input amssym.def \input amssym
\normalbaselineskip=14pt plus.2pt minus.1pt
\font\sixrm=cmr6 \font\sixi=cmmi6 \font\sixsy=cmsy6 \font\sixbf=cmbx6
\font\sixmsa=msam6 \font\sixmsb=msbm6 \font\sixeufm=eufm6
\font\eightrm=cmr8 \font\eighti=cmmi8 \font\eightsy=cmsy8 \font\eightbf=cmbx8
\font\eighttt=cmtt8 \font\eightit=cmti8 \font\eightsl=cmsl8 \font\eightex=cmex8
\font\eightmsa=msam8 \font\eightmsb=msbm8 \font\eighteufm=eufm8
\font\ninerm=cmr9 \font\ninei=cmmi9  \font\ninesy=cmsy9 \font\ninebf=cmbx9
\font\ninett=cmtt9 \font\ninesl=cmsl9 \font\nineit=cmti9 \font\nineex=cmex9
\font\ninemsa=msam9 \font\ninemsb=msbm9 \font\nineeufm=eufm9
\font\sevenit=cmti7 \font\sevenex=cmex7
\font\fiveeurm=eurm5\font\sixeurm=eurm6\font\seveneurm=eurm7
\font\eighteurm=eurm8\font\nineeurm=eurm9\font\teneurm=eurm10
\font\bold=cmbx10 scaled\magstep1 \font\ser=cmssi10
\font\bm=cmmib10 \font\bms=cmbsy10 \newskip\ttglue
\newfam\eurmfam \textfont\eurmfam=\teneurm \scriptfont\eurmfam=\seveneurm
\scriptscriptfont\eurmfam=\fiveeurm

\def\loadBigten#1{\catcode`\@=11\def\TpT{scaled#1}
\font\Fiverm=cmr5 \TpT \font\Fivei=cmmi5 \TpT \font\Fivesy=cmsy5 \TpT
\font\Fivebf=cmbx5 \TpT \font\Fivemsa=msam5 \TpT \font\Fivemsb=msbm5 \TpT
\font\Fiveeufm=eufm5 \TpT \font\Fiveeurm=eurm5 \TpT
\font\Sevenrm=cmr7 \TpT \font\Seveni=cmmi7 \TpT \font\Sevensy=cmsy7 \TpT
\font\Sevenbf=cmbx7 \TpT \font\Sevenit=cmti7 \TpT \font\Eightrm=cmr8 \TpT
\font\Sevenmsa=msam7 \TpT \font\Sevenmsb=msbm7 \TpT \font\Seveneufm=eufm7 \TpT
\font\Seveneurm=eurm7 \TpT
\font\Tenrm=cmr10 \TpT \font\Teni=cmmi10 \TpT \font\Tensy=cmsy10 \TpT
\font\Tenbf=cmbx10 \TpT \font\Tenit=cmti10 \TpT \font\Tentt=cmtt10 \TpT
\font\Tensl=cmsl10 \TpT \font\Tenex=cmex10 \TpT \font\Tenmsa=msam10 \TpT
\font\Tenmsb=msbm10 \TpT \font\Teneufm=eufm10 \TpT \font\Teneurm=eurm10 \TpT
\font\bm=cmmib10 \TpT \font\bms=cmbsy10 \TpT
\def\Tenbig#1{{\hbox{$\left#1\vbox to8.5pt{}\right.\n@space$}}}
\def\Bigpoint{\def\rm{\fam0\Tenrm}
\textfont0=\Tenrm \scriptfont0=\Sevenrm \scriptscriptfont0=\Fiverm
\textfont1=\Teni \scriptfont1=\Seveni \scriptscriptfont1=\Fivei
\textfont2=\Tensy \scriptfont2=\Sevensy \scriptscriptfont2=\Fivesy
\textfont3=\Tenex \scriptfont3=\Tenex \scriptscriptfont3=\Tenex
\textfont\itfam=\Tenit \def\it{\fam\itfam\Tenit}%
\textfont\slfam=\Tensl \def\sl{\fam\slfam\Tensl}%
\textfont\ttfam=\Tentt \def\tt{\fam\ttfam\Tentt}%
\textfont\bffam=\Tenbf \scriptfont\bffam=\Sevenbf
 \scriptscriptfont\bffam=\Fivebf \def\bf{\fam\bffam\Tenbf}%
\textfont\msafam=\Tenmsa\scriptfont\msafam=\Sevenmsa
 \scriptscriptfont\msafam=\Fivemsa
\textfont\msbfam=\Tenmsb \scriptfont\msbfam=\Sevenmsb
 \scriptscriptfont\msbfam=\Fivemsb
\textfont\eufmfam=\Teneufm \scriptfont\eufmfam=\Seveneufm
 \scriptscriptfont\eufmfam=\Fiveeufm
\textfont\eurmfam=\Teneurm \scriptfont\eurmfam=\Seveneurm
 \scriptscriptfont\eurmfam=\Fiveeurm
\tt \ttglue=.5em plus.25em minus.15em
\setbox\strutbox=\hbox{\vrule height8.5pt depth3.5pt width0pt}
\let\sc=\Eightrm \let\big=\Tenbig \rm} \catcode`\@=12} 

\def\tenpoint{\def\rm{\fam0\tenrm}
\textfont0=\tenrm \scriptfont0=\sevenrm \scriptscriptfont0=\fiverm
\textfont1=\teni \scriptfont1=\seveni \scriptscriptfont1=\fivei
\textfont2=\tensy \scriptfont2=\sevensy \scriptscriptfont2=\fivesy
\textfont3=\tenex \scriptfont3=\tenex \scriptscriptfont3=\tenex
\textfont\itfam=\tenit \def\it{\fam\itfam\tenit}%
\textfont\slfam=\tensl \def\sl{\fam\slfam\tensl}%
\textfont\ttfam=\tentt \def\tt{\fam\ttfam\tentt}%
\textfont\bffam=\tenbf \scriptfont\bffam=\sevenbf
 \scriptscriptfont\bffam=\fivebf \def\bf{\fam\bffam\tenbf}%
\textfont\msafam=\tenmsa\scriptfont\msafam=\sevenmsa
 \scriptscriptfont\msafam=\fivemsa
\textfont\msbfam=\tenmsb \scriptfont\msbfam=\sevenmsb
 \scriptscriptfont\msbfam=\fivemsb
\textfont\eufmfam=\teneufm \scriptfont\eufmfam=\seveneufm
 \scriptscriptfont\eufmfam=\fiveeufm
\textfont\eurmfam=\teneurm \scriptfont\eurmfam=\seveneurm
 \scriptscriptfont\eurmfam=\fiveeurm
\tt \ttglue=.5em plus.25em minus.15em
\setbox\strutbox=\hbox{\vrule height8.5pt depth3.5pt width0pt}
\let\sc=\eightrm \let\big=\tenbig \rm} 

\def\ninepoint{\def\rm{\fam0\ninerm}
\textfont0=\ninerm \scriptfont0=\sixrm \scriptscriptfont0=\fiverm
\textfont1=\ninei \scriptfont1=\sixi \scriptscriptfont1=\fivei
\textfont2=\ninesy \scriptfont2=\sixsy \scriptscriptfont2=\fivesy
\textfont3=\nineex \scriptfont3=\nineex \scriptscriptfont3=\nineex
\textfont\itfam=\nineit \def\it{\fam\itfam\nineit}%
\textfont\slfam=\ninesl \def\sl{\fam\slfam\ninesl}%
\textfont\ttfam=\ninett \def\tt{\fam\ttfam\ninett}%
\textfont\bffam=\ninebf \scriptfont\bffam=\sixbf
 \scriptscriptfont\bffam=\fivebf \def\bf{\fam\bffam\ninebf}%
\textfont\msafam=\ninemsa\scriptfont\msafam=\sixmsa
 \scriptscriptfont\msafam=\fivemsa
\textfont\msbfam=\ninemsb \scriptfont\msbfam=\sixmsb
 \scriptscriptfont\msbfam=\fivemsb
\textfont\eufmfam=\nineeufm \scriptfont\eufmfam=\sixeufm
 \scriptscriptfont\eufmfam=\fiveeufm
\textfont\eurmfam=\nineeurm \scriptfont\eurmfam=\sixeurm
 \scriptscriptfont\eurmfam=\fiveeurm
\tt \ttglue=.5em plus.25em minus.15em
\setbox\strutbox=\hbox{\vrule height7.75pt depth2.75pt width0pt}%
\let\sc=\sevenrm \let\big=\ninebig \rm}

\def\eightpoint{\def\rm{\fam0\eightrm}
\textfont0=\eightrm \scriptfont0=\sixrm \scriptscriptfont0=\fiverm
\textfont1=\eighti \scriptfont1=\sixi \scriptscriptfont1=\fivei
\textfont2=\eightsy \scriptfont2=\sixsy \scriptscriptfont2=\fivesy
\textfont3=\eightex \scriptfont3=\eightex \scriptscriptfont3=\eightex
\textfont\itfam=\eightit \def\it{\fam\itfam\eightit}%
\textfont\slfam=\eightsl \def\sl{\fam\slfam\eightsl}%
\textfont\ttfam=\eighttt \def\tt{\fam\ttfam\eighttt}%
\textfont\bffam=\eightbf \scriptfont\bffam=\sixbf
 \scriptscriptfont\bffam=\fivebf \def\bf{\fam\bffam\eightbf}%
\textfont\msafam=\eightmsa\scriptfont\msafam=\sixmsa
 \scriptscriptfont\msafam=\fivemsa
\textfont\msbfam=\eightmsb \scriptfont\msbfam=\sixmsb
 \scriptscriptfont\msbfam=\fivemsb
\textfont\eufmfam=\eighteufm \scriptfont\eufmfam=\sixeufm
 \scriptscriptfont\eufmfam=\fiveeufm
\textfont\eurmfam=\eighteurm \scriptfont\eurmfam=\sixeurm
 \scriptscriptfont\eurmfam=\fiveeurm
\tt \ttglue=.5em plus.25em minus.15em
\setbox\strutbox=\hbox{\vrule height7pt depth2pt width0pt}%
\let\sc=\sixrm \let\big=\eightbig \rm}

\def\sevenpoint{\def\rm{\fam0\sevenrm}
\textfont0=\sevenrm \scriptfont0=\fiverm \scriptscriptfont0=\fiverm
\textfont1=\seveni \scriptfont1=\fivei \scriptscriptfont1=\fivei
\textfont2=\sevensy \scriptfont2=\fivesy \scriptscriptfont2=\fivesy
\textfont3=\sevenex \scriptfont3=\sevenex \scriptscriptfont3=\sevenex
\textfont\itfam=\sevenit \def\it{\fam\itfam\sevenit}%
\textfont\bffam=\sevenbf \scriptfont\bffam=\fivebf
 \scriptscriptfont\bffam=\fivebf \def\bf{\fam\bffam\sevenbf}%
\tt \ttglue=.5em plus.25em minus.15em
\setbox\strutbox=\hbox{\vrule height6.5pt depth1.5pt width0pt}%
\let\sc=\fiverm \let\big=\sevenbig \rm}

\catcode`\@=11
\def\tenbig#1{{\hbox{$\left#1\vbox to8.5pt{}\right.\n@space$}}}
\def\ninebig#1{{\hbox{$\textfont0=\tenrm\textfont2=\tensy
\left#1\vbox to7.25pt{}\right.\n@space$}}}
\def\eightbig#1{{\hbox{$\textfont0=\ninerm\textfont2=\ninesy
\left#1\vbox to6.5pt{}\right.\n@space$}}}
\def\sevenbig#1{{\hbox{$\textfont0=\eightrm\textfont2=\eightsy
\left#1\vbox to5.5pt{}\right.\n@space$}}}
\catcode`\@=12

\def\captionpoint{\eightpoint\baselineskip=11pt} \def\footpoint{\captionpoint}
\def\normalpoint{\tenpoint \normalbaselines} \normalpoint
\def\foot#1#2{\footnote{\spaceskip=0pt$^{#1}$\spaceskip=0pt}{{\footpoint #2}%
\normalpoint\openup1\jot}} \def\br{\hfil\break} \def\rarr{\rightarrow}
\def\scrscr{\scriptscriptstyle} \def\ul{\underline} 
\def\undertilde#1{\smash{\hbox{\raise0.91666pt\hbox{$\mathop{#1}\limits_{\raise
2pt\hbox{$\scr\sim$}}$}}}\vphantom{\ul{#1}}\vphantom{#1}}
\def\sundertilde#1{\smash{\hbox{\hskip-1pt\hbox{$\scrscr\mathop{#1}\limits_{
\raise2pt\hbox{$\scrscr\sim$}}$}}}\vphantom{\ul{#1}}\vphantom{#1}}
 
 \def\hence{\Rightarrow}
 \def\begincaption{\medskip\openup-1\jot\eightpoint}
\def\endcaption{\tenpoint\openup1\jot\leftskip=0pt\rightskip=0pt}
\def\caption#1#2{\message{#1}\begincaption\leftskip=15true mm\rightskip=15true
mm\vbox{\halign{\vtop{\parindent=0pt\parskip=0pt\strut##\strut}\cr{\bf#1}\quad
#2\cr}}\endcaption}    \mathchardef\crss"0202\def\cross{\hbox{\lower.2ex\hbox{$\crss$}}}
\def\dsp{\displaystyle} \def\scr{\scriptstyle} 
 \def\ul{\underline}
\def\sqr#1#2#3{{\vbox{\hrule height.#2pt \hbox{\vrule width.#2pt
height#1pt\kern#1pt\vrule width.#2pt}\hrule height.#2pt}\hbox{\hskip.#3em}}}
 \def\et{\eta}
\def\tick{\hbox{\lower.37ex\hbox{$\scrscr\vee$}\hskip-.293em{\sevenit/}}}
\def\lsim{\mathop{\hbox{${\lower3.8pt\hbox{$<$}}\atop{\raise0.2pt\hbox{$\sim$}}
$}}} \def\gsim{\mathop{\hbox{${\lower3.8pt\hbox{$>$}}\atop{\raise0.2pt\hbox{$
\sim$}}$}}} \def\goesas{\mathop{\sim}\limits}

 \def\hs{\hbox{--}} \def\Hs#1{\hs\vphantom{#1}}
\def\W#1{^{\raise2pt\hbox{$\scrscr#1$}}} \def\Y#1{^{\raise2pt\hbox{$\scr#1$}}}
\def\X#1{_{\lower2pt\hbox{$\scrscr#1$}}} \def\Z#1{_{\lower2pt\hbox{$\scr#1$}}}
\def\ns#1{_{\hbox{\sevenrm #1}}} \def\z#1{_{\lower1pt\hbox{$\scr#1$}}}
\def\Ns#1{\Z{\hbox{\sevenrm #1}}} 
\def\w#1{\;\hbox{#1}\;} 
\def\inftyp{{\hbox to 4.30556pt{\hfil}\infty\hbox to2pt{\hfil}\prime}}
\mathchardef\hash="015D \def\Lap{\nabla} \def\const{\hbox{const}}
\def\Cop{\hbox{$C$\hskip-.7em{\raise.52ex\hbox{{\sixbf/}}}\hskip.17em}}
\def\Oop{\hbox{$O$\hskip-.7em{\raise.52ex\hbox{{\sixbf/}}}\hskip.17em}}
 \def\Rop{\hbox{$I$\hskip-0.32em$R$}}
 \def\Zop{\hbox{{\ser Z\hskip-.4em Z}}}
\def\al{\alpha}\def\be{\beta}\def\de{\delta}
\def\ee{\varepsilon}\def\ka{\kappa}\def\th{\theta}\def\ph{\phi}\def\PH{\Phi}
\def\varch{{\raise.4516ex\hbox{$\chi$}}}\def\la{\lambda}
\def\si{\sigma}\def\SI{\Sigma}\def\pt{\partial}
\def\DE{\Delta}\def\ta{{\tau}}\def\OO{{\rm O}}\def\e{{\rm e}}
\def\OM{\Omega}\def\dd{{\rm d}}
\def\LA{\Lambda} \def\ps{\psi}\def\PS{\Psi} 
\def\hence{\Rightarrow} \def\B#1{\hbox{\bm#1}}

 \def\Bom{\B{\char33}} 
  
  \def\CB#1{\hbox{\bms#1}}
\def\LAP{\CB{\char114}}  \def\GA{\Gamma}

\def\Mi#1{{\cal M}^{#1}}

 \def\PL#1{Phys.\
Lett.\ {\bf#1}}\def\CMP#1{Commun.\ Math.\ Phys.\ {\bf#1}} \def\PRL#1{Phys.\
Rev.\ Lett.\ {\bf#1}}\def\AP#1#2{Ann.\ Phys.\ (#1) {\bf#2}} \def\PR#1{Phys.\
Rev.\ {\bf#1}}\def\CQG#1{Class.\ Quantum Grav.\ {\bf#1}}\def\NP#1{Nucl.\ Phys.\
{\bf#1}}\def\GRG#1{Gen.\ Relativ.\ Grav.\ {\bf#1}} \def\JMP#1{J.\ Math.\ Phys.\
{\bf#1}}\def\PTP#1{Prog.\ Theor.\ Phys.\ {\bf#1}} 
\def\PRS#1{Proc.\ R. Soc.\ Lond.\ {\bf#1}}\def\NC#1{Nuovo Cimento {\bf#1}}
\def\JP#1{J.\ Phys.\ {\bf#1}} \def\IJMP#1{Int.\ J. Mod.\ Phys.\ {\bf #1}}
\def\MPL#1{Mod.\ Phys.\ Lett.\ {\bf #1}} 
\def\AIHP#1{Ann.\ Inst.\ H. Poincar\'e {\bf#1}}

 \def\frac#1#2{{\textstyle{#1\over#2}}}
\def\Der#1#2{{#1\hphantom{#2}\over#1#2}}\def\Deriv#1#2#3{{#1#3\over#1#2}}
\def\DDer#1#2#3{{#1^{#3}\hphantom{#2}\over#1#2^{#3}}}
\def\UD#1#2{^{#1}_{\hphantom{#1}#2}} \def\DU#1#2{_{#1}^{\hphantom{#1}#2}}

\def\eqname#1#2{\xdef #1{(\secsym\the\meqno)}\writedef{#1\leftbracket#1}%
\global\advance\meqno by1#2#1\eqlabeL#1}
\def\eqn#1{\eqname#1\eqno} \def\Eqn#1{\eqname#1{}} \def\ref{\the\refno\nref}
\def\nref#1{\xdef#1{\the\refno}\writedef{#1\leftbracket#1}%
\ifnum\refno=1\immediate\openout\rfile=refs.tmp\fi
\global\advance\refno by1\chardef\wfile=\rfile\immediate
\write\rfile{\noexpand\item{#1.\ }\reflabeL{#1\hskip.31in}\pctsign}\findarg}

\newdimen\refminspace \refminspace=36pt \def\refskip{\parindent=20pt}

%

\def\fig#1{\dfig#1 #1}
\ifx\epsfbox\UnDeFiNeD\message{(NO epsf.tex, FIGURES WILL BE IGNORED)}
\def\dfig#1{\xdef#1{Fig.~\the\figno}%
\ifnum\figno=1\immediate\openout\ffile=figs.tmp\fi\chardef\wfile=\ffile%
\global\advance\figno by1}
\def\nfig#1{\writedef{#1\leftbracket \noexpand#1}%
\immediate\write\ffile{\noexpand\medskip\noexpand\item{\noexpand #1 }
\reflabeL{#1\hskip.55in}\pctsign}\findarg}
\def\ifig#1#2#3{\nfig#1{#2}} \def\Ifig#1#2#3{\dfig#1\nfig#1{#2}}
 \def\Pfig#1#2#3{\dfig#1\nfig#1{#2}}
\def\listfigs{\par\bigbreak\bigskip 
\vskip0pt plus\refminspace\penalty-400\vskip0pt plus-\refminspace
\immediate\closeout\ffile{\parindent=40pt
\baselineskip14pt\leftline{{\bf Figure Captions}}\nobreak\medskip
\leftskip=7truemm\escapechar=` \input figs.tmp\par}
\leftskip=0pt}

\else \message{(FIGURES WILL BE INCLUDED)}
\def\dfig#1{\xdef#1{Fig.~\the\figno}\global\advance\figno by1}
\def\mfig#1#2#3#4{\goodbreak#4\centerline{#3}%
\smallskip\centerline{\vbox{\baselineskip12pt \leftskip=36pt\rightskip=36pt
\captionpoint \noindent{\bf #1:} #2}}\endinsert \normalpoint}
\def\ifig#1#2#3{\mfig#1{#2}{#3}{\midinsert}}

\def\Ifig#1#2#3{\dfig#1\ifig#1{#2}{#3}}
\def\Pfig#1#2#3{\dfig#1\ifig#1{#2}{#3}}
\def\listfigs{} \fi

\def\title#1{\halign{\bold\centerline{##}\hfil\cr#1\crcr}\vskip5mm}

\def\abstract{\vskip15mm{\bf\centerline{Abstract}}\vskip8mm \leftskip=1true
cm\rightskip=1true cm}
